%% file: latex.tex
\documentclass[12pt]{article}

\usepackage{amsmath,amssymb,latexsym,theorem,epsfig,amscd}
\usepackage[all]{xy}
\CompileMatrices

\newcommand{\be}{\begin{equation}}
\newcommand{\ee}{\end{equation}}

\newcommand{\rank}{{\text{ rank} }}

\newcommand{\cp}[1]{{\mathbb P}^{#1}}
\newcommand{\op}[1]{\operatorname{#1}}
\newcommand{\ocp}[1][]{{\mathcal O}_{\cp{1}}{#1}}
\newcommand{\oy}[1][]{{\mathcal O}_B{#1}}
\newcommand{\oz}[1][]{{\mathcal O}_{B^{'}}{#1}}
\newcommand{\ox}[1][]{{\mathcal O}_X{#1}}

\newcommand{\les}[8]{\xymatrix{       &      & ...  \ar[r]  &  {#1}    \ar@{->} `r[d] `[l] `^dl[dlll]  `^dr/14pt[dll]    [dll] \\
&  {#2} \ar[r] & {#3} \ar[r] & {#4}  \ar `r/10pt[d] `[l]  `^dl[dlll]  `^dr/14pt[dll]   [dll] \\ 
& {#5} \ar[r]  & {#6} \ar[r] & {#7}  \ar `r/10pt[d] `[l]  `^dl[dlll]  `^dr/14pt[dll]   [dll] \\
&  {#8} \ar[r] & ... & & }}

\newcommand{\ts}[4]
{\xymatrix{
0 \ar[r] & {#1} \ar[r] & {#2} \ar[r] & {#3} \ar[r] & 0 \\  
         &            & {#4} \ar[u] \ar@{.>}[ul]^{\alpha} \ar@{.>}[ur]_{\beta}  &            & \\
         &            &  0  \ar[u] &            &
}}

{\theorembodyfont{\rmfamily} }
{\theorembodyfont{\rmfamily} }

\numberwithin{equation}{section}

\begin{document}

\include{chapter1}

\end{document}

%% file: chapter1.tex
\begin{titlepage}

\vspace{-5cm}

\title{
   \hfill{\normalsize hep-th/0303020}\\ \vspace{-0.3cm}
   \hfill{\normalsize UPR-1016-T} \\[1em]
   {\LARGE Invariant Homology on Standard Model Manifolds}
\\
[1em] }
\author{
     Burt A. Ovrut$^1$, Tony Pantev$^2$ and
Ren\'e Reinbacher$^1$ \\[0.5em]
   {\normalsize $^1$Department of Physics, University of Pennsylvania} \\[-0.4em] {\normalsize Philadelphia, PA 19104--6396}\\
   {\normalsize $^2$Department of Mathematics, University of Pennsylvania} \\[-0.4em]
   {\normalsize Philadelphia, PA 19104--6395, USA}\\ }
\date{}
\maketitle
\begin{abstract}
\noindent
Torus-fibered Calabi-Yau threefolds $Z$, with base $d{\mathbb P}_9$ and fundamental group $\pi_1(Z)={\mathbb Z}_2 \times {\mathbb Z}_2 $, are reviewed. It is shown that $Z=X/({\mathbb Z}_2 \times {\mathbb Z}_2)$, where $X=B \times_{\cp{1}} B^{'}$ are elliptically fibered Calabi-Yau threefolds that admit a freely acting ${\mathbb Z}_2 \times {\mathbb Z}_2$ automorphism group. $B$ and $B^{'}$ are rational elliptic surfaces, each with a ${\mathbb Z}_2 \times {\mathbb Z}_2$ group of automorphisms. It is shown that the  ${\mathbb Z}_2 \times {\mathbb Z}_2$ invariant classes of curves of each surface have four generators which produce, via the fiber product, seven ${\mathbb Z}_2 \times {\mathbb Z}_2$ invariant generators in $H_4(X,{\mathbb Z})$. All invariant homology classes are computed explicitly. These descend to produce a rank seven homology group $H_4(Z,{\mathbb Z})$ on $Z$. The existence of these homology  classes on $Z$ is essential to the construction of anomaly free, three family standard-like models with suppressed nucleon decay in both weakly and strongly coupled heterotic superstring theory.

\end{abstract}

\thispagestyle{empty}

\end{titlepage}

\section{Introduction}

In \cite{hw1,hw2,w1}  Ho\v rava  and Witten showed that the simplest vacuum of strongly coupled heterotic superstring theory consists of an eleven dimensional bulk space bounded on either end of the eleventh dimension by a ten dimensional orbifold fixed plane. Each fixed plane supports an $E_8$ supergauge theory on its worldvolume. This theory was compactified on Calabi-Yau threefolds to produce, at low energy, an effective five dimensional bulk space bounded in the fifth dimension by two three-branes. When the Calabi-Yau space supports a non-trivial $E_8$ gauge instanton with structure group $G \subset E_8$, the gauge group on that three-brane will no longer be $E_8$. Rather, it is reduced to a group $H$ which is the commutant of $G$ in $E_8$. In addition to these ``end-of-the-world'' three-branes, there are, generically, five-branes in the bulk space which are wrapped on a holomorphic curve in the Calabi-Yau threefold. This five-dimensional theory with three-branes is called heterotic M-theory. It is a fundamental paradigm for the so called ``brane-world'' theories of particle physics. The basic construction of heterotic M-theory was presented in \cite{losw1,losw2,low1}.

To produce phenomenologically relevant models of particle physics on one of the three-branes, called the ``observable brane'', it is essential to construct the appropriate gauge instantons on the Calabi-Yau threefold. This is similar to  constructing  instantons on a flat ${\mathbb R}^4$ manifold, using for example, the ADHM method \cite{adhm}. Now, however, instantons satisfying the hermitian Yang-Mills equations must be found on a curved Calabi-Yau threefold, a much more challenging problem. The basic ideas for such a construction were presented in \cite{fmw1,fmw2} for elliptically fibered Calabi-Yau threefolds using the equivalent concept of stable, holomorphic vector bundles. Extending this work, it was shown in \cite{low2,dlow} that phenomenologically relevant grand unified theories (GUTs) can, indeed, be constructed on the observable brane. This is accomplished by compactifying Ho\v rava-Witten theory on elliptically fibered Calabi-Yau threefolds with trivial fundamental group that support holomorphic vector bundles with structure group $G=SU(n)$. Choosing $n$ to be $5$ and $4$, for example, leads to GUT groups $H$ of $SU(5)$ and $SO(10)$ respectively.

Constructing  holomorphic  vector bundles that break the $E_8$ gauge theory to standard-like models is more difficult. However, it was shown in \cite{dopw-i,dopw-ii,dopw-iii,dopw-iv} that three-family models with gauge group $H=SU(3)_C\times SU(2)_L \times U(1)_Y$ do exist in heterotic M-theory. This was accomplished as follows. First, elliptically fibered Calabi-Yau threefolds with non-trivial fundamental group ${\mathbb Z}_2$ were constructed. It was found to be convenient to choose the base surface of these threefolds to be $B=d{\mathbb P}_9$. See also \cite{schoenCY}. Second, new methods were introduced for producing stable, holomorphic vector bundles with structure group $G=SU(5)$ over these Calabi-Yau spaces. Such instantons break the $E_8$ group to $SU(5)$, as mentioned above. Now, however, the non-trivial ${\mathbb Z}_2$ fundamental group allows topologically stable flat bundles, that is, ${\mathbb Z}_2$ Wilson lines, which break $SU(5)$ down to $SU(3)_C\times SU(2)_L \times U(1)_Y$. Importantly, by sufficiently restricting the Calabi-Yau threefolds, it was shown that new homology classes can appear that allow the theory to be anomaly free and to admit three families of quarks and leptons.

Although very compelling in many ways, standard-like models constructed from $G=SU(5)$ instantons and ${\mathbb Z}_2$ Wilson lines have a potential problem. It is well-known that superstring vacua associated with GUT theories with gauge group $SU(5)$ may allow nucleon decay more rapid than the experimental bounds. Although the nucleon decay rate can be reduced by imposing certain discrete symmetries and by superstring effects, such vacua do not exhibit a natural mechanism for suppressing nucleon decay. This problem was emphasized by Witten \cite{w2,w3} and discussed in \cite{rt}. The question of nucleon decay has motivated the authors of this paper to attempt to construct alternative standard-like models of particle physics in heterotic M-theory based on $G=SU(4)$ instantons and ${\mathbb Z}_2 \times {\mathbb Z}_2$ Wilson lines. As mentioned above, $SU(4)$ instantons break $E_8$ to an $SO(10)$ GUT group, which is further broken to a standard-like model by a ${\mathbb Z}_2 \times {\mathbb Z}_2$ Wilson line. The key point is that the gauge group of this standard-like model now contains an unbroken $U(1)_{B-L}$ factor, which naturally suppresses nucleon decay.

This program was begun in \cite{opr}, where the requisite torus-fibered Calabi-Yau threefolds with fundamental group ${\mathbb Z}_2 \times {\mathbb Z}_2$ were constructed. Before proceeding with the construction of $SU(4)$ instantons on these manifolds, however, it is essential that one compute the transformation laws of certain elements of the Calabi-Yau homology ring under the ${\mathbb Z}_2 \times {\mathbb Z}_2$ automorphism group. This is necessary in order to establish  the existence and properties of homology classes that are invariant under ${\mathbb Z}_2 \times {\mathbb Z}_2$. As discussed in \cite{dopw-i,dopw-ii,dopw-iii,dopw-iv}, such invariant classes are essential to produce three-family, anomaly free standard-like models. This rather technical, but very important, step in our program is carried out in this paper in detail.

To be specific, in Section~\ref{Z} we briefly review the construction of torus-fibered Calabi-Yau threefolds with  non-trivial fundamental group ${\mathbb Z}_2 \times {\mathbb Z}_2$. It is shown that such threefolds have the form $Z=X/({\mathbb Z}_2 \times {\mathbb Z}_2)$, where $X$ is an elliptically fibered Calabi-Yau threefold that is the fiber product of two rational elliptic surfaces $B$ and $B^{'}$. Section~\ref{rat} is devoted to a detailed review of the properties of restricted surfaces $B$ that admit at least a ${\mathbb Z}_2 \times {\mathbb Z}_2$ automorphism group. In addition, the relevant involutions $(-1)_B$, $\alpha_B$, $t_{e_6}$ and $t_{e_4}$ on $B$, as well as the ${\mathbb Z}_2 \times {\mathbb Z}_2$ generating automorphisms $\tau_{B1}$ and $\tau_{B2}$, are defined and discussed. In \cite{opr} and Section~\ref{rat}, it is shown that generic $B=d{\mathbb P}_9$ surfaces allowing an involution $\alpha_B$ are elliptically fibered with projection map $\beta: B \to \cp{1}$ and are four-fold covers of a surface $Q=\cp{1}\times \cp{1}$. However, the strong restrictions on $B$ necessary to allow a ${\mathbb Z}_2 \times {\mathbb Z}_2$ automorphism group, also produce extra structure on these surfaces. In Section~\ref{double}, it is shown that $B$ can, alternatively, be expressed as a two-fold cover of another surface $\bar{Q}=\cp{1}\times \cp{1}$. Using this result, we demonstrate in Section~\ref{ruled} that the restricted surfaces are ``ruled''. That is, these surfaces have a second fibration with projection $\delta: B \to \cp{1}$ whose generic fiber is not elliptic but, rather, the projective space $\cp{1}$. These, admittingly technical, results play an important role. They allow one to identify, and give the properties of, a canonical set of generators of $H_2(B,{\mathbb Z})$, which we denote by $e_i,\;i=1,\dots,9$ and $l$. This is done in  Section~\ref{basis}. Sections~\ref{-1}, \ref{alpha} and  \ref{trans} are devoted to computing the explicit transformation laws of these generators under the involutions $(-1)_B$, $\alpha_B$, and $t_{e_6}$, $t_{e_4}$ respectively.  These results are then composed to produce the action of the ${\mathbb Z}_2\times {\mathbb Z}_2$ automorphism group on the canonical set of generators. This is carried out in Section~\ref{comp}, where an explicit table of the transformations of the generators $e_1,\dots,e_9$ and $l$ of $H_2(B,{\mathbb Z})$ under all relevant involutions is presented. Having found the ${\mathbb Z}_2\times {\mathbb Z}_2$ action on $H_2(B,{\mathbb Z})$, we can search for the curve classes that are invariant under this action. In Section~\ref{comp}, we show that these are generated by four invariant classes, which are presented explicitly. Finally, in Section~\ref{XZ}, using the construction outlined in Section~\ref{Z}, we ``lift'' the four invariant classes on each of $B$ and $B^{'}$ to the fiber product manifold $X$. By construction, these classes are in $H_4(X,{\mathbb Z})$. It is shown that $H_4(X,{\mathbb Z})$ has seven independent homology generators which are invariant under the ${\mathbb Z}_2\times {\mathbb Z}_2$ automorphism group on $X$. These classes are, of course, constructed from the lift of invariant classes on $B$ and $B^{'}$. We explicitly exhibit these seven generators. Clearly, they survive the modding out of $X$ by the ${\mathbb Z}_2\times {\mathbb Z}_2$ action. Therefore, the torus-fibered Calabi-Yau threefolds $Z=X/({\mathbb Z}_2\times {\mathbb Z}_2)$ with fundamental group $\pi_1(Z)={\mathbb Z}_2\times {\mathbb Z}_2$ admit a rank seven  homology group $H_4(Z,{\mathbb Z})$, as required to produce anomaly free, three family standard-like models.

Having established these requisite results, we can turn to the problem of constructing stable, holomorphic vector bundles with  structure group $G=SU(4)$ on the Calabi-Yau threefolds $Z$. This will be accomplished in \cite{dopr-i}. A detailed discussion of the associated particle physics will appear in \cite{dopr-iii,dlor}. A more precise mathematical discussion of the entire program is also in preparation \cite{dopr-ii}. We emphasize to the reader that, although much of our discussion was within the context of the strongly coupled heterotic superstring, that is, M-theory, our results are equally applicable to weakly coupled heterotic vacua.

\section{The Calabi-Yau Threefolds $Z$}\label{Z}

To make the paper self contained and to fix notation, we recall in this section the construction of torus fibered Calabi-Yau threefolds $Z$ with fundamental group 
\be
\pi_1(Z)={\mathbb Z}_2 \times {\mathbb Z}_2.
\ee
For a more complete exposition, the reader is referred to \cite{opr}. To construct torus fibered Calabi-Yau spaces $Z$ with non-trivial first homotopy group, one starts with simply connected elliptically fibered Calabi-Yau threefolds $X$. Denote the projection map by
\be
\pi: X \to B^{'}.
\ee
The generic fiber of $\pi$ is isomorphic to a torus $T^2$. Furthermore, there exists a zero section $\sigma: B^{'}\to X$ which fixes a unique point on each fiber. This point acts as the identity of an Abelian group, turning $T^2$ into an elliptic curve. In our construction, the base $B^{'}$ is chosen to be isomorphic to a rational elliptic surface $d{\mathbb P}_9$. Being a rational elliptic surface, $B^{'}$ is itself elliptically fibered over $\cp{1}$. We denote its projection map by
\be
\beta^{'}: B^{'}\to {\mathbb P}^{1'}.
\ee
It can be shown, see for example \cite{opr}, that every elliptically fibered Calabi-Yau threefold $X$ over a base $B^{'}$ which is isomorphic to $d{\mathbb P}_9$ is actually a fiber product  of two rational elliptic surfaces $B$ and $B^{'}$. That is,
\be
X=B\times_{\cp{1}}B^{'},
\ee
where
\be
B\times_{\cp{1}}B^{'}=\{(b,b^{'})\in B \times B^{'} | \beta(b)=\beta^{'}(b^{'})\}
\ee
and $\beta: B \to \cp{1}$ and $\beta^{'}:B^{'} \to \mathbb{P}^{1'}$ are the projection maps of the two rational elliptic surfaces. Note that hidden in the definition of $B\times_{\cp{1}}B^{'}$ is an isomorphism $i: \cp{1}\to {\mathbb P}^{1'}$ which identifies these spaces. The fiber product structure allows us to construct involutions $\tau_X$ on $X$ in terms of involutions $\tau_{B}$ and $\tau_{B^{'}}$ on $B$ and $B^{'}$ respectively.

As proven in \cite{opr}, when $B^{}$ and $B^{'}$ are each restricted to a two-parameter subset of $d{\mathbb P}_9$ surfaces, there exists a freely acting automorphism group ${\mathbb Z}_2 \times {\mathbb Z}_2$ on $X$. This group is generated by two  commuting involutions $\tau_{X1}$ and $\tau_{X1}$ defined as
\be\label{invX}
\tau_{X1}=\tau_{B1}\times_{\cp{1}}\tau_{B^{'}1},\;\;\;\;\tau_{X2}=\tau_{B2}\times_{\cp{1}}\tau_{B^{'}2},
\ee
where $\tau_{Bi}$ and $\tau_{B^{'}i}$ for $i=1,2$ are involutions on $B$ and $B^{'}$. It follows that one can define a smooth quotient manifold
\be
Z=X/({\mathbb Z}_2 \times {\mathbb Z}_2).
\ee
It was proven in \cite{opr, dopr-ii} that $Z$ is indeed a Calabi-Yau threefold, albeit one with  no global sections. Hence, it is only a torus fibration and is not elliptically fibered.
Clearly,  to understand these Calabi-Yau threefolds $Z$, it is essential to understand the properties of the Calabi-Yau manifolds $X$ and the action of the automorphism group ${\mathbb Z}_2 \times {\mathbb Z}_2$. 
However, it follows from the above discussion that the properties of  $X$ and its automorphism group  are determined by the choice of the rational elliptic surfaces $B $ and $B^{'}$, the involutions $\tau_{Bi}$ and $\tau_{B^{'}i}$ for $i=1,2$ on these surfaces and the specification of an identification map $i: \cp{1}\to {\mathbb P}^{1'}$. We turn, therefore, to the properties of rational elliptic surfaces and the classification of their involutions.

\section{Rational Elliptic Surfaces}\label{rat}

We begin this section by considering generic rational elliptic surfaces.
As already mentioned, a rational elliptic surface $B$ is an elliptic fibration over $\cp{1}$. We denote the projection map by
\be\label{beta}
\beta: B \to \cp{1}.
\ee
The generic fiber of $\beta$ is isomorphic to a torus $T^{2}$. Furthermore, there is a zero section 
\be
e: \cp{1}\to B
\ee
which fixes a unique zero point on each fiber, turning it into an elliptic curve. 
In addition,  there exists a second projection map, denoted by $\beta_2$ in \cite{opr}, such that
\be
\beta_2: B \to \cp{2}.
\ee
The pre-image of $\beta_2$ for a generic point $p \in \cp{2}$, that is $\beta^{-1}_2(p)$, consists of one point. However, at nine points, which we denote by $\omega_a$ for $a=1,\dots,9$,
\be
\beta_2^{-1}(\omega_a)\equiv e_a\cong \cp{1}.
\ee
Hence, $B$ is a blow-up of $\cp{2}$ at nine points. This description of $B$ is very useful in  determining several geometric properties of the surface. We find that the generators for the homology classes of $H_2(B,{\mathbb Z})$ are the nine blow-up $\cp{1}$ lines  $e_1,\dots,e_9$, also called exceptional divisors, and the pull-back $\beta_2^{-1}(l)$ of a line $l$ in $\cp{2}$.  We simply denote $\beta_2^{-1}(l)$ by $l$ as well. That is,  the lattice $H_2(B,{\mathbb Z})$ is spanned by these ten elements.
Since $B$ is connected, it follows that  $H_0(B,{\mathbb Z})\cong {\mathbb Z}$ and, hence,  $H_0(B,{\mathbb Z})$ is generated by a point. Additionally, since $B$ is compact and orientable, $H_4(B,{\mathbb Z})\cong {\mathbb Z}$. Using, for example, the Van Kampen theorem, one can show that $H_1(B,{\mathbb Z})$ and $H_3(B,{\mathbb Z})$ vanish. Therefore, the topological Euler characteristic of $B$ is given by
\be
\chi(B)=\sum_{i=0}^{4} (-1)^{i} \rank H_{i}(B,{\mathbb Z})=1+10+1=12.
\ee
It is easy to calculate the canonical bundle $K_B=\wedge^2(T^{*}B)$, where we have denoted the holomorphic cotangent bundle of $B$ by $T^{*}B$. The result is
\be\label{can}
K_B=\beta_2^{*}K_{\cp{2}}\otimes {\mathcal O}_B(\sum_{i=1}^{9}e_i)={\mathcal O}_B(-3l+\sum_{i=1}^{9}e_i).
\ee

As shown in \cite{opr}, these geometric properties, derived using the projection $\beta_2 : B \to \cp{2}$, have consequences for the fibration $\beta: B \to \cp{1}$.
The first consequence is that one can consider the exceptional divisors $e_1,\dots,e_9$ as sections of this fibration. For concreteness, we choose the zero section to be
\be
e=e_9.
\ee
It was shown in \cite{opr} that the class of a generic fiber of (\ref{beta}) is given by
\be\label{df}
f=3l-\sum_{i=1}^9 e_i.
\ee
Therefore, (\ref{can}) can be written as
\be
K_B={\mathcal O}_B(-f).
\ee
Since the Euler characteristic of a smooth torus vanishes, the fact that $\chi(B)=12$ requires the occurrence of singular fibers of (\ref{beta}). For generic surfaces $B$, these singular fibers are of Kodaira type \cite{kodaira-casIII} $I_1$ with $\chi(I_1)=1$. Hence, there are twelve such fibers. 

Having discussed the basic properties of generic rational elliptic surfaces $B$, we can proceed to determine all possible involutions on them. As shown in \cite{opr}, all involutions
\be
\tau_B: B \to B
\ee
induce an involution on the base $\cp{1}$, namely, $\tau_{\cp{1}}:\cp{1}\to \cp{1}$, obeying 
\be
\beta\circ \tau_B = \tau_{\cp{1}}\circ \beta.
\ee
Hence, $\tau_B$ maps each fiber $f$ to some possibly different fiber $\tau_B(f)$. Since involutions $\tau_{\cp{1}}:\cp{1}\to \cp{1}$ are either the identity map $\tau_{\cp{1}}=id$ or a non-trivial map with two fixed points, where the fixed points determine $\tau_{\cp{1}}$ uniquely, the induced action on the base $\cp{1}$ gives a crude classification of involutions on $B$. Let us first consider  those involutions which act as the identity on the base.

One such involution is canonically given for any elliptic fibration, namely
\be
(-1)_B : B \to B.
\ee
It is defined as follows. Since the zero section $e$ fixes a unique zero on each fiber, which we denote by $e$ as well, each fiber forms an Abelian group. We will denote the operation of addition on each fiber by  ${\stackrel{.}{+}}$ and the operation of inverse on any element $a$ by ${\stackrel{.}{-}}a$. The restriction of $(-1)_B$ to a fiber maps each point $a$ of that fiber to its Abelian group inverse. Therefore, we obtain
\be
(-1)_B|_f (a)= {\stackrel{.}{-}} a.
\ee
Since the choice of the identity element $e$ is determined by the global section $e$, $(-1)_B$ is defined globally. Furthermore, $(-1)_B$ induces an automorphism on the global sections of $B$. That is
\be
(-1)_B : \Gamma(B)\to \Gamma(B),
\ee
where $\Gamma(B)$ denotes the set of global sections of the fibration $\beta: B\to \cp{1}$. More specifically, for any section $\xi \in \Gamma(B)$
\be\label{3.15}
(-1)_B(\xi)={\stackrel{.}{-}}\xi,
\ee
where ${\stackrel{.}{-}}\xi$ is defined as follows. Consider any point $p \in \cp{1}$ and let $\xi(p)=a$. Then the section ${\stackrel{.}{-}}\xi$ is defined by $({\stackrel{.}{-}}\xi)(p)={\stackrel{.}{-}}a$.

Another automorphism of $B$ which acts trivially on the base $\cp{1}$ is given, for any section $\xi$, by the translation $t_{\xi}$.  If we denote the point of  intersection of $\xi$ with a fiber $f$ by $\xi$ as well, then the action of $t_\xi$ is given by
\be
t_{\xi}|_f: a \to a{\stackrel{.}{+}}\xi
\ee
for each point $a$ on $f$. Again, the action $t_\xi$ is globally defined and induces an automorphism on the global sections of $B$, namely
\be\label{3.17}
t_\xi (s) = s{\stackrel{.}{+}}\xi
\ee 
for any section $s$ in $\Gamma(B)$. For $p \in \cp{1}$ and $s(p)=a$, $s{\stackrel{.}{+}}\xi$ is defined by $(s{\stackrel{.}{+}}\xi)(p)=a{\stackrel{.}{+}}\xi$. Note that (\ref{3.15}) and (\ref{3.17}) induce an Abelian group structure on $\Gamma(B)$. Since this structure is defined in terms of the group law on each fiber, we continue to denote addition and inverse on $\Gamma(B)$ by  ${\stackrel{.}{+}}$ and ${\stackrel{.}{-}}$.\footnote{It is helpful at this point to discuss our notation in more detail. On any fibered surface $B$, a section $\xi \in \Gamma(B)$ will determine a homology class, also denoted by $\xi$, in $H_2(B,{\mathbb Z})$. There is a natural notion of addition and inverse in $H_2(B,{\mathbb Z})$ which we denote everywhere in this paper by $+$ and $-$. For generic fibered surfaces, however, there is no group law on $\Gamma(B)$. When $B$ is elliptically fibered, a notion of addition and inverse on $\Gamma(B)$ does exist. To distinguish this from the group law on $H_2(B,{\mathbb Z})$, we denote the group law on $\Gamma(B)$ by ${\stackrel{.}{+}}$ and ${\stackrel{.}{-}}$.} It is clear, that $t_\xi$ is generically not an involution. A necessary condition for it to be an involution is that the section $\xi$ intersect each fiber at a point of order two, that is, $\xi{\stackrel{.}{+}}\xi=e$. There are four such points on each fiber, which  we denote by $e,e^{'},e^{''}$ and $e^{'''}$. Of course, the identity element $e$ is one such point. Observe that  points of order two are also fixed under the action of $(-1)_B$. For example
\be
(-1)_B (e^{'})={\stackrel{.}{-}}e^{'}=(e^{'}{\stackrel{.}{+}}e^{'}){\stackrel{.}{-}}e^{'}=e^{'}.
\ee
However, the above condition is not sufficient to guarantee that  global sections $\xi$ with $t_{\xi}^{2}=id$  exist. In fact, for generic elliptic surfaces $B$ they do not. They can occur, however, in a restricted subset of surfaces $B$. We will use such restricted surfaces to construct the Calabi-Yau threefolds $X$.

A further type of involution $\tau_B$ on $B$ which induces a trivial action on the base $\cp{1}$ is given by 
\be
\tau_B=t_{\xi}\circ (-1)_B.
\ee
However, since it is impossible to lift these involutions to a fixed point free involution on $X$, we will not consider them further.

We now discuss involutions on $B$ which act non-trivially on the base $\cp{1}$. By a theorem of \cite{dopw-i}, each such involution can be written as 
\be
\tau_B=t_\xi\circ \alpha_B,
\ee
where $\alpha_B : B \to B$ is an involution of $B$ leaving the zero section $e$ fixed and inducing a non-trivial action on $\cp{1}$. That is, $\tau_{\cp{1}}\circ \beta=\beta\circ \alpha_B$ where $\tau_{\cp{1}}$ is a non-trivial involution on $\cp{1}$ with two fixed points, which we call $0$ and $\infty$. The section $\xi$ must have the property that
\be\label{xi}
\alpha_B(\xi)=(-1)_B(\xi).
\ee
As shown in \cite{opr}, a generic rational elliptic surface $B$ does not admit such involutions $\alpha_B$. However, there is five dimensional sub-family in the eight dimensional family of rational elliptic surfaces which does allow such involutions. Furthermore, it was shown in \cite{dopw-i} that each member of this five dimensional sub-family has a rank four lattice of sections $\xi$ fulfilling condition (\ref{xi}).

As described in \cite{opr}, $\alpha_B$ leaves two fibers of $B$ stable, namely $f_0=\beta^{-1}(0)$ and $f_\infty = \beta^{-1}(\infty)$. Furthermore, it was shown that $\alpha_B$ acts as the identity map on $f_0$ and as $(-1)_B$ on $f_\infty$. Hence, the double cover
\be\label{3.22}
\kappa: B \to B/\alpha_B
\ee
has the image of $f_0$ and the image of the four points of order two of the fiber $f_\infty$ under $\kappa$ as its branch locus. Hence, the image of $f_\infty$, that is  $f_\infty/\alpha_B \subset B/\alpha_B$, contains four isolated $A_1$ surface singularities of $B/\alpha_B$. 
\begin{figure}
\begin{center}
\input{figure3.pstex_t} 
\end{center}
\caption{The bidegree $(2,4)$ curve $M=T\cup r$ in $Q=\cp{1}\times \cp{1}$. The projections $p_i: Q \to \mathbb{P}_i^1, i=1,2$ are explicitly shown.}
\label{gen-M} 
\end{figure}
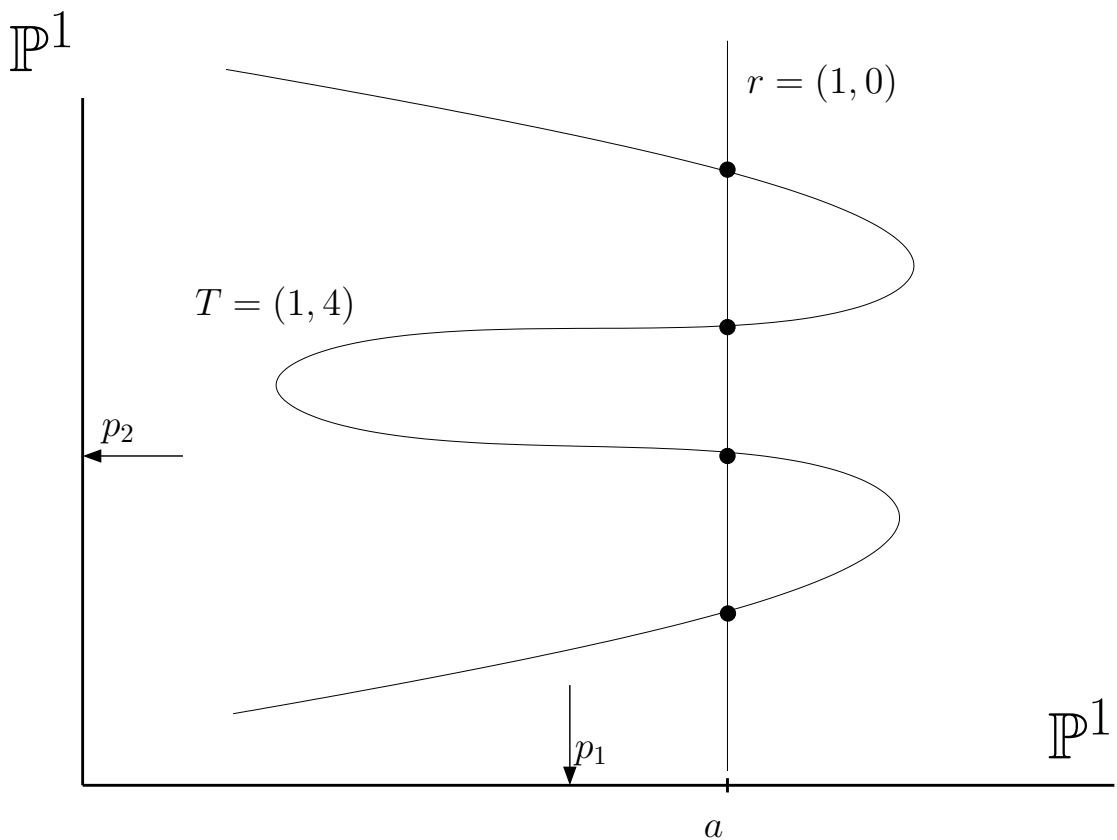
These can be resolved by blowing up and one obtains an $I_0^{*}$ fiber.
It was shown in \cite{opr} that, for generic $B$ in the five parameter family of rational elliptic surfaces, the quotient $B/\alpha_B$ can be described as the double cover of $Q=\cp{1}\times \cp{1}$. That is
\be\label{3.23}
\pi: B/\alpha_B \to Q,
\ee
where the branch locus $M $ of $\pi$ consists of the union of  a $(1,4)$ curve $T$ and a $(1,0)$ curve $r$. These are shown in  Figure~\ref{gen-M}, where  the two natural projections of $Q$, $p_i : Q \to \cp{1}_i,\;i=1,2$ to both $\cp{1}$ factors of $Q$ are also defined.

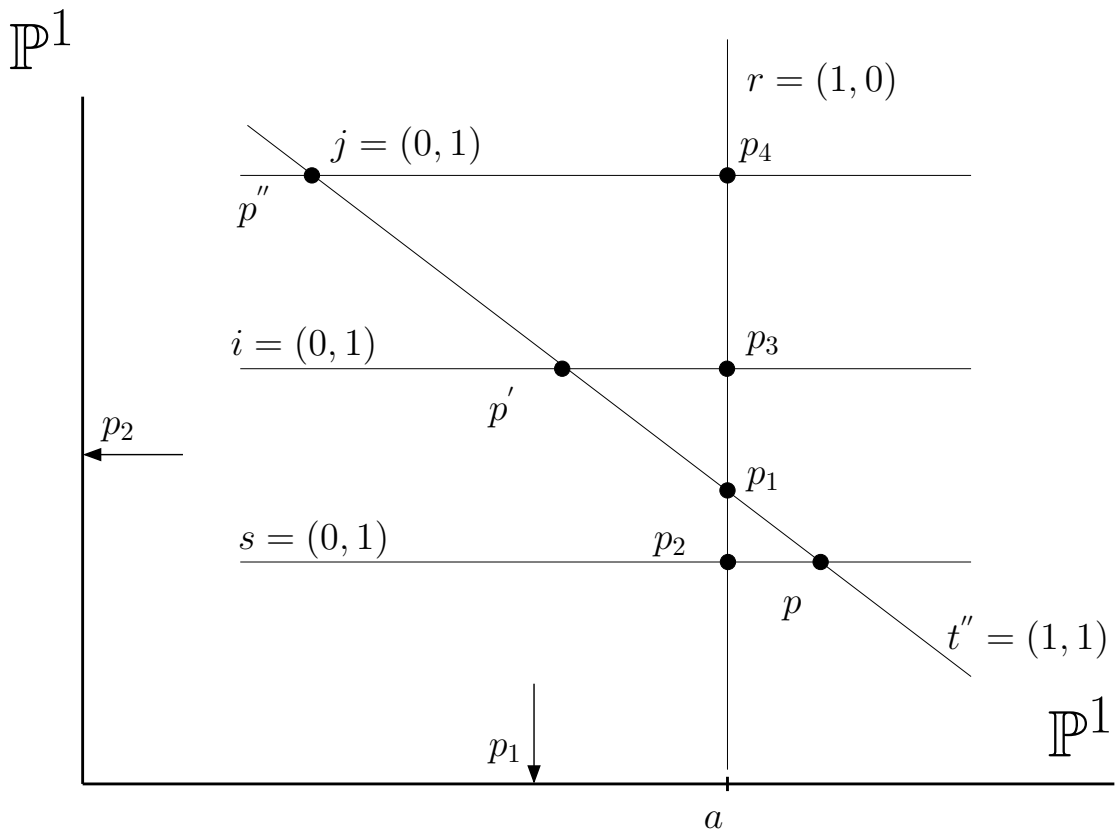
\begin{figure}
\begin{center}
\input{figure11.pstex_t} 
\end{center}
\caption{A schematic representation of curve $M=t^{''}\cup s\cup i \cup j \cup r$ in $Q=\cp{1}\times \cp{1}$.}
\label{2df} 
\end{figure}

Having described all possible involutions on rational elliptic surfaces $B$, we now further restrict ourselves to a two dimensional sub-family of rational elliptic surfaces. These were used in \cite{opr} to construct Calabi-Yau threefolds $X$ with two freely acting involutions. Since this two dimensional family is a subset of the five dimensional family mentioned above, these surfaces continue to admit  involutions $\alpha_B$.
However, for these restricted surfaces, the curve $T$ splits into $T=t^{''}\cup s\cup i\cup j$, where $t^{''}$ is a $(1,1)$ curve and $s,i$ and $j$ are $(0,1)$ rulings. See Figure~\ref{2df} for a pictorial description of them.
Note that  $t^{''}$ intersects each of $r,s,i$ and $j$ at one point. Call these points $p_1,p,p^{'}$ and $p^{''}$ respectively. Furthermore, $s,i$ and $j$ each intersect $r$ at a single point, which  we denote by as $p_2,p_3$ and $p_4$. 

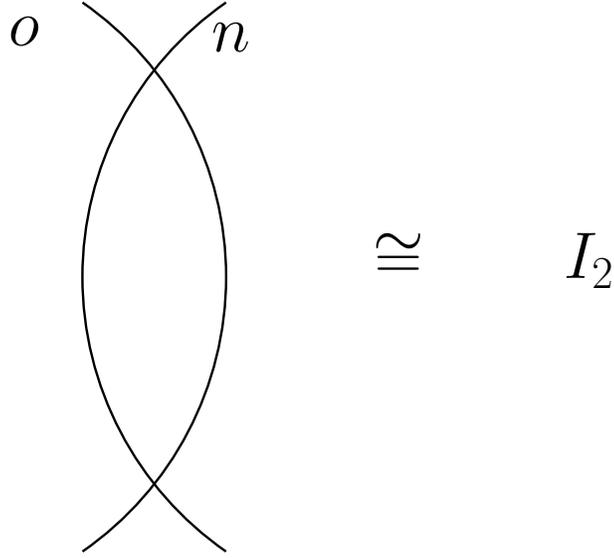
\begin{figure}[!ht]
\begin{center}
\input{figure8.pstex_t} 
\end{center}
\caption{The reducible fiber $I_2$, where $o$ denotes the proper transform and $n$ the exceptional divisor.}
\label{I2} 
\end{figure}

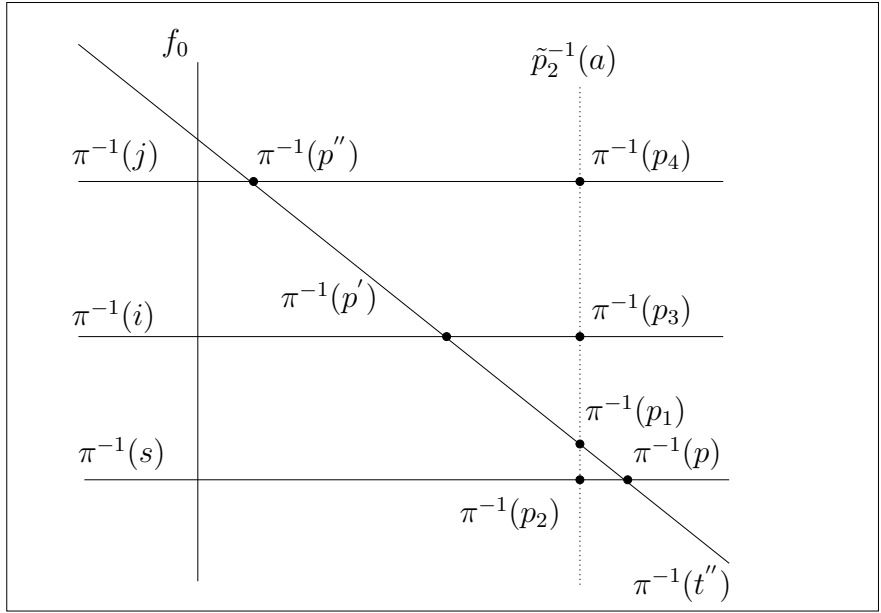
\begin{figure}[!ht]
\begin{center}
\input{sectWM.pstex_t} 
\end{center}
\caption{The elliptic fibration $W_M$ with the four distinguished sections.}
\label{four-s} 
\end{figure}

What is the double cover  of $Q$ corresponding to this restricted branch locus? For maximal clarity, we will here denote this double cover by $W_M$, relating it to $B/\alpha_B$ only later. We continue to indicate the covering map by $\pi: W_M \to Q$. Let us consider the composition map $\tilde{p}_1=p_1\circ \pi$. Clearly, we have the projection map
\be\label{tildep}
\tilde{p}_1: W_M \to \cp{1}.
\ee
The fiber of $\tilde{p}_1$, for a generic point  $x \in \cp{1}$, is the double cover of the fiber of projection $p_1$ at this point, branched over the four points where this fiber intersects the branch locus $M$, namely, at its intersection with $t^{''},s,j$ and $i$. It was shown in \cite{opr} that this generic  fiber of $\tilde{p}_1$ is isomorphic to a torus. Hence,  $W_M$
is a torus fibration. What is the double cover of the fiber corresponding to the point $p_1(r)=a$? Since $r$ is in the branch locus itself, $\tilde{p}^{-1}_1(a)$ is a double $\cp{1}$ line. Furthermore, since $r$ intersects the other components of the branch curve $M$, namely, $t^{''},s,i$ and $j$ at  $p_1,p_2,p_3$ and $p_4$ respectively, $W_M$ has four $A_1$ surface singularities on $\tilde{p}_1^{-1}(a)$ located at  $\pi^{-1}(p_i)$ for $i=1,\dots,4$. Recall from \cite{opr} that resolving these singularities would produce an $I_0^{*}$ fiber.
In addition to $\tilde{p}_1^{-1}(a)$, there are three other fibers of $\tilde{p}_1$ which contain singular points of $W_M$. Singular points  occur when different components of the branch locus of $W_M $ intersect. Clearly, the remaining intersection  points are in the fibers ${p}_1^{-1}(p_1(p)),\;{p}_1^{-1}(p_1(p^{'}))$ and ${p}_1^{-1}(p_1(p^{''}))$. As explained in \cite{opr}, the double covers of these three fibers in $W_M $  are singular, each containing an $A_1$ surface singularity.  After resolving the singularities, each of these fibers consists of two components, the proper transform of the original fiber and an exceptional divisor which is isomorphic to $\cp{1}$. This is called an $I_2$ fiber in the Kodaira classification and is sketched in Figure~\ref{I2}.
To conclude, the fibration $(\ref{tildep})$ has four singular fibers, one containing four $A_1$ singularities and the remaining three with one $A_1$ singularity each. Resolving these singularities, we obtain a  smooth surface, which we denote by $\widehat{W_M}$, with  one $I_0^{*}$ fiber and  three $I_2$ fibers. Recall from \cite{opr} that $\chi(I_0^{*})=6$ and $\chi(I_2)=2$. Therefore $\chi(\widehat{W_M})=12$. 
Furthermore, the fibration $\tilde{p}_1 : W_M \to \cp{1}$ has four special sections, namely $\pi^{-1}(s),\pi^{-1}(i)$, $ \pi^{-1}(j)$ and $\pi^{-1}(t^{''})$. These sections, each of which is isomorphic to $\cp{1}$, are described graphically in Figure~\ref{four-s}. Choosing one of them, which we take to be   $\pi^{-1}(s)$, as the zero section completes the proof that $\widehat{W_M}$ is a rational elliptically fibered surface with one $I_0^{*}$ fiber and three $I_2$. As discussed in \cite{opr}, $W_M$ is its Weierstrass model. We can  summarize the projections $\pi, p_1$ and $\tilde{p}_1$ in the diagram
\be
\xymatrix{
W_M \ar[r]^-{\pi} \ar[d]_-{\tilde{p}_1} & Q \ar[d]^-{{p}_1} \ar@{}[drr]|-{.} &&\\
{\mathbb P}^{1} \ar[r]_{\op{id}} & {\mathbb P}^{1} & & \\
} 
\ee
Recall from \cite{opr} that  one can choose a smooth fiber, which we call $f_0$, in $W_M$ and construct the double cover of $W_M$ branched over $f_0$ and the four singular points in $\tilde{p}_1^{-1}(a)$. One then obtains the double cover  of $W_M$, which we denote by $\bar{B}$ with the cover map
\be
\bar{\kappa}: \bar{B} \to W_M.
\ee
This double covering map  $\bar{\kappa}$ induces an involution  $\alpha_{\bar{B}}$ on $\bar{B}$, which acts non-trivially as $\tau_{\cp{1}}$ on the base $\cp{1}$. Defining 
\be
sq: \cp{1} \to \cp{1}/\tau_{\cp{1}}
\ee
and noting that $\cp{1}/\tau_{\cp{1}}\cong \cp{1}$, we obtain
\be\label{d2}
\xymatrix{
\bar{B} \ar[r]^-{\bar{\kappa}} \ar[d]_-{\bar{\beta}} & W_M \ar[d]^-{\tilde{p}_1} \ar@{}[drr]|-{,} &&\\
{\mathbb P}^{1} \ar[r]_{\op{sq}} & {\mathbb P}^{1} & & \\
}
\ee
which defines the fibration $\bar{\beta}: \bar{B} \to \cp{1}$. It is important to note that $\bar{B}$, the double cover of $W_M$, is not yet the surface $B$. The reason is the following.

Let us  recall from \cite{opr} some of the properties of $\bar{B}$. First, from the diagram~(\ref{d2}) it is easy to see that a generic fiber of $W_M$ has two disjoint pre-images in $\bar{B}$. These form two fibers of the fibration $\bar{\beta}: \bar{B} \to \cp{1}$ and get exchanged by the involution $\alpha_{\bar B}$. Hence, only the fibers $\bar{\kappa}^{-1}(f_0)$ and $\bar{\kappa}^{-1}(\tilde{p}_1^{-1}(a))$ are stable in $\bar{B}$. We denote these fibers  by $f_0$ and $f_\infty$ respectively. Of course, these are the fibers over the fixed points $0$ and $\infty$ of $\tau_{\cp{1}}$, the involution on $\cp{1}$ induced by $\alpha_{\bar{B}}$. Although $\tilde{p}_1^{-1}(a)$ in $W_M$ contains four surface singularities of type $A_1$, its pre-image $f_\infty$  in $\bar{B}$  is smooth. This is explained in \cite{opr} and follows from the fact that the pre-images of the four singularities of $\tilde{p}_1^{-1}(a) $ in $W_M$ are fixed under the covering involution $\alpha_{\bar B}$. Be that as it may, $\bar{B}$ does have singular fibers. Recall that $W_M$ had three additional singular fibers containing an $A_1$ surface singularity. Since these fibers are generically not $f_0$ and $\tilde{p}_1^{-1}(a)$ in $W_M$, their pre-images each  consists of two fibers in $\bar{B}$. Clearly, each such fiber contains an $A_1$ surface singularity. Therefore, there are six fibers in $\bar{B}$ containing a surface singularity of type $A_1$. We conclude that the surface ${\bar B}$ is singular, whereas the rational elliptic surface $B$ is smooth. Therefore, $B$ differs from $\bar{B}$, being obtained from it by blowing up each of the six $A_1$ singularities, as we will discuss below. 

What are the pre-images of the four special sections of $W_M$? First consider $\pi^{-1}(s)$. Recall that $\pi^{-1}(s)$  intersects the fiber $f_0$ in $W_M$ and contains the singular point $\pi^{-1}(p_2)$. Hence, it intersects the branch locus of $\bar{B}$ in two points and its pre-image $\bar{\kappa}^{-1}(\pi^{-1}(s))$ is the double cover of $\pi^{-1}(s)$. Since $\pi^{-1}(s)\cong \cp{1}$, it follows that  $\bar{\kappa}^{-1}(\pi^{-1}(s))$ is isomorphic to $\cp{1}$ as well. Furthermore, since $\bar{\kappa}^{-1}(\pi^{-1}(s))$ intersects each fiber of $\bar{B}$ once, it is a section of the fibration $\bar{\beta}: \bar{B} \to \cp{1}$. Note that $\bar{\kappa}^{-1}(\pi^{-1}(s))$ also contains two $A_1$ surface singularities, namely the  pre-image $\bar{\kappa}^{-1}(\pi^{-1}(p))$. 
The pre-images under $\bar{\kappa}$ of $\pi^{-1}(i)$ and $\pi^{-1}(j)$ have similar properties. They form sections of $\bar{B}$ each containing two $A_1$ surface singularities, namely the two points   $\bar{\kappa}^{-1}(\pi^{-1}(p^{'}))$ and the two points $\bar{\kappa}^{-1}(\pi^{-1}(p^{''}))$ respectively. The pre-image $\bar{\kappa}^{-1}(\pi^{-1}(t^{''}))$ in $\bar{B}$, the double cover of $\pi^{-1}(t^{''})$ branched over $\pi^{-1}(p_1)$ and $f_0\cap \pi^{-1}(t^{''})$, is isomorphic to $\cp{1}$ as well. Clearly, it is also a section of $\bar{B}$. However, unlike  the other three  sections $\bar{\kappa}^{-1}(\pi^{-1}(s))$, $\bar{\kappa}^{-1}(\pi^{-1}(i))$ and  $\bar{\kappa}^{-1}(\pi^{-1}(j))$, $\bar{\kappa}^{-1}(\pi^{-1}(t^{''}))$ contains all six $A_1$ surface singularities. This follows from the fact that  $\pi^{-1}(t^{''})$ contains $\pi^{-1}(p),\;\pi^{-1}(p^{'})$ and $\pi^{-1}(p^{''})$.

Having discussed the singular surfaces $\bar{B}$, we finally can construct their resolution, the rational elliptic surfaces $B$ that  we wish to consider. If we denote by $B$  the blow-up of $\bar{B}$  at the six $A_1$ singularities and let $c : B \to \bar{B}$ be the projection map, then  clearly
\be\label{blow}
\xymatrix{
B \ar[r]^{c} \ar[d]_-{\beta} & \bar{B} \ar[d]^-{\bar{\beta}} \ar@{}[drr]|-{,} &&\\
{\mathbb P}^{1} \ar[r]_{\op{id}} & {\mathbb P}^{1} & & \\
} 
\ee
which defines the fibration $\beta: B \to \cp{1}$ as $\beta=\bar{\beta}\circ c$. Combining diagrams (\ref{d2}) and (\ref{blow}), we see that
\be
\kappa: B \to W_M
\ee
is the covering of $W_M$ by $B$ where $\kappa=\bar{\kappa}\circ c$. We denote by $\alpha_B$ the induced involution on $B$ of $\alpha_{\bar B}$. 

What are the fibers of $B$? As described in \cite{opr}, each singular fiber with an $A_1$ surface singularity can be resolved into an $I_2$ fiber. Hence, $B$ has six $I_2$ fibers, namely, the pre-images of $\kappa^{-1}(\tilde{p}_1^{-1}(p_1(p)))$, $\kappa^{-1}(\tilde{p}_1^{-1}(p_1(p^{'})))$ and $\kappa^{-1}(\tilde{p}_1^{-1}(p_1(p^{''})))$. Let us denote the exceptional divisors corresponding to $\kappa^{-1}(\tilde{p}_1^{-1}(p_1(p^{})))$ by $n_1$ and $n_2$, the exceptional divisors corresponding to $\kappa^{-1}(\tilde{p}_1^{-1}(p_1(p^{'})))$ by $n_3$ and $n_4$ and the exceptional divisors corresponding to $\kappa^{-1}(\tilde{p}_1^{-1}(p_1(p^{''})))$ by $n_5$ and $n_6$. The associated proper transforms are denoted  $o_1,\dots,o_6$. Hence, the six $I_2$ fibers in $B$ have the structure  $n_i\cup o_i,\;i=1,\dots,6$.

Next, consider the proper transform of the four distinguished sections in $\bar{B}$. Clearly, they are still sections of $B$. Following the notation of \cite{opr}, we denote the proper transform of $\kappa^{-1}(\pi^{-1}(s))$ by $e_9$, the proper transform of $\kappa^{-1}(\pi^{-1}(i))$ by $e_6$ and the proper transform of $\kappa^{-1}(\pi^{-1}(j))$ by $e_4$.  From the distribution of the singularities  at the image of these sections in $\bar{B}$, as described above, it is clear that $e_9$ intersects $n_1$ and $n_2$, $e_6$ intersects $n_3$ and $n_4$ and $e_4$ intersects $n_5$ and $n_6$. We will show in the next section that the proper transform of $\kappa^{-1}(\pi^{-1}(t^{''}))$ is $e_4{\stackrel{.}{+}}e_6$.

At this point, we want to make an important technical comment. As discussed above, $\bar{B}$ is the singular four-fold cover of $Q$. $B$, on the other hand, is a smooth surface obtained by blowing up the singularities in $\bar{B}$. The exact relationship is expressed in (\ref{blow}). However, in most of the remaining discussion in this paper, one can readily distinguish whether a surface is $\bar{B}$ or its blow-up $B$ from context. With this in mind, we find that it leads to enormous simplification of our notation if we simply use the symbol $B$ to denote both $B$ and $\bar{B}$. In keeping with this notation, we will also denote both sets of mappings $\bar{\beta}, \bar{\kappa}, \alpha_{\bar B}$ and $\beta, \kappa, \alpha_B$ by $\beta, \kappa, \alpha_B$. We do this henceforth. With this simplified notation, we can identify
\be
W_M\cong B/\alpha_B,
\ee
as in (\ref{3.22}) and (\ref{3.23}). In a few places when it is helpful to specify the surface and the associated mappings exactly, we will do so.

As in \cite{opr}, we choose $e_9$ to be the zero section of the rational elliptic surface $B$. Furthermore, as we will show in the next section, $e_4$ and $e_6$ are sections of $B$ which intersect each fiber at a point of order two. Hence
\be
(-1)_B (e_9)=e_9,\;\;\;\;\;(-1)_B (e_6)=e_6,\;\;\;\;\;(-1)_B (e_4)=e_4.
\ee
It is clear from the construction that $e_9$, $e_6$ and  $e_4$ are invariant under $\alpha_B$. Hence, they fulfill condition (\ref{xi}). Note, in addition, that $e_4{\stackrel{.}{+}}e_6$ is a section of $B$ which intersects each fiber at a point of order two.  It follows that it too satisfies (\ref{xi}). Using these sections, we can define the two involutions
\be\label{inv}
\tau_{B1}=t_{e_6}\circ \alpha_B,\;\;\;\;\;\;\tau_{B2}=t_{e_4}\circ \alpha_B.
\ee
Using the fact that 
\be
\tau_{B1}\circ \tau_{B2}=t_{e_4 {\stackrel{.}{+}} e_6}
\ee
and
\be
t_{e_4 {\stackrel{.}{+}} e_6}^2=id,
\ee
it was shown in \cite{opr} that $\tau_{B1}$ and $\tau_{B2}$ commute.
It was explained in \cite{opr} that one can lift these commuting involutions to freely acting involutions on the Calabi-Yau threefolds $X$. Henceforth, we will assume (\ref{inv}) to be the generators of the $\mathbb{Z}_2 \times \mathbb{Z}_2$ automorphism group on the rational elliptic surface $B$.

We choose the rational elliptic surface $B^{'}$ to be within the same two dimensional family. We denote the components of its $I_2$ fibers by $n_i^{'}$ and $o_i^{'},\;i=1,\dots,6$. The zero section is denoted by $e^{'}\equiv e_9^{'}$ and the other sections of order two by $e_6^{'}$, $e_4^{'}$ and $e_4^{'}{\stackrel{.}{+}}e_6^{'}$. The generators of the $\mathbb{Z}_2 \times \mathbb{Z}_2$ automorphism group on $B^{'}$ are then given by
\be
\tau_{B^{'}1}=t_{e_6^{'}}\circ \alpha_B^{'},\;\;\;\;\;\;\tau_{B^{'}2}=t_{e_4^{'}}\circ \alpha_B^{'}.
\ee

This concludes our review of rational elliptic surfaces. As discussed in Section~\ref{Z}, the fiber product $X=B \times_{\cp{1}} B^{'}$ is a Calabi-Yau threefold and admits the two commuting involutions
\be
\tau_{X1}=\tau_{B1}\times_{\cp{1}}\tau_{B^{'}1},\;\;\;\;\;\;\tau_{X2}=\tau_{B2}\times_{\cp{1}}\tau_{B^{'}2}.
\ee
They can be shown \cite{opr} to generate a freely acting automorphism group $\mathbb{Z}_2 \times \mathbb{Z}_2$ on $X$. Hence, the quotient manifold
\be
Z=X/(\mathbb{Z}_2 \times \mathbb{Z}_2)
\ee
is smooth and, as shown in \cite{opr,dopr-ii}, is  a Calabi-Yau threefold.

\section{$B$ as a Double Cover of $\bar{Q}=\cp{1}\times \cp{1}$}\label{double}

In this section, we begin the process of computing  the transformation laws of the elements of $H_4(X,{\mathbb Z})$ of the Calabi-Yau threefolds $X$ under the automorphism group ${\mathbb Z}_2 \times {\mathbb Z}_2$.  To do this, we need to find the transformation laws of the second homology groups  of the rational elliptic surfaces $B$ and $B^{'}$ under the respective ${\mathbb Z}_2 \times {\mathbb Z}_2$ automorphisms.

Let us consider a rational elliptic surface within the two dimensional sub-family described in the previous section. To find the transformation laws of $H_2(B,{\mathbb Z})$ under ${\mathbb Z}_2\times {\mathbb Z}_2$, we must first identify a set of generating  elements of  $H_2(B,{\mathbb Z})$. For a generic surface, $B$ is a blow-up of $\cp{2}$ at nine separated points. Hence, for a generic $B$ one can choose  the nine exceptional divisors $e_1,\dots,e_9$ and $l$, the pre-image of a line in $\cp{2}$, as a set of generators for $H_2(B,{\mathbb Z})$. Note that each of these exceptional curve is isomorphic to $\cp{1}$ and, hence, irreducible. Furthermore, the intersection numbers of $e_1,\dots,e_9$ and $l$ are given by
\be\label{intersection}
e_i\cdot e_j =-\delta_{ij},\;\;\;e_j\cdot l=0,\;\;\;l\cdot l=1,\;\;\;i,j=1,\dots,9.
\ee
However, the rational elliptic surfaces we are going to consider are not generic. Rather, they are chosen to be in a restricted two dimensional sub-family. It can be shown that one effect of this restriction is to  collide together several of the nine points in $\cp{2}$. That is, not all of the nine points are distinct. The blow-ups of these degenerate points continue to contain exceptional curves $e_i$, but their properties and  linear dependence are more complicated. It turns out that for surfaces $B$ in the two dimensional sub-family, $l$ and the nine curves $e_1,\dots e_9$ with intersection numbers (\ref{intersection}) still form a set of generators of $H_2(B,{\mathbb Z})$. However, these curves are no longer  necessarily irreducible,  that is, they may consists of several components. This will be discussed in detail in Section~\ref{basis}.  This fact greatly complicates the process of determining their transformation laws under the ${\mathbb Z}_2\times {\mathbb Z}_2$ automorphisms. So much so, that it becomes necessary to use properties of rational elliptic surfaces not discussed in \cite{opr} or in the previous section.

To find the  transformation laws of $e_1,\dots e_9$ under ${\mathbb Z}_2\times {\mathbb Z}_2$, it is important to notice that, in addition to the fibration $\beta: B \to \cp{1}$ discussed in \cite{opr} and Section~\ref{rat}, $B$ admits a second fibration 
\be
\delta: B \to \cp{1},
\ee
where the generic fiber of $\delta$ is isomorphic to $\cp{1}$ and not to a torus. This second fibration is very helpful in determining the ${\mathbb Z}_2\times {\mathbb Z}_2$ transformation laws since, first of all, its fibers turn out all to be invariant under the involution  $\alpha_B$ and, secondly,  all curves $e_1,\dots e_9$ with exception of $e_7$ are contained in some of its singular fibers.  This fibration is described in great detail in Section~\ref{ruled}. However, as a prerequisite, we need to prove the following property of $B$. Recall from Section~\ref{rat} that $B$ is a four-fold cover, with mapping $\pi\circ \kappa$, of $Q=\cp{1}\times \cp{1}$.  In this section, we will show that, viewed differently, $B$ can actually be expressed as a double cover of another $\cp{1}\times \cp{1}$ surface, which we denote by $\bar{Q}$. This result will then allow us to construct the fibration  $\delta: B \to \cp{1}$.

\begin{figure}[!ht]
{\Large
\[
\xymatrix@C=-0.6in@R=1.5in{
&B \ar[dr]^-{\kappa^{'}} \ar[dl]_-{\kappa} \ar[d]_-{\psi} & \\
B/\alpha_B\cong W_M \ar[dr]_-{\pi} & B/(-1)_B\cong\bar{Q} \ar[d]_-{\Delta} & W_{M^{'}}\cong B/( (-1)_B\circ\alpha_B) \ar[dl]^-{\pi^{'}}\\
& Q \cong W_M/(-1)_{W_M}\cong \bar{Q}/\alpha_{\bar{Q}}\cong W_{M^{'}}/(-1)_{W_{M^{'}}}& 
}
\]
}
\caption{Three different ways to consider $B$ as a degree four cover of $Q$.}
\label{tricom}
\end{figure}
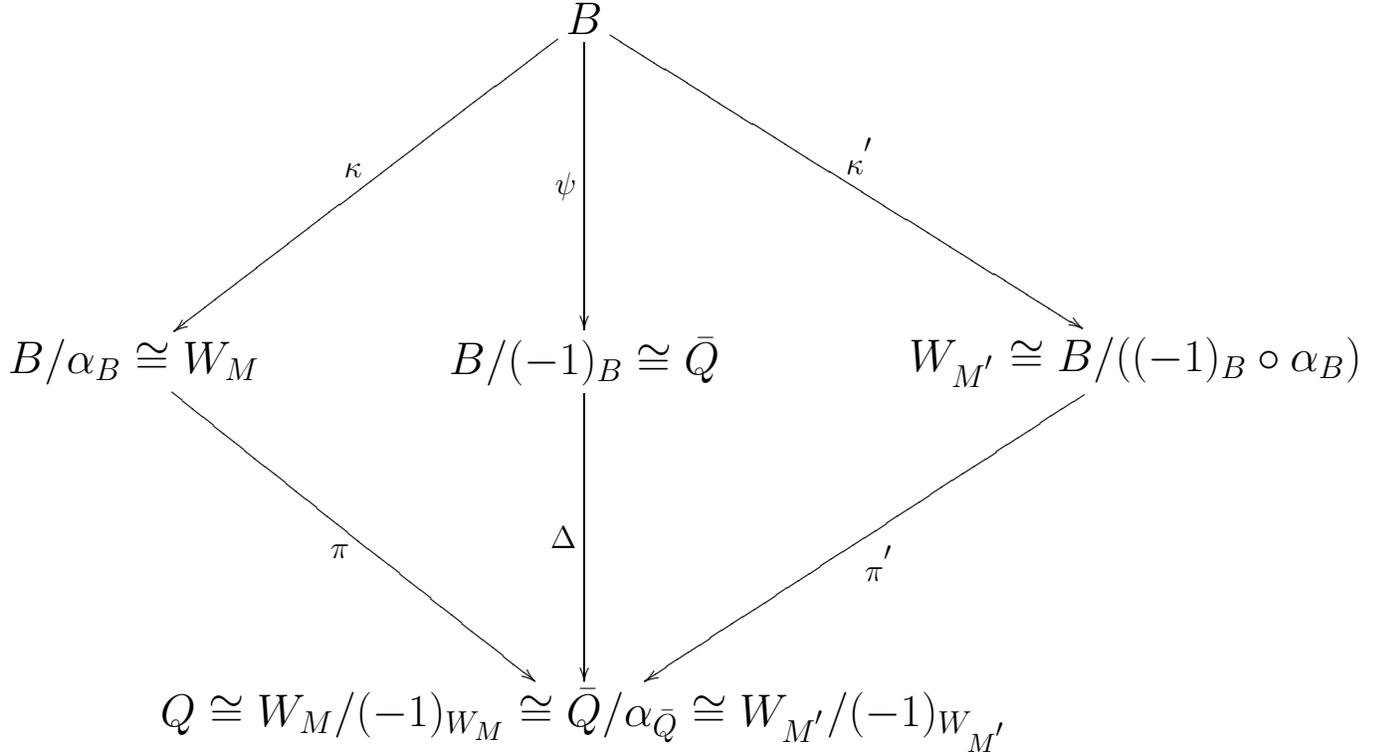

We begin by analyzing the covering map  $\pi: W_M\to Q$ more closely. Recall from Section~\ref{rat} that $B$ is a double cover of $W_M$ with covering map $\kappa: B \to W_M$. By construction, the involution on $B$ which is associated with this covering is $\alpha_B$. That is, $W_M\cong B/\alpha_B$. One refers to $\alpha_B$ as the covering involution of the map $\kappa$. Next we need to analyze  the covering involution on $W_M$ associated with the map $\pi: W_M \to Q$? Determining this will afford us deeper insight into the structure of $B$. To find this involution, first remember that
\be\label{pi}
\xymatrix{
W_M \ar[r]^{\pi} \ar[d]_{\tilde{p}_1} &  Q  \ar[d]^{{p}_1} \ar@{}[drr]|-{,} &&\\
{\mathbb P}^{1} \ar[r]_{\op{id}} & {\mathbb P}^{1} & & \\
} 
\ee

\noindent
where the branch locus of $\pi$ in $Q$ is $M=s\cup i \cup j\cup t^{''}\cup r$. In addition, recall that the branch locus contains the image of the zero section, $\pi^{-1}(s)$, of $W_M$. It is clear from this and the structure of diagram~(\ref{pi}) that  the covering involution on $W_M$ of $\pi$ must act as the identity on the base $\cp{1}$ and leave the zero section point-wise fixed. There is a unique  involution on $W_M$ with this property. To see this, consider the involution $(-1)_B$ on $B$ defined in Section~\ref{rat}. Clearly, this induces a unique involution  $(-1)_{W_M}$ on $W_M$ under the covering map $\kappa: B \to W_M$. It is straightforward to see that $(-1)_{W_M}$ has the required properties and, hence, is the covering involution for $\pi: W_M \to Q$. That is
\be
Q\cong W_M/(-1)_{W_M}.
\ee
Combining this result with the discussion in Section~\ref{rat}, we see that $B$ is a four-fold cover of $Q$ given by
\be\label{l}
\xymatrix{
B\ar[r]^-{\kappa} &  W_M\cong B/\alpha_B \ar[r]^-{\pi}  &  Q\cong W_M/(-1)_{W_M}. 
} 
\ee
Note that  we have first modded out by $\alpha_B$ and then by $(-1)_{W_M}$. This is shown graphically as the left hand routing in Figure~\ref{tricom}.

Expressing $B$ as the four-fold covering of $Q$  in this manner leads to the following important insight. Namely, note that it should not matter whether  we first  mod out by $\alpha_B$ and then by $(-1)_{W_M}$, the involution induced on $W_M$ from $(-1)_B$, as in (\ref{l}),  or first mod out by $(-1)_B$ and then by the induced involution of $\alpha_B$ on $B/(-1)_B$. Either approach represents $B$ as the same four-fold covering of $Q$. Let us specify this second decomposition of the covering more concretely. First, we denote
\be
B/(-1)_B\cong \bar{Q}
\ee
and the covering map associated with involution $(-1)_B$ by 
\be
\psi : B \to \bar{Q}.
\ee
Now denote the induced involution of $\alpha_B$ on $\bar{Q}$ by $\alpha_{\bar{Q}}$. Modding out $\bar{Q}$ by $\alpha_{\bar{Q}}$ must produce a surface isomorphic to $Q$. That is,
\be
\bar{Q}/\alpha_{\bar{Q}}\cong Q.
\ee
We write the associate covering map as 
\be
\Delta: \bar{Q}\to Q.
\ee
Combining these results, we see that the four-fold covering of $Q$ by $B$ can be expressed  in an alternative manner to (\ref{l}) given by
\be\label{m}
\xymatrix{
B\ar[r]^-{\psi} &  \bar{Q}\cong B/(-1)_B \ar[r]^-{\Delta}  &  Q\cong \bar{Q}/\alpha_{\bar{Q}}. 
} 
\ee
Here, we have first modded out by $(-1)_B$ and then by $\alpha_{\bar{Q}}$, which is induced from $\alpha_B$ on $B$. This is shown graphically as the middle routing in Figure~\ref{tricom}.
Expressions (\ref{l}) and (\ref{m}) are not the only ways of writing $B$ as the four-fold cover of $Q$. A third way is as follows. First mod out by the composite involution  $(-1)_B\circ \alpha_B$ on $B$ to produce the surface 
\be
B/((-1)_B\circ \alpha_B)\cong W_{M^{'}}
\ee
and the covering map
\be
\kappa^{'}:B \to W_{M^{'}}.
\ee
Now denote the induced involution of $(-1)_B$ on $W_{M^{'}}$ by $(-1)_{W_{M^{'}}}$. Modding out $W_{M^{'}}$ by $(-1)_{W_{M^{'}}}$ must produce a surface isomorphic to $Q$. That is,
\be
W_{M^{'}}/(-1)_{W_{M^{'}}}\cong Q.
\ee
We write the associated covering map as
\be
\pi^{'}:W_{M^{'}} \to Q.
\ee
Combining these results, we see that
\be\label{r}
\xymatrix{
B\ar[r]^-{\kappa^{'}} &  W_{M^{'}}\cong B/((-1)_B\circ\alpha_B) \ar[r]^-{\pi^{'}}  &  Q\cong W_{M^{'}}/(-1)_{W_{M^{'}}}, 
} 
\ee
where we have first modded out by $(-1)_B\circ \alpha_B$ and then by $(-1)_{W_{M^{'}}}$, which is induced from $(-1)_B$ on $B$. This is shown graphically as the right hand routing in Figure~\ref{tricom}.

\begin{figure}[!ht]
{\Large
\[
\xymatrix@C=2.5in@R=1.5in{ & B
\ar[ld]|-{\boxed{\epsfig{file=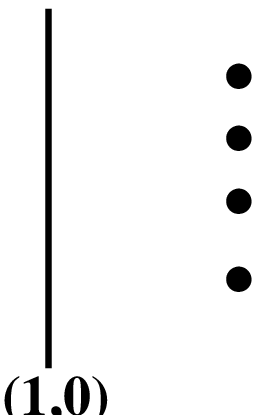,height=.75in}}} 
\ar[d]|-{\boxed{\epsfig{file=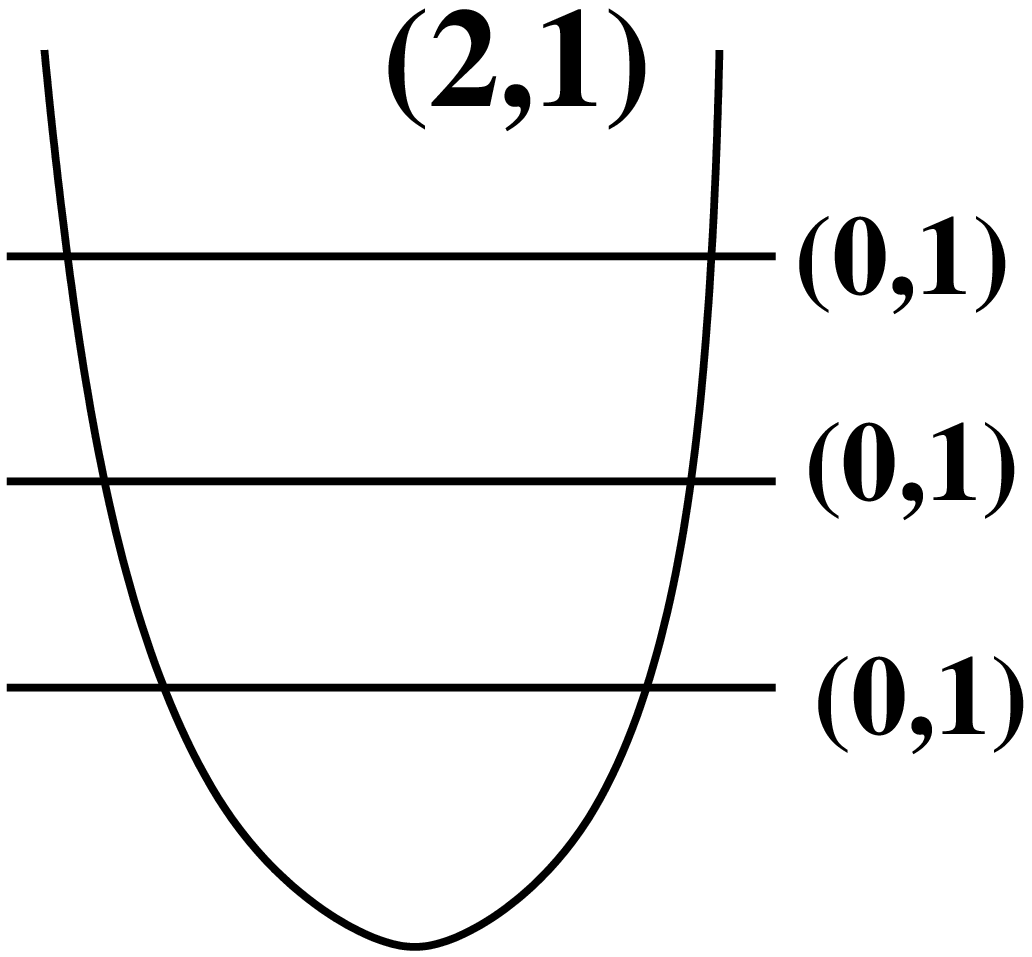,height=.75in}}}
\ar[rd]|-{\boxed{\epsfig{file=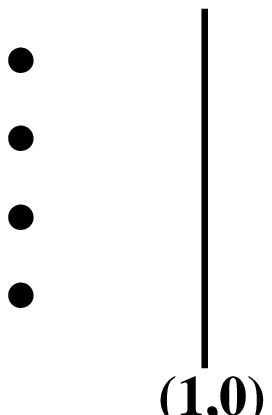,height=.75in}}}  &\\
W_{M} \ar[rd]|-{\boxed{\epsfig{file=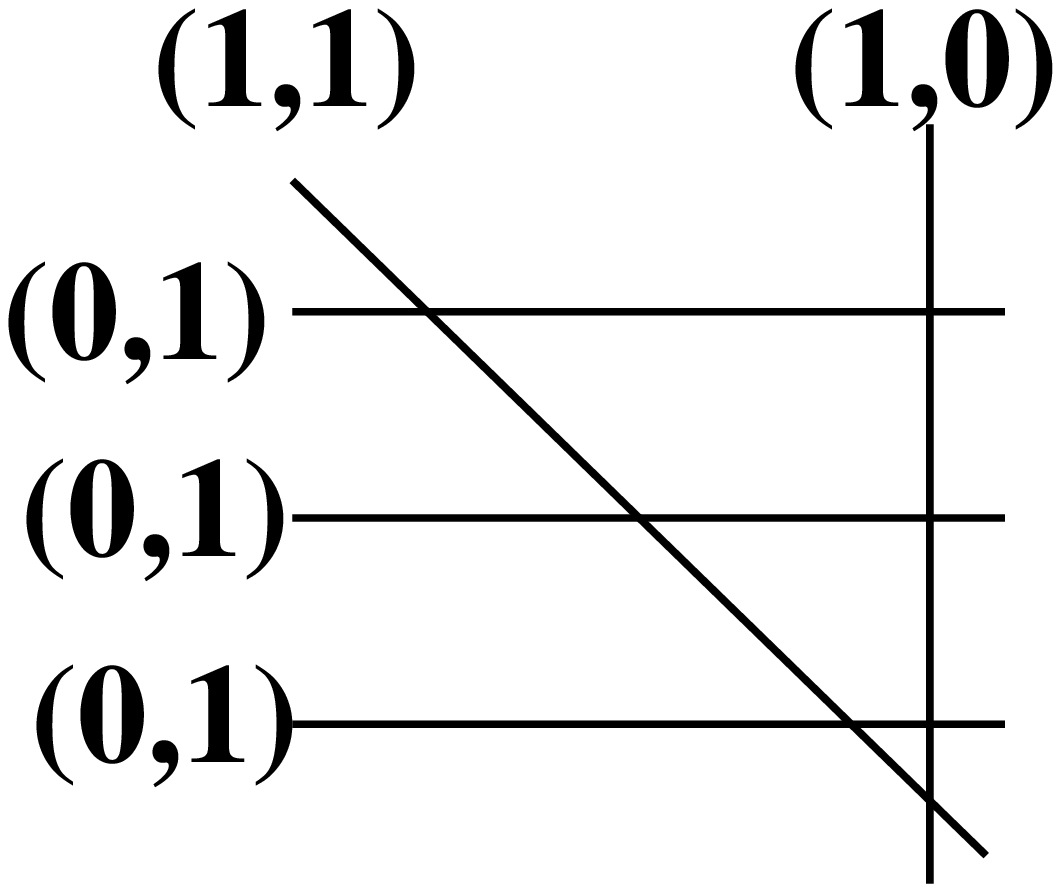,height=.75in}}}  &  
{\bar{Q}} \ar[d]|-{\boxed{\epsfig{file=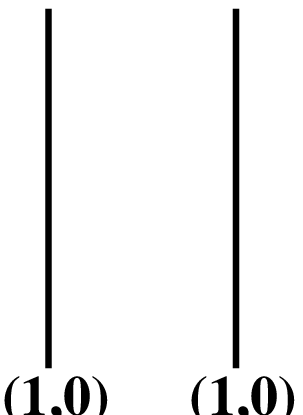,height=.75in}}} & 
W_{M'} \ar[ld]|-{\boxed{\epsfig{file=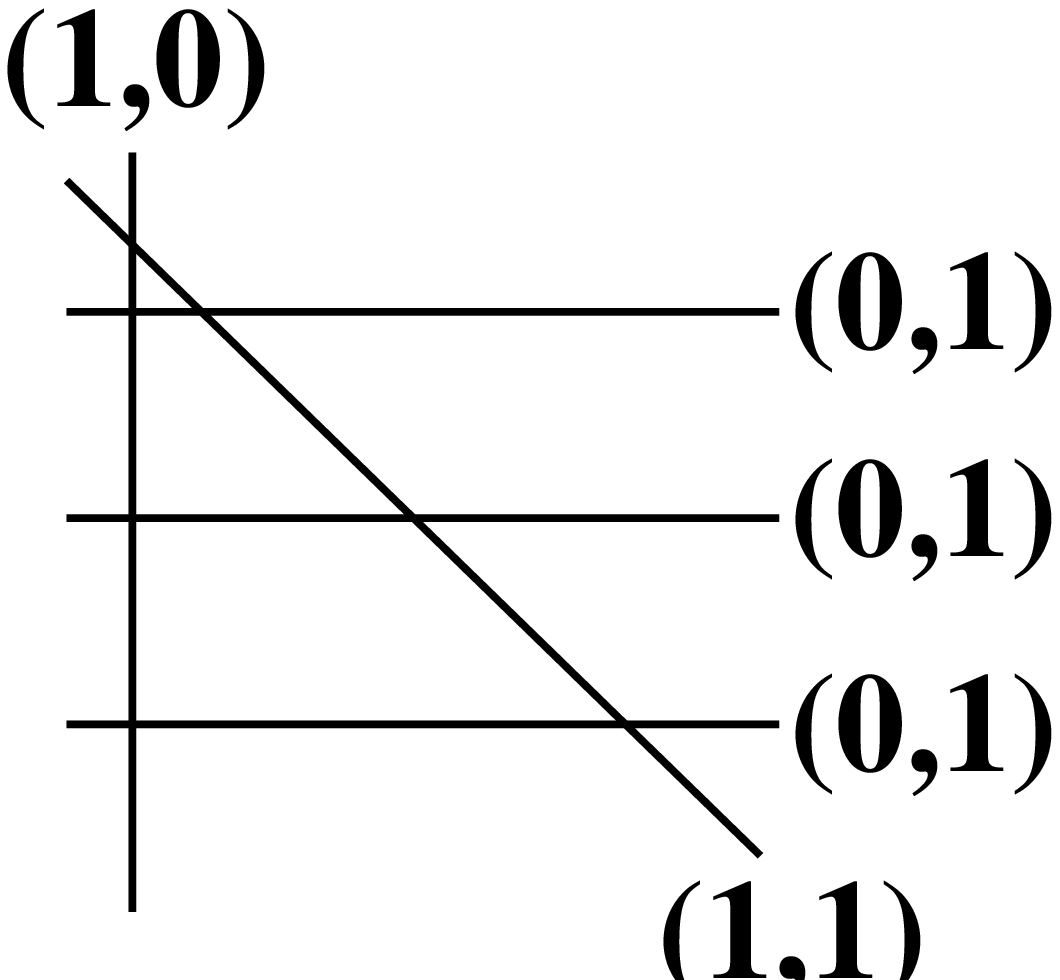,height=.75in}}}\\ 
& Q &
}
\]
}
\caption{$W_{\beta}$ as a double cover of a quadric}
\label{tri-fg} 
\end{figure}

What are the branch loci of $\kappa,\; \pi$ in (\ref{l}), $\psi,\;\op{\Delta}$ in (\ref{m}) and $\kappa^{'},\; \pi^{'}$ in (\ref{r})? The branch loci of $\kappa$ and $\pi$ were discussed in the previous section and are pictured in the left hand routing of Figure~\ref{tri-fg}. Let us now discuss the branch loci  for $\kappa^{'}$ and $\pi^{'}$, beginning with  $\kappa^{'}$. The covering involution for $\kappa^{'}$ is $(-1)_B \circ \alpha_B$. Since $\alpha_B$ acts as the identity map on $f_0$ and as $(-1)_B$ on $f_\infty$, the composition $(-1)_B\circ \alpha_B$ acts as $(-1)_B$ on $f_0$ and as the identity map on $f_\infty$. Hence, the branch locus for $\kappa^{'}$ in $W_M^{'}$  consists of the image of the four fixed points under $(-1)_B$ on $f_0$ and the fiber $f_\infty$. Hence, $W_M^{'}$ has similar properties to $W_M$, one fiber  $f_0$ which contains four $A_1$ surface singularities and which can be resolved into an $I_0^{*}$ fiber, and three singular fibers which can be resolved into $I_2$ fibers. The branch locus in $Q$ of the double cover $\pi^{'}$ now clearly consists of the three $(0,1)$ rulings $s,i$ and $j$, the $(1,1)$ curve $t^{''}$ and the image $\pi^{'}(f_0)$ of the fiber $f_0$ in $W_M^{'}$, which is a $(1,0)$ ruling that we denote by $r^{'}$. That explains the right hand routing of Figure~\ref{tri-fg}. Finally, we consider the branch loci of $\psi $ and $\Delta$. Let us start with $\Delta$. To proceed,  we need to invoke a mathematical theorem which states that the branch locus of $\Delta$ in ${Q}$ consists of the union of the branch loci of $\pi$ and $\pi^{'}$ in $Q$ minus their intersection. Hence, the branch locus of $\Delta$ in ${Q}$ consists of the two $(1,0)$ rulings $r$ and $r^{'}$. Now consider the covering map $\psi$. To find its branch locus, we first must elucidate the structure of surface $\bar{Q}$ more explicitly. To do this, recall that $Q=\cp{1}\times\cp{1}$ has two natural projections $p_i:Q\to \cp{1}_i$ for $i=1,2$ to the $i$th $\cp{1}$ component. Choose any point $x \in \cp{1}_2$ and consider the fiber $\Delta^{-1}(p_2^{-1}(x))$ in $\bar{Q}$. It is the double cover the fiber $p_2^{-1}(x)$, which is isomorphic to $\cp{1}$, branched over the two points where $p^{-1}_2(x)$ intersects the rulings $r$ and $r^{'}$. Hence $\Delta^{-1}(p_2^{-1}(x))$ is isomorphic to $\cp{1}$. It follows that $\bar{Q}$ is isomorphic to $\cp{1}\times \cp{1}$. That is
\be
\bar{Q}=\cp{1}\times \cp{1}.
\ee
We denote the two natural projections of $\bar{Q}$ by $\bar{p}_i:\bar{Q}\to \cp{1}_i$ for $i=1,2$.  Of course, $\bar{Q}$ is smooth since the branch locus of $\bar{Q}$ in $Q$ is smooth.
With this information, we can now determine  the branch locus of the  double cover $\psi$. It is simply the pre-image under $\Delta$ of the curves $s,i,j$ and $t^{''}$. Since each of the rulings $s,i$ and $j$ pulls back to a unique fiber of $\bar{p}_2:\bar{Q}\to \cp{1}$, they are $(0,1)$ rulings in $\bar{Q}$. The pre-image of $t^{''}$ is isomorphic to $\cp{1}$ as well, since $t{''}$ intersects the branch locus for $\Delta$ in two points, namely $t^{''}\cap r$ and $t^{''}\cap r^{'}$. However, $t^{''}$ intersects the generic fiber of $p_2 : Q\to \cp{1}$ at a point not contained in the branch locus of $Q$. Therefore, its pre-image $\op{\Delta}^{-1}(t^{''})$ intersects the fiber $\bar{p}_2^{-1}(x)$ in two points. It follows that $\Delta^{-1}(t^{''})$ is a bidegree $(2,1)$ curve in $\bar{Q}$. The branch loci for $\Delta$ and $\psi$ are described in the middle routing of Figure~\ref{tri-fg}. This finishes our explanation of the branch loci of $B$ as a degree four cover of $Q$. We have shown that there are three different ways of describing it, a fact that will prove to be important later when, for a chosen set of generating elements of $H_2(B,{\mathbb Z})$, we try to compute their ${\mathbb Z}_2 \times {\mathbb Z}_2$ transformation laws. 

We will finish this section with the proof that the sections $e_4$ and $e_6$  of $\beta : B \to \cp{1}$ intersect each fiber $f$ in a point of order two. This fact turned out to be crucial in \cite{opr} to construct commuting involutions on $B$. First, it follows from the commutativity of Figure~\ref{tricom} that
\be
\psi^{-1}\circ \Delta^{-1}=\kappa^{-1}\circ \pi^{-1}.
\ee 
But, we have shown in Section~\ref{rat} that 
\be\label{sec}
\kappa^{-1}\circ \pi^{-1}(s),\;\;\;\;\kappa^{-1}\circ \pi^{-1}(i),\;\;\;\;\kappa^{-1}\circ \pi^{-1}(j),\;\;\;\;\kappa^{-1}\circ \pi^{-1}(t^{''})
\ee
correspond to sections of $\beta : B \to \cp{1}$. Hence, 
\be
\psi^{-1}\circ \Delta^{-1}(s),\;\;\;\;\psi^{-1}\circ \Delta^{-1}(i),\;\;\;\;\psi^{-1}\circ \Delta^{-1}(j),\;\;\;\;\psi^{-1}\circ \Delta^{-1}(t^{''})\;\;\;\;
\ee
correspond to sections of $\beta : B \to \cp{1}$. Secondly, recall from above that $\Delta^{-1}(s),\;\Delta^{-1}(i)$, $\Delta^{-1}(j)$ and $\Delta^{-1}(t^{''})$ are in the branch locus of 
\be
\psi: B \to \bar{Q}\cong B/(-1)_B.
\ee
Hence, the sections (\ref{sec}) are point-wise fixed under $(-1)_B$. Of course, this has to hold for the section $e_9$ corresponding to  $\kappa^{-1}\circ \pi^{-1}(s)$, since we had chosen $e_9$ to be the zero section. In addition, this proves the invariance under $(-1)_B$ of the sections $e_6$ and $e_4$ corresponding to $\kappa^{-1}\circ \pi^{-1}(i) $ and $\kappa^{-1}\circ \pi^{-1}(j)$ respectively. Hence, these sections intersect each fiber $f$ at a point of order two. Finally, since $e_4{\stackrel{.}{+}}e_6$ is the last possible  section invariant under $(-1)_B$, this proves that $e_4{\stackrel{.}{+}}e_6$ corresponds to $\kappa^{-1}\circ \pi^{-1}(t^{''})$. 

Studying the branch locus of $\psi : B \to \bar{Q}$ in $\bar{Q}$, we see that the branch curve has six singularities. Hence, without resolving, $B$ will have six $A_1$ type singularities at the pre-image of these points. Thus, each section $e_9, e_4$ and $e_6$ contains two of these singularities. This is consistent with  the explanation in the previous section. Furthermore, the section  $e_4{\stackrel{.}{+}}e_6$  contains all six singularities.

\section{$B$ as a Ruled Surface}\label{ruled}

In this section, we will show that $B$ is a ruled surface. A ruled surface is a two dimensional fibration whose generic fiber is not a torus $T^2$ but, rather,  $\cp{1}$. Expressing  $B$ as  a ruled surface  will  be very helpful in determining the transformation laws of the elements of $H_2(B,{\mathbb Z})$ under the ${\mathbb Z}_2\times {\mathbb Z}_2 $ automorphism group. Recall that B is elliptically fibered with respect to the mapping $\beta: B \to \cp{1}$. This mapping is defined by $sq\circ\beta= p_1\circ\pi\circ\kappa$ where 
\be
\xymatrix{
sq\circ \beta: B \ar[r]^-{ \kappa} & W_M \ar[r]^{\pi}& Q \ar[r]^{p_1}& \cp{1}, \\
}
\ee
where $p_1: Q \to \cp{1}$ is the natural projection of $Q=\cp{1}\times \cp{1}$ onto its first $\cp{1}$ factor.

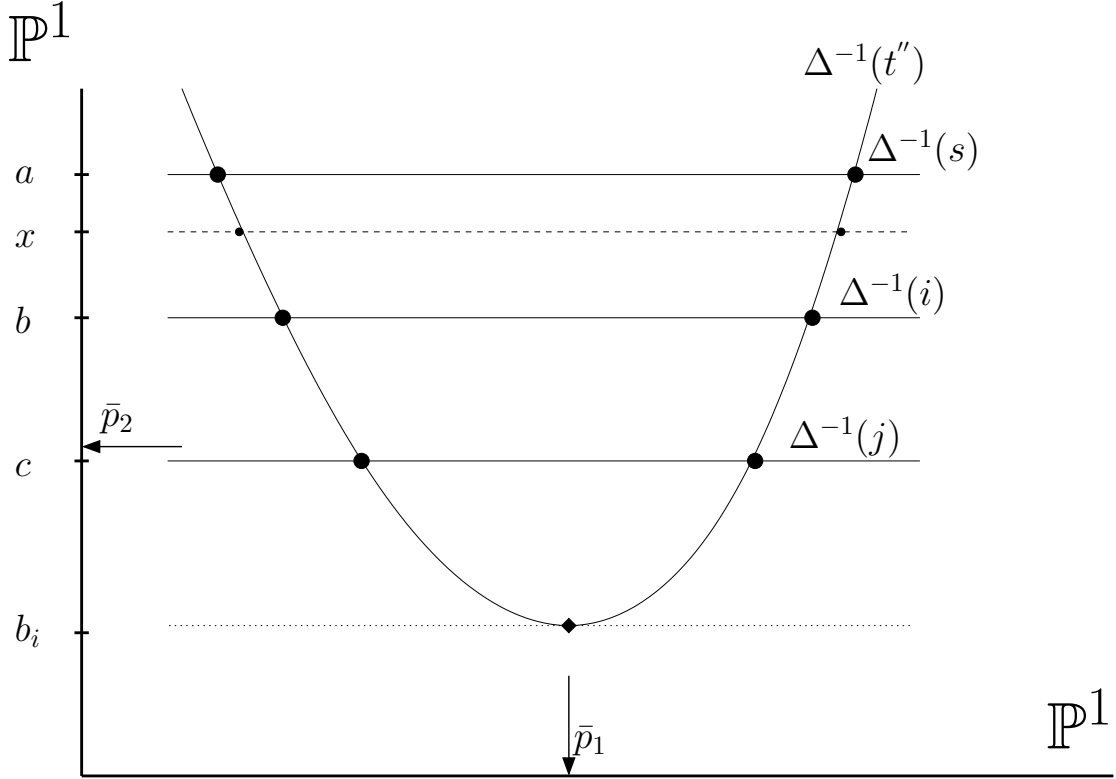
\begin{figure}[!ht]
\begin{center}
\input{psibranch.pstex_t} 
\end{center}
\caption{A schematic representation of the branch locus of $\psi$ in $\bar{Q}=\cp{1}\times \cp{1}$. In addition, a fiber over a generic point $x$ represented by a dashed line and  a fiber over a branch point $b_i$ are shown.}
\label{psibranch} 
\end{figure}

Now recall that there is a second natural projection, namely, $p_2: Q \to \cp{1}$ where $\cp{1}$ is the second $\cp{1}$ factor of $Q$. Clearly, then, we can construct a second fibration of $B$,
\be\label{delta1}
\delta: B \to \cp{1},
\ee
where 
\be
\delta=p_2\circ \pi\circ \kappa
\ee
and
\be
\xymatrix{
\delta: B \ar[r]^-{ \kappa} & W_M \ar[r]^{\pi}& Q \ar[r]^{p_2}& \cp{1}. \\
}
\ee
What is the generic fiber of the mapping $\delta$? To answer this, it is very helpful to use an alternative way that $B$ covers $Q$, namely, the covering through the surface $\bar{Q}=\cp{1}\times \cp{1}$ given in (\ref{m}). This is expressed as the middle routing in  Figure~\ref{tri-fg}. Using the commutativity of Figure~\ref{tricom}, we can write
\be
\delta=p_2\circ \Delta \circ \psi,
\ee
where now
\be
\xymatrix{
\delta: B \ar[r]^-{ \psi} & \bar{Q} \ar[r]^{\Delta}& Q \ar[r]^{p_2}& \cp{1}. \\
}
\ee
We showed in the previous section that for a generic $x\in \cp{1}_2 \subset Q$,  $\Delta^{-1}(p_2^{-1}(x))\cong \cp{1}$. Furthermore, note that if we identify $\cp{1}_2 \subset Q $ with $\cp{1}_2 \subset \bar{Q} $ then $\bar{p}_2^{-1}(x)=\Delta^{-1}(p_2^{-1}(x))$. For a generic point $x\in \cp{1}_2 \subset \bar{Q}$, the fiber $\bar{p}_2^{-1}(x)$ intersects the branch locus of $\psi$ in  $\bar{Q}$ only at two points, each contained in the bidegree $(2,1)$ curve $\Delta^{-1}(t^{''})$. This is indicated by the dashed line in Figure~\ref{psibranch}. Therefore, the double cover $\psi^{-1}(\bar{p}_2^{-1}(x))$ of  $\Delta^{-1}(t^{''})$ is two $\cp{1}$ lines identified at two points and, hence,  is isomorphic to $\cp{1}$. But
\be
\cp{1}\cong \psi^{-1}(\bar{p}_2^{-1}(x))=(p_2\circ \Delta \circ \psi)^{-1}(x)=\delta^{-1}(x)
\ee
is the generic fiber of (\ref{delta1}). We conclude that $B$ can be written as the fibration $\delta: B \to \cp{1}$ whose generic fiber is $\cp{1}$ and, hence, $B$ is a ruled surface. Note that the fiber of $\delta$ is a degree four cover of the fiber of $p_2$ or, equivalently, a degree two cover of the fiber of $\bar{p}_2$. Adopting the second point of view will simplify many calculations.

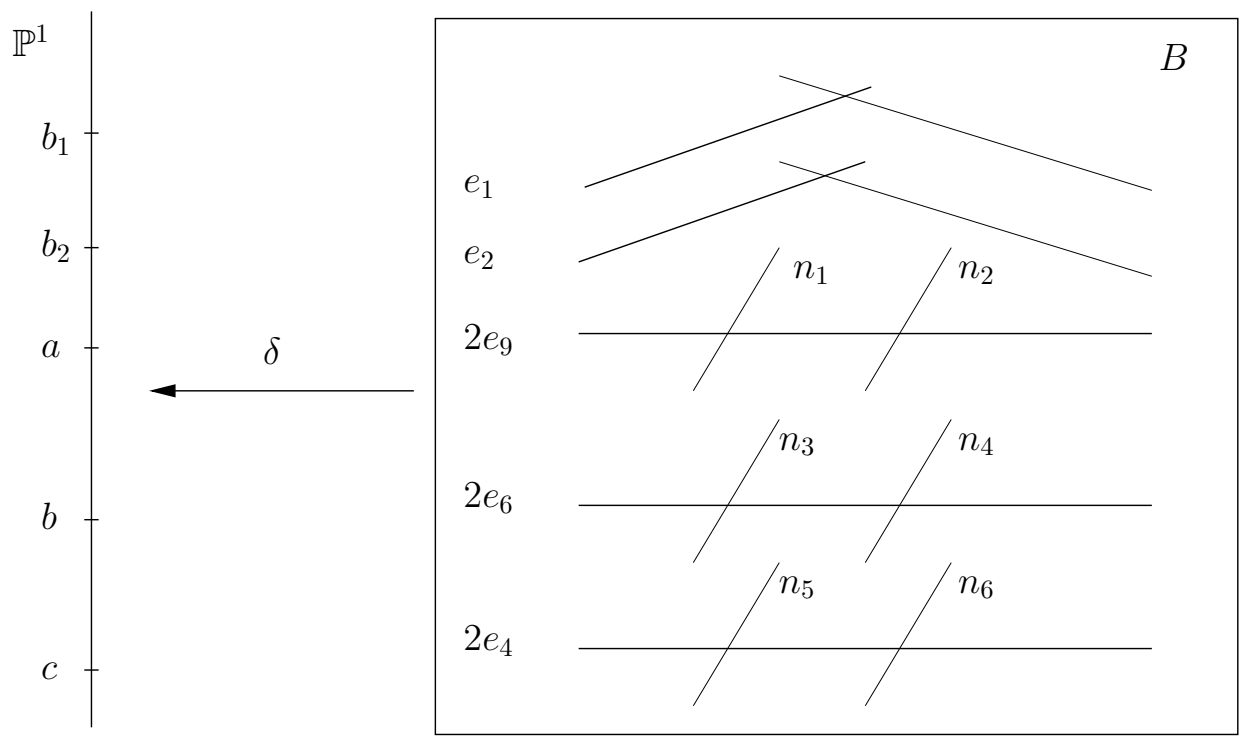
\begin{figure}[!ht]
\begin{center}
\input{deltafibers.pstex_t} 
\end{center}
\caption{The singular fibers of $\delta: B \to \cp{1}$.}
\label{delta} 
\end{figure}

What are the singular fibers of $\delta: B \to \cp{1}$? First consider the three bidegree $(0,1)$ curves  $\Delta^{-1}(s)$,\; $\Delta^{-1}(i)$ and  $\Delta^{-1}(j)$ in the branch locus of $\psi$ in $\bar{Q}$. These are shown in Figure~\ref{psibranch}, where we have defined $\bar{p}_2(\Delta^{-1}(s))=a$, $\bar{p}_2(\Delta^{-1}(i))=b$ and $\bar{p}_2(\Delta^{-1}(j))=c$.  Note that each of these curves is a $\cp{1}$  that intersects $\Delta^{-1}(t^{''})$ at two points. Hence, the pre-image under $\psi$ in $B$ of each of these points must be blown-up to obtain a smooth surface. It follows that the pre-image under $\psi$ of each $(0,1)$ curve is a double $\cp{1}$ line intersecting two exceptional divisors in two distinct points. As discussed previously, we can identify these fibers as
\be
e\cup (n_1 \cup n_2)=\delta^{-1}(a),\;\;\;\;e_6\cup (n_3 \cup n_4)=\delta^{-1}(b),\;\;\;\;e\cup (n_5 \cup n_6)=\delta^{-1}(c)
\ee
respectively. They are shown explicitly in Figure~\ref{delta}.

Are there other singular fibers of $\delta: B \to \cp{1}$? Consider the  $(2,1)$ curve $\Delta^{-1}(t^{''})$ in $\bar{Q}$. Clearly, under the projection $\bar{p}_1: \bar{Q}\to \cp{1}_1$ this curve is a degree one cover of $\cp{1}$ and, hence, 
\be\label{4.9}
\chi(\Delta^{-1}(t^{''}))=\chi(\cp{1})=2.
\ee
However, under the projection $\bar{p}_2: \bar{Q}\to \cp{1}$, $\Delta^{-1}(t^{''})$ is a degree two cover of $\cp{1}$. Then the Riemann-Hurwitz formula states that
\be\label{4.10}
\chi(\Delta^{-1}(t^{''}))=2\chi(\cp{1}\setminus\{b_i\})+A,
\ee
where $\{b_i\}$ for $i=1,\dots,A$ are the branch points of the simple ramification points of $\Delta^{-1}(t^{''})$ over $\cp{1}_2$. By definition, above each $b_i \in \cp{1}_2$, $\Delta^{-1}(t^{''})$ intersects $\bar{p}_2^{-1}(b_i)$ in only one point, not in two. At this point, the ramification point $R_i$, the fiber $\bar{p}_2^{-1}(b_i)$ intersects $\Delta^{-1}(t^{''})$ transversally. This is shown in Figure~\ref{psibranch}.
Therefore, the pre-image under $\psi$ of each of $\bar{p}_2^{-1}(b_i)$  consists of two distinct $\cp{1}$ lines in $B$ intersecting in a point. Note, however, that this point is smooth in $B$. How many such fibers are there? Combining equations (\ref{4.9}) and (\ref{4.10}) we find
\be
A=2.
\ee
We conclude that there are two such singular fibers, $\delta^{-1}(b_i)$ for $i=1,2$. Although not immediately obvious, we will show in the next section that one $\cp{1}$ component of $\delta^{-1}(b_1)$ can be identified with $e_1$, whereas one $\cp{1}$ component of $\delta^{-1}(b_2)$ is $e_2$. These two singular fibers are described in Figure~\ref{delta}.

In summary, we have shown that, in addition to the fibration $\beta: B\to \cp{1}$ discussed in \cite{opr} and Section~\ref{rat}, there is a second way to fiber $B$, namely, $\delta: B\to \cp{1}$. The $\delta$ fibers are generically $\cp{1}$ and, hence, $B$ is a ruled surface. This second fibration has five
singular fibers, all shown explicitly in Figure~\ref{delta}. This fibration will be very helpful in identifying the set of generators of $H_2(B,{\mathbb Z})$, which  will be done in the next section.

\section{Generators of $H_2(B,{\mathbb Z})$}\label{basis}

Recall that a generic rational elliptic surfaces $B$ is a $\cp{2}$  blown-up by a $\cp{1}$ at each of nine separate points. Furthermore, each of these $\cp{1}$ blown-up curves is an irreducible, exceptional divisor of $B$ with self-intersection $-1$. We denote these nine exceptional divisors by $e_1,\dots,e_9$, and have  shown that $e_1,\dots,e_9$ and $l$, where $l$ is the pre-image of a line in $\cp{2}$, form a set of generators of $H_2(B,{\mathbb Z})$. However, the rational elliptic surfaces $B$ that we wish to consider are not generic. Rather, they form a restricted two-dimensional sub-family of such surfaces. One can show that, in this case, some of the nine points in $\cp{2}$ coalesce. This greatly obscures the construction of a surface $B$ in this two dimensional sub-family as  a blow-up of a $\cp{2}$. In fact, as argued in \cite{opr} and Section~\ref{rat}, it is more convenient to obtain these surfaces by different means, namely, as the four-fold cover of $Q$ with a specific branch locus. Be that as it may, for surfaces in this two dimensional sub-family it is now essential that we exactly express $B$ as a blow-up of $\cp{2}$. In doing this, we will explicitly identify nine divisors with self-intersection $-1$, which we again denote as $e_1,\dots,e_9$, and give their structure in $B$. These nine divisors, along with the pre-image of  a line $l$ in $\cp{2}$, will again form the set of generators of $H_2(B,{\mathbb Z})$. To proceed, we find it essential to use the $\cp{1}$ fibration $\delta: B \to \cp{1}$ presented in the previous section.

To identify the rational elliptic surfaces $B$ as a blow-up of $\cp{2}$, we need to invoke the so-called Castelnuovo-Enriques criterion. This involves a process called blow-down, the inverse of blowing-up. The Castelnuovo-Enriques criterion states that for any surface containing a $-1$ curve, that is, a curve isomorphic to $\cp{1}$ of self-intersection $-1$, one can blow down this curve and obtain a smooth surface. For more details on the Castelnuovo-Enriques criterion see \cite{gh}.
To apply this process to the $d{\mathbb P}_9$ surfaces of our two dimensional sub-family,  we must identify nine $-1$ curves and successively blow them down. To do this, recall  that the surface $B$ can be considered as a ruled surface, that is
\be
\delta : B \to \cp{1}
\ee
with the generic fiber of $\delta$ being isomorphic to $\cp{1}$. We will denote the  class of such a fiber by $F$. As discussed in Section~\ref{ruled}, this fibration has five singular fibers, each graphically described in Figure~\ref{delta}.  First, we consider the two singular fibers each of which consists of two $\cp{1}$ components which intersect in a single point. Choose one of these two fibers and denote its components by $c_1$ and $c_2$. Then, as homology classes
\be
F=c_1 + c_2.
\ee
Note that here and henceforth, $+$ and $-$ denote the group operations in the homology group $H_2(B,{\mathbb Z})$. These must not be confused with the addition and inverse of elements in $\Gamma(B)$, defined in Section~\ref{rat}, which, to distinguish them, were denoted using ${\stackrel{.}{+}}$ and ${\stackrel{.}{-}}$.
Recalling that for any fiber class $F^2=0$, and using $c_1\cdot c_2=1$, we find
\be
0=F^2=(c_1+ c_2)^2=c_1^2+2+c_2^2.
\ee
Since the components $c_1$ and $c_2$ are completely symmetric in $B$, it follows that
\be
c_1^2=c_2^2=-1.
\ee
Therefore, this singular fiber  contains two $-1$ curves. By the  Castelnuovo-Enriques criterion explained above, we can blow down either of them. Choose one component, for example $c_1$, and shrink it to a point. What is the self-intersection of the remaining component? Since this component  now forms  a complete fiber of the fibration induced by $\delta$ on the blown down surface, its self-intersection is zero. Hence, there is no longer any $-1$ curve in this fiber.  Of course, the second fiber of $\delta : B \to \cp{1}$ consisting of two $\cp{1}$ components intersecting at a point did not get changed in the process of  blowing down. Therefore, it still admits two $-1$ curves, one of which can now be blown down. Hence, we have identified two $-1$ curves in $B$ which can be blown down simultaneously. 

We now show that there are additional $-1$ curves in the remaining three singular fibers of $\delta : B \to \cp{1}$. Start with the fiber containing the double component $e_9$ and the two exceptional divisors $n_1$ and $n_2$. For a pictorial representation, we refer the reader to Figure~\ref{delta}. For this fiber we can write
\be
F=2e_9+n_1+n_2.
\ee
Recall that we obtained $n_1$ and $n_2$ by blowing up two $A_1$ surface singularities. Hence, blowing them down would not lead to a smooth surface. Therefore, by the Castelnuovo-Enriques criterion, neither exceptional divisor can  be a $-1$ curve. It can be shown, however, that, since they arise by blowing up $A_1$ surface singularities, their self-intersection number is $-2$. 
Using this, and the fact that $e_9$ intersects each of $n_1$ and $n_2$ in one point, we find that
\be
0=(2e_9+n_1+n_2)^2=4e_9^2-2-2+2(2+2)
\ee
and, hence
\be
e_9^2=-1.
\ee
Therefore, we can blow down $e_9$ and obtain a smooth surface. What are the self-intersection numbers of the images of $n_1$ and $n_2$ in the blown down surface? Since this induced fiber in the blown down surface consists of two $\cp{1}$ components which intersect at one point, we are in the same situation as described above and can conclude that the image of each  of $n_1$ and $n_2$ in the blown down surface self-intersects as $-1$. Hence, we can pick one of them and blow it down.

Of course, the reasoning applied to the fiber $2e_9+n_1+n_2$ can be applied to the fibers $2e_6+n_3+n_4$ and $2e_4+n_5+n_6$. Therefore, the three singular fibers each containing a double component and two exceptional divisors allow us to blow down six $-1$ curves. Taking the previous two singular fiber of $\delta : B \to \cp{1}$ into account as well, we have found eight $-1$ curves which can be simultaneously blown down. But since $B\cong d\mathbb{P}_9$,  we need to find one additional $-1$ curve to obtain a $\cp{2}$. Here, we will simply state a result proven in \cite{dopr-i}. 
Recall the map
\be
\psi: B \to \bar{Q},
\ee  
which is a double cover branched over the three $(0,1) $ rulings $\Delta^{-1}(s)$, $\Delta^{-1}(i)$, $\Delta^{-1}(j)$ and the $(2,1)$ curve $\Delta^{-1}(t^{''})$. Consider a $(1,1)$ curve in $\bar{Q}$ passing through one of the two intersection points of each of $\Delta^{-1}(s)\cap \Delta^{-1}(t^{''}) $, $\Delta^{-1}(i)\cap \Delta^{-1}(t^{''}) $ and  $\Delta^{-1}(j)\cap \Delta^{-1}(t^{''}) $. Specifically, we choose this curve to pass through the points whose pre-image under $\psi$ blows-up to $n_2,\;n_4$ and $n_6$ respectively. It can be shown that such a curve in $\bar{Q} $ exists and that its proper transform in $B$ is a $-1$ curve which is a section of the fibration $\delta : B \to \cp{1}$.  We denote this $-1$ curve by $e_7$. 

We are finally in a position to construct our restricted surface $B$ as a blow-up of $\cp{2}$. In the process, we will specifically identify, and give the properties of, nine $-1$ curves in $B$. To do this, begin with one such surface $B$.
First, consider the two singular fibers of $\delta : B \to \cp{1}$ each consisting of two components. Denote by $e_1$ the component of one of these fibers which does not intersect $e_7$ and by $e_2 $ the component of the other fiber which does not intersect $e_7$. Blow them down. Next, blow the $-1$ curves $e_9,\;e_6$ and $e_4$ down. Continuing, blow the images of the curves $n_1,n_3$ and $n_5$ down. Note that none of the curves $e_9, e_6, e_4, n_1, n_3$ or $n_5$ intersects $e_7$. Finally, upon blowing down $e_7$ we obtain $\cp{2}$. Of the curves $e_1, e_2, e_4, e_6, e_7, e_9, n_1, n_3$ and $ n_5$, all are $-1$ curves with the exception of $n_1, n_3$ and $ n_5$. In addition, some of the $e$-type and $n$-type curves intersect. This motivates us to define
\be\label{def1}
e_3=e_4+n_5,\;\;\;e_5=e_6+n_3,\;\;\;e_8=e_9+n_1.
\ee
It is straightforward to verify that each of $e_3, e_5$ and $e_8$ are $-1$ curves and that $e_1,\dots,e_9$ are all non-intersecting.  To reiterate, if we start with a restricted surface $B$ and blow down the curves $e_1, e_2, e_4, e_6, e_9, n_1, n_3, n_5$ and $e_7$ we obtain a $\cp{2}$. Reversing the process, we have constructed $B$ as the blow-up of a $\cp{2}$ surface, as desired.

If we denote by $l$ the pull back of a curve on $\cp{2}$, then $e_1,\dots,e_9$ and $l$ form a set of generators for $H_2(B,{\mathbb Z})$ with the same intersection numbers as given in (\ref{intersection}). That is,
\be
H_2(B,{\mathbb Z})={\mathbb Z}l\oplus(\oplus_{i=1}^9 {\mathbb Z}e_i).
\ee
Therefore, we have successfully described a set of generators of the curve homology of $B$ in terms of the nine $-1$ curves and the pre-image of a $\cp{1}$ line in $\cp{2}$. Note that this motived our original choice in \cite{opr} and Section~\ref{rat} to name the two non-zero sections of order two as $e_4$ and $e_6$. 

In this basis, the components of the $I_2$ fibers can be written as
\be\label{def}
\begin{split}
n_{1} & = e_{8} - e_{9} \\
o_{1} & = f - e_{8} + e_{9} \\
n_{2} & = l - e_{7} - e_{8} - e_{9} \\
o_{2} & = 2l - e_{1} - e_{2} - e_{3} - e_{4} - e_{5} - e_{6}\\
n_{3} & = e_{5} - e_{6} \\
o_{3} & = f - e_{5} + e_{6} \\
n_{4} & = l - e_{7} - e_{6} - e_{5} \\
o_{4} & = 2l - e_{1} - e_{2} - e_{3} - e_{4} - e_{8} - e_{9}\\
n_{5} & = e_{3} - e_{4} \\
o_{5} & = f - e_{3} + e_{4} \\
n_{6} & = l - e_{7} - e_{4} - e_{3} \\
o_{6} & = 2l - e_{1} - e_{2} - e_{5} - e_{6} - e_{8} - e_{9}.\\
\end{split}
\ee
where $f$ is the fiber class of the elliptic fibration $\beta: B \to \cp{1}$.
Let us make a remark on these equations. The expressions for $n_1, n_3$ and $n_5$ follow easily from (\ref{def1}). To understand the expressions for $n_2, n_4$ and $n_6$,  observe the following. Take, for example, $n_2$. It is the component of the singular fiber of $\delta: B \to \cp{1}$ intersecting $e_7$ which is left after blowing down $e_9,\; e_7$ and $n_1$. Hence, the image of $n_2$ in $\cp{2}$ must be $l$. It follows that  $n_2$ in $B$ is given by $l$ reduced by the exceptional divisors $2e_9,\;e_7,$ and $n_1$. That is
\be
n_2=l-2e_9-e_7-n_1,
\ee
which is the expression given in (\ref{def}). The equations for $n_4$ and $n_6$ are derived in a similar way.  Once we have defined $n_1,\dots,n_6$, the proper transforms $o_1,\dots,o_6$ are given by  $o_i=f-n_i,\;i=1,\dots,6$.  Recall from Section~\ref{rat} that the fiber class $f$ is $f=3l-\sum_{i=1}^9 e_i$. Note that the section $e_7$ intersects  $n_2, n_4, n_6$ and $o_1, o_3, o_5$. 
Finally, even through $e_1,\dots,e_9$ and $l$ form a canonical set of generators, the set $e_1, e_2, e_4, e_6$, $ e_7, e_9, n_1, n_3, n_5$ and $l$ also generates  $H_2(B,{\mathbb Z})$. We will use these two sets of generators interchangeably depending on the context.

\section{Transformations under $(-1)_B$}\label{-1}

In the following sections, we will determine the transformation laws of the elements of $H_2(B,\mathbb{Z})$ under the involutions $\tau_{B1}$ and $\tau_{B2}$ defined in Section~\ref{rat}.
To determine these transformations, we will use both the elliptic fibration $\beta: B \to \cp{1}$, whose fiber we continue to call $f$, and the ruling $\delta : B \to \cp{1}$, whose fiber we continue to call $F$. We will also need all three descriptions of  $B$ shown in Figure~\ref{tri-fg}. In this section, for completeness and to establish methodology,  we will determine the induced action of $(-1)_B$ on $H_2(B,\mathbb{Z})$ .

To begin, observe that 
\be\label{-f}
(-1)_B (f)=f.
\ee
This is clear since $(-1)_B$ acts fiber-wise on the elliptic fibration. In addition, note that
\be\label{-F}
(-1)_B (F)=F.
\ee
This follows from the fact that $(-1)_B$ is the covering involution for $ \psi : B \to \bar{Q}$ and $F=\delta^{-1}(x)$ is the double cover of the $\cp{1}$ line $\bar{p}_2^{-1}(x)$, where $\bar{p}_2$ is the second natural projection of $\bar{Q}$, namely, $\bar{p}_2: \bar{Q}\to \cp{1}_2$.

Now consider the two singular fibers of $\delta :B \to \cp{1}$ containing the sections $e_1$ and $e_2$ respectively. Choose one of them, say the fiber containing $e_1$. As discussed above, this fiber can be written as
\be
F=e_1+c_2,
\ee
where $c_2$ intersects $e_7$. Since this fiber arises as the pre-image under $\psi$ of a single $\cp{1}$ line which intersects the branch locus of $\psi$ in $\bar{Q}$ at a single point, it follows that $e_1$ and $c_2$ must be exchanged under $(-1)_B$. That is,
\be
c_2=(-1)_B(e_1).
\ee
Exactly the same statement is true for the singular fiber containing $e_2$. Therefore, the homology class of the fiber $F$ can be written as
\be\label{1}
F=e_i +(-1)_B (e_i),\;i=1,2.
\ee
Note that this is consistent with the expression (\ref{-F}). Now consider a generic fiber of $\delta: B \to \cp{1}$. On might imagine that $l$ would be the class of this fiber. However, it is clear from the previous discussion that the generic fiber must intersect $e_7$ at a single point. Using the intersections  $l\cdot e_7=0$ and $e_7^2=-1$, we see that the fiber class can be written as 
\be\label{2}
F=l-e_7.
\ee
Note that it follows from  (\ref{-F}) that
\be\label{l-e}
(-1)_B(l-e_7)=l-e_7.
\ee
From (\ref{1}) and (\ref{2}), we conclude  that
\be
(-1)_B (e_i)=l-e_7-e_i,\;i=1,2. 
\ee

Next consider the six exceptional divisors $n_i,\; i=1,\dots,6$. Since the image under $\psi$ of the six surface singularities in  $\bar{B}$ are contained in the branch locus of $\psi$ in $\bar{Q}$, it follows that the six exceptional divisors $n_i,\; i=1,\dots,6$ are fixed under $(-1)_B$. That is
\be\label{-n}
(-1)_B(n_i)=n_i,\;i=1,\dots,6.
\ee
Recall that the proper transforms are  defined to be $o_i=f-n_i$ for $ i=1,\dots,6$. Then, it follows from (\ref{-f}) and (\ref{-n}) that
\be
(-1)_B(o_i)=o_i,\;i=1,\dots,6.
\ee
The sections $e_4, e_6$ and $e_9$ are invariant under $(-1)_B$ by definition. Applying these results to the sections $e_3, e_5$ and $e_8$ defined in (\ref{def1}) using the linearity of $(-1)_B$, the invariance of $e_4, e_6$ and $e_9$ and (\ref{-n}), we find, for example, that
\be
(-1)_B(e_3)=(-1)_B(e_4 +n_5)=e_4 +n_5=e_3.
\ee
Similarly, it follows that
\be
(-1)_B(e_5)=e_5,\;\;\;\;\;(-1)_B(e_8)=e_8.
\ee
Finally, using $o_2$ given in (\ref{def}), the fact that $(-1)_B(o_2)=o_2$ and the invariance of $e_3, e_4, e_5$ and $e_6$ under $(-1)_B$, it follows that
\be\label{-l}
(-1)_B(l)=2l-(e_1+e_2+e_7).
\ee
Furthermore, from (\ref{l-e}) and (\ref{-l}) we find 
\be
(-1)_B(e_7)=l-(e_1+e_2).
\ee
This completes the analysis for the action of $(-1)_B$ on the set of generators  $e_1,\dots,e_9$ and $l$ of $H_2(B,{\mathbb Z})$. These results are summarized in Table~\ref{tab(-1)}. The transformation of an arbitrary class in $H_2(B,{\mathbb Z})$ follows immediately from these results.

\bigskip
\begin{table}[!ht]
\begin{center}
\begin{tabular}{|l||l|l|l|l|l|l|l|l|l|l|}\hline 
  & $e_1$ & $e_2$ & $e_3$ & $e_4$ & $e_5$ & $e_6$ & $e_7$ & $e_8$ & $e_8$ & $l$ \\ \hline
$(-1)_{B}$  &$l-e_7-e_1$&$l-e_7-e_2$&$e_3$&$e_4$&$e_5$&$e_6$& $l-e_1-e_2$& $e_8$&$e_9$&$2l-(e_1+e_2+e_7)$ \\ \hline
\end{tabular}
\end{center}
\caption{The action of $(-1)_{B}$ on the canonical set of generators of $H_{2}(B,{\mathbb Z})$.}\label{tab(-1)}
\end{table}

\section{The Action of $\alpha_B$}\label{alpha}

In this section, we begin our analysis of the transformations of  the elements of $H_2(B,{\mathbb Z})$ under $\tau_{B1}$ and $\tau_{B2}$.  Recalling from (\ref{inv}) that $\tau_{B1}=t_{e_6}\circ \alpha_B$ and $\tau_{B2}=t_{e_4}\circ \alpha_B$, we start by finding the action of $\alpha_B$ on $H_2(B,{\mathbb Z})$. First, observe that
\be\label{af}
\alpha_B(f)=f.
\ee
This follows from the fact that $\alpha_B$ preserves the fiber of the elliptic fibration $\beta: B \to \cp{1}$. The commutativity of Figure~\ref{tri-fg} implies that $\delta^{-1}(x)=\kappa^{-1}(\pi^{-1}(p_2^{-1}(x)))$. Since $\alpha_B$ is the covering involution for map $\kappa$, this shows that the  fiber class of $\delta : B \to \cp{1}$ is also invariant under $\alpha_B$. That is,
\be\label{aF}
\alpha_B(F)=F.
\ee

Let us first discuss the generic representation of $F$ given in (\ref{2}). Then this and (\ref{aF}) imply that under $\alpha_B$
\be
\alpha_B(l-e_7)=l-e_7.
\ee
Now, consider the singular fiber $2e_9+n_1+n_2$ of $\delta: B \to \cp{1}$. Note that $\alpha_{\bar B}$ exchanges the two surface singularities in  $\bar{B}$ which blow up into $n_1$ and $n_2$. Hence, $\alpha_B$ exchanges $n_1$ and $n_2$. That is, 
\be\label{n1}
\alpha_B(n_1)=n_2,\;\;\;\;\;\alpha_B(n_2)=n_1.
\ee
Since, by (\ref{aF}), $2e_9+n_1+n_2$ must be invariant under $\alpha_B$, it follows that
\be
\alpha_B(e_9)=e_9.
\ee
Similar remarks hold for the singular fibers $2e_6+n_3+n_4$ and $2e_4+n_5+n_6$. We see, therefore,  that
\be\label{n3}
\alpha_B(n_3)=n_4, \;\;\;\;\;\alpha_B(n_4)=n_3
\ee
and  
\be\label{n4}
\alpha_B(n_5)=n_6, \;\;\;\;\;\alpha_B(n_6)=n_5.
\ee
Combining this with the invariance of both singular fibers under $\alpha_B$, we conclude that
\be\label{a6}
\alpha_B(e_6)=e_6,\;\;\;\;\;\alpha_B(e_4)=e_4.
\ee
Next consider  the proper transform $o_i=f-n_i$ for $ i=1,\dots,6$. Then  it follows from (\ref{af}), (\ref{n1}), (\ref{n3}) and (\ref{n4}) that
\be
\alpha_B(o_{2j-1})=o_{2j},\;\;\;\;\;\;\alpha_B(o_{2j})=o_{2j-1}
\ee
for $j=1,2,3$. It is easy to read off the transformation laws for  $e_3, e_5$ and $e_8$ defined in (\ref{def1}). For example, using (\ref{n4}), (\ref{a6}) and (\ref{def}) we see that
\be\label{a3}
\alpha_B(e_3)=\alpha_B(e_4 +n_5)=e_4 +n_6=l-e_3-e_7.
\ee
Similarly, we have
\be\label{a5}
\alpha_B(e_5)=\alpha_B(e_6 +n_3)=e_6 +n_4=l-e_5-e_7
\ee
and
\be\label{a8}
\alpha_B(e_8)=\alpha_B(e_9 +n_1)=e_9 +n_2=l-e_7-e_8.
\ee

Let us now discuss the transformation law of the sections $e_1$ and $e_2$. Each of these is  contained  in one of the two singular fibers of $\delta : B \to \cp{1}$ consisting of two components. Consider either one of these singular fibers. Since it must be stable under $\alpha_B$, $\alpha_B$ can either exchange its two components or leave them invariant. To answer this question, we must analyze these two fibers of $\delta$ more closely. By definition, these fibers are double covers of the $(0,1)$ rulings in $\bar{Q}$ which intersect the $(2,1) $ curve $\Delta^{-1}(t^{''})$ tangentially. It is easy to see that these $(0,1)$ rulings in $\bar{Q}$ are double covers of the two $(0,1)$ rulings in ${Q}$ with pass  through either $r^{'}\cap t^{''}$ or $r\cap t^{''}$. Let us assume that $e_1$ is contained in the fiber of $\delta$ corresponding to the degree four cover of the $(0,1)$ ruling in ${Q}$ which passes through the point $r^{'}\cap t^{''}$. It follows that $e_2$ is contained in the fiber of $\delta$ corresponding to the degree four cover of the $(0,1)$ ruling in ${Q}$ which passes through the point $r\cap t^{''}$. From the commutativity of  Figure~\ref{tri-fg}, we can discuss these two fibers of $\delta$ using the relation $\delta^{-1}(x)=\kappa^{-1}(\pi^{-1}(p_2^{-1}(x)))$. To determine the transformation law of $e_1$, consider the $(0,1)$ ruling in ${Q}$ which passes through $r^{'}\cap t^{''}$. This ruling intersects the branch divisor of the map $\pi$ at two different points. Hence, its double cover in $W_M$ is a smooth $\cp{1}$. This curve intersects the branch divisor of $\kappa $ at a single point in $f_0$ in $W_M$. Therefore, its pre-image in $B$ consists of two components which are exchanged by the covering involution for $\kappa$ and, hence, by $\alpha_B$. Since one of the two components is $e_1$,
\be
F= e_1+\alpha_B(e_1)
\ee
represents the class of a full fiber of $\delta$. It follows from this and (\ref{2}) that
\be
e_1+\alpha_B(e_1)=l-e_7,
\ee
or equivalently
\be\label{a1}
\alpha_B(e_1)=l-e_1-e_7.
\ee
What about the fiber of $\delta$  containing $e_2$? Consider the  $(0,1)$ ruling in ${Q}$ which passes through $r\cap t^{''}$. This is the only point where this ruling intersects the branch locus of $\pi$ in $Q$. Hence, the pre-image of this ruling in $W_M$ consists of two components, both isomorphic to $\cp{1}$. What is the pre-image of these two $\cp{1}$ lines in $B$? Since each component intersects the branch locus of $\kappa$ in two points, namely its intersection with $f_0$ and the point $\pi^{-1}(p_1)$ in $W_M$, the pre-image of each component is a smooth $\cp{1}$. Clearly, these components are not exchanged under the covering involution of $\kappa$ and, hence, under $\alpha_B$. One of the components we called $e_2$. Therefore,
\be\label{a2}
\alpha_B(e_2)=e_2.
\ee  

Finally,  solving the equations $\alpha_B(o_1)=o_2$ and $\alpha_B(l-e_7)=l-e_7$ using (\ref{df}), (\ref{def}), (\ref{a6}), (\ref{a3}), (\ref{a5}), (\ref{a1}) and (\ref{a2}),  we obtain 
\be
\alpha_B(l)=3l-e_1-e_3-e_5-2e_7-e_8
\ee
and
\be
\alpha_B(e_7)=2l-e_1-e_3-e_5-e_7-e_8.
\ee
This concludes our calculation of the transformation laws of the canonical set of generators of $H_2(B,\mathbb{Z})$ under $\alpha_B$. The results are summarized in Table~\ref{tab(al)}. The action of $\alpha_B$ on any class in $H_2(B,\mathbb{Z})$ follows immediately from these results.

\bigskip
\begin{table}[!ht]
\begin{center}
\begin{tabular}{|l||l|l|l|l|l|l|l|l|l|l|}\hline 
  & $e_1$ & $e_2$ & $e_3$ & $e_4$ & $e_5$  \\ \hline
$\alpha_{B}$  &$l -e_7-e_1 $&$e_2$&$ l-e_3-e_7$&$e_4$&$l-e_ 5-e_7$ \\ \hline\hline
& $e_6$ & $e_7$ & $e_8$ & $e_9$ & $l$ \\ \hline
$\alpha_{B}$ &$e_6 $&$ 2l-(e_1+e_3+e_5+e_7+e_8)$&$l-e_7-e_8  $ &$e_9$ & $3l-(e_1+e_3+e_5+2e_7+e_8)$\\ \hline
\end{tabular}
\end{center}
\caption{The action of $\alpha_{B}$ on the canonical  set of generators of $H_{2}(B,{\mathbb Z})$.}\label{tab(al)}
\end{table}

\section{The Action of $t_{e_6}$ and $t_{e_4}$}\label{trans}

In this section, we complete the analysis of the transformations of the elements of $H_{2}(B,{\mathbb Z})$ under $\tau_{B1}$ and $\tau_{B2}$ by computing the actions of $t_{e_6}$ and $t_{e_4}$. We will proceed in several steps, first computing the transformations of the $I_2$ fiber components $n_i$ and $o_i$ for $i=1,\dots,6$ and then using these results to find the transformations of the canonical set of generators of $H_{2}(B,{\mathbb Z})$.

\subsection{The Action of $t_{e_6}$ and $t_{e_4}$ on Fiber Components}

Recall that $t_{e_6}$ and $t_{e_4}$ preserve the fibers of the elliptic fibration $\beta : B \to \cp{1}$. That is 
\be\label{t6f}
t_{e_6}(f)=f,\;\;\;\;\;t_{e_4}(f)=f.
\ee
As discussed in Section~\ref{rat}, any surface $B$ in the restricted two parameter family has six single $I_2$ fibers. Each such $I_2$ fiber is composed of two $\cp{1}$ components, $n$ and $o$, intersecting at two singular points. This is shown in Figure~\ref{I2}. That is, the $I_2$ singular fibers of $B$ can be written as $n_i\cup o_i,\,i=1,\dots,6$. The relationship of these fiber components to the canonical set of generators $e_1,\dots,e_9$ and $l$ of $H_{2}(B,{\mathbb Z})$ was given in (\ref{def}). It follows from these remarks and (\ref{t6f}) that the action of $t_{e_6}$ and $t_{e_4}$ either exchanges the two components of an $I_2$ fiber or leaves them invariant. The remainder of this sub-section is devoted to deciding which of these possibilities occur.

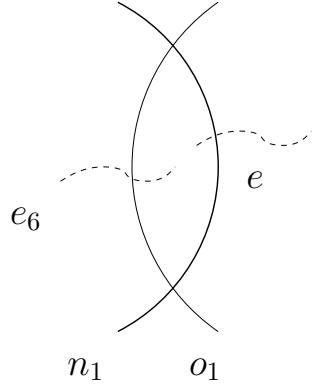
\begin{figure}[!ht]
\begin{center}
\input{intersect1.pstex_t} 
\end{center}
\caption{A schematic representation of the intersection of $e$ and $ e_6$ with $n_1$ and $o_1$.}
\label{tn1} 
\end{figure}

We begin our analysis with the first $I_2$ fiber, defined as $n_1\cup o_1$, and consider the action on it of $t_{e_6}$. Recall that each smooth fiber of $\beta : B \to \cp{1}$ admits an Abelian group structure whose identity element is the point of  intersection of the fiber with the zero section $e=e_9$.  It can be shown that, after removing the two singular points, each  $I_2$ fiber also carries an Abelian group structure. Using (\ref{def}) and the intersection numbers of the canonical set of generators of $H_{2}(B,{\mathbb Z})$, we see that
\be
 n_1\cdot e=1,\;\;\;\;\;o_1\cdot e=0. 
\ee
Hence, the identity element is in the $n_1$ component. In addition, note that
\be
n_1\cdot e_6=0,\;\;\;\;\;o_1\cdot e_6=1. 
\ee
These results are represented pictorially in  Figure~\ref{tn1}.
Now consider a specific point on $n_1$, which we choose to be the identity element  $n_1\cap e$. Translating this point by $t_6$ gives
\be
t_6(n_1\cap e)=(n_1\cup o_1)\cap e_6=o_1\cap e_6,
\ee
which is a point in $o_1$. If follows the $t_{e_6}$ must exchange $n_1$ and $o_1$. That is
\be
t_{e_6}(n_1)=o_1,\;\;\;\;\;t_{e_6}(o_1)=n_1.
\ee
It is straightforward to see that $t_{e_6}$ will exchange the components of any $I_2$ fiber for which $e$ and $e_6$ intersect different components, as in Figure~\ref{tn1}.

\begin{figure}[!ht]
\begin{center}
\input{intersect2.pstex_t} 
\end{center}
\caption{A schematic representation of the intersection of $e$ and $ e_6$ with $n_5$ and $o_5$.}
\label{tn5} 
\end{figure}
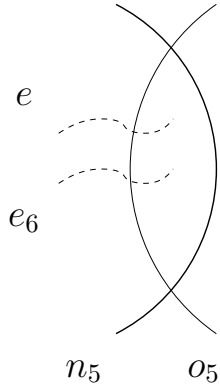

What happens for an $I_2$ fiber for which $e$ and $e_6$ intersect the same component? As an example, consider  the fifth $I_2$ fiber defined as $n_5 \cup o_5$. Using (\ref{def}) and the intersection numbers of the canonical set of generators of $H_{2}(B,{\mathbb Z})$, it is easy to show that
\be
n_5 \cdot e= n_5\cdot e_6=0
\ee
and 
\be
o_5 \cdot e= o_5\cdot e_6=1.
\ee
Here, the identity element is in the $o_5$ component.  These results are represented in  Figure~\ref{tn5}.
Again, consider the zero element $o_5\cap e$. Translating this point by $t_{e_6}$ gives
\be
t_{e_6}(o_5\cap e)=(n_5\cup o_5)\cap e_6= o_5\cap e_6,
\ee
which remains a point in $o_5$. We conclude that $t_{e_6}$ leaves $n_5$ and $o_5$ stable. That is 
\be
t_{e_6}(n_5)=n_5,\;\;\;\;\;t_{e_6}(o_5)=o_5.
\ee
Clearly, this will be the case for any $I_2$ fiber for which $e$ and $e_6$ intersect the same component, is in Figure~\ref{tn5}.

It is easy now to derive the remaining transformation laws of $n_i$ and $o_i$ for $i=1,\dots,6$ under $t_{e_6}$ and $t_{e_4}$. From (\ref{def}) one can read off the intersection numbers of the various $I_2$ fiber components with $e, e_6$ and $e_4$. By the above remarks, it is then straightforward to determine the transformation laws of the fiber components under $t_{e_6}$ and $t_{e_4}$. We summarize our findings in Table~\ref{tI2}, where we also specify the action of the  compositions $\tau_{B1}= t_{e_6}\circ \alpha_{B}$ and $\tau_{B2}= t_{e_4}\circ \alpha_{B}$ on $n_i$ and $o_i$ for $i=1,\dots,6$. To do the latter, we combine the results of this section with those of Section \ref{alpha}. 

\bigskip
\bigskip
\begin{table}[!ht]
\begin{center}
\begin{tabular}{|l||c|c|c|c|c|c|c|c|} \hline
  & $e_9$ &$ e_6$ &$ e_4$ & $t_{e_6}$ & $t_{e_4}$ & $\alpha_B$ & $\tau_{B1}=t_{e_6}\circ \alpha_{B} $ & $\tau_{B2}=t_{e_4}\circ \alpha_{B} $ \\ \hline \hline
$n_1$ & $1$ & $0$ & $0$ & $o_1$ & $o_1$ & $n_2$ &  $o_2$ & $o_2$ \\ \hline
$n_2$ & $1$ & $0$ & $0$ & $o_2$ & $o_2$ & $n_1$ &  $o_1$ & $o_1$ \\ \hline
$n_3$ & $0$ & $1$ & $0$ & $o_3$ & $n_3$ & $n_4$ &  $o_4$ & $n_4$ \\ \hline
$n_4$ & $0$ & $1$ & $0$ & $o_4$ & $n_4$ & $n_3$ &  $o_3$ & $n_3$ \\ \hline
$n_5$ & $0$ & $0$ & $1$ & $n_5$ & $o_5$ & $n_6$ &  $n_6$ & $o_6$ \\ \hline
$n_6$ & $0$ & $0$ & $1$ & $n_6$ & $o_6$ & $n_5$ &  $n_5$ & $o_5$ \\ \hline
$o_1$ & $0$ & $1$ & $1$ & $n_1$ & $n_1$ & $o_2$ &  $n_2$ & $n_2$ \\ \hline
$o_2$ & $0$ & $1$ & $1$ & $n_2$ & $n_2$ & $o_1$ &  $n_1$ & $n_1$ \\ \hline
$o_3$ & $1$ & $0$ & $1$ & $n_3$ & $o_3$ & $o_4$ &  $n_4$ & $o_4$ \\ \hline
$o_4$ & $1$ & $0$ & $1$ & $n_4$ & $o_4$ & $o_3$ &  $n_3$ & $o_3$ \\ \hline
$o_5$ & $1$ & $1$ & $0$ & $o_5$ & $n_5$ & $o_6$ &  $o_6$ & $n_6$ \\ \hline
$o_6$ & $1$ & $1$ & $0$ & $o_6$ & $n_6$ & $o_5$ &  $o_5$ & $n_5$ \\ \hline
\end{tabular}
\end{center}
\caption{The intersection of the components of the singular fibers with the sections $e, e_6$ and $e_4$, and their transformation under the automorphisms  $t_{e_6}$, $t_{e_4}$, $\alpha_B$, $\tau_{B1}$ and $\tau_{B2}$. }\label{tI2}
\end{table}

\subsection{Transformations of the Canonical Generators under $t_{e_6}$}

Having found the action of $t_{e_6}$ and $t_{e_4}$ on the fiber components, we now use these results to compute the  transformation laws of the canonical set of generators of $H_2(B,{\mathbb Z})$ under each translation. In this sub-section, we consider the action of $t_{e_6}$. Some of these laws are easy to read off, others are more complicated. Let us begin with the easy ones.
First, it follows from the definition of a translation, and the fact that $e_9=e$ is the zero section, that
\be\label{4.1}
t_{e_6}(e_9)=e_9{\stackrel{.}{+}}e_6=e_6.
\ee
Furthermore, since $e_6$ is a section of order two, we have
\be\label{4.2}
t_{e_6}(e_6)=e_6{\stackrel{.}{+}}e_6=e_9.
\ee
With help of (\ref{4.1}), (\ref{4.2}) and Table~\ref{tI2}, it is now straightforward to determine the transformation laws of $e_5$ and $e_8$. Using, in addition, (\ref{def1}), the definition of $o_i=f-n_i$ for $i=1,\dots,6$ and (\ref{def}), we find that
\be
t_{e_6}(e_{5}) = t_{e_6}(e_{6} + n_{3}) = e_6 + o_3 = e_9+f-n_3 = f-e_5+e_6+e_9
\ee
and
\be
t_{e_6}(e_{8}) = t_{e_6}(e_{9} + n_{1}) = e_6 + o_1 = e_6+f-n_1 = f+e_6-e_8+e_9.\ee

To find the transformation laws of $e_1, e_2, e_4$ and $e_7$ under $t_{e_6}$, however, we need to exploit the relationship between elements in $H_2(B,{\mathbb Z})$ and line bundles which, until now, was unnecessary. First, by Poincare duality, we have a canonical isomorphism
\be
H_2(B,{\mathbb Z})\cong H^2(B,{\mathbb Z}).
\ee
Furthermore, it can be shown that on a  simply connected smooth surface
\be
H^2(B,{\mathbb Z})\cong Pic(B),
\ee
where $Pic(B)$ is the space of all line bundles up to isomorphism on $B$. Hence, we obtain
\be
H_2(B,{\mathbb Z})\cong Pic(B).
\ee
More explicitly, for any element $\xi \in H_2(B,{\mathbb Z})$, we have
\be
\xi \to {\mathcal O}_B(\xi).
\ee
What is the relationship between an involution  $\tau: B \to B$ on $H_2(B,{\mathbb Z})$ and its action on $Pic(B)$? To answer this, consider a class $\xi \in H_2(B,{\mathbb Z})$ and its associated line bundle ${\mathcal O}_B(\xi) \in Pic(B)$. If $\tau$ is an involution on $B$, then we can  define  the pull-back bundle, $\tau^{*}{\mathcal O}_B(\xi)$, under $\tau$ which satisfies the commutative diagram 
\be
\xymatrix{
\tau^{*}{\mathcal O}_B(\xi) \ar[r]^-{\pi_2}\ar[d]_-{\pi^{'}} & {\mathcal O}_B(\xi)\ar[d]^-{\pi}  \ar@{}[drr]|-{.} && \\
B \ar[r]^-{\tau} &  B &&
}
\ee
Map $\pi_2$ takes the fiber $\pi^{'-1}(p) \subset \tau^{*}{\mathcal O}_B(\xi)$ to the fiber $\pi^{-1}(\tau(p)) \subset {\mathcal O}_B(\xi)$ for all points $p\in B$. As is, the bundle $\tau^{*}{\mathcal O}_B(\xi)$ is abstractly defined. However, one can identify $\tau^{*}{\mathcal O}_B(\xi)$ more concretely, as we now show. We will assume that the class in $H_2(B,{\mathbb Z})$ can be represented by a curve, which we also call $\xi$. This implies that ${\mathcal O}_B(\xi)$ has a section 
\be
s: B \to {\mathcal O}_B(\xi)
\ee
whose vanishing locus is on $\xi$, that is, $s(\xi)=0$. Now consider the map
\be
s^{'}: B \to \tau^{*}{\mathcal O}_B(\xi)
\ee
defined by
\be
\pi_2\circ s^{'}=s\circ \tau.
\ee
Then, it is clear that $s^{'}$ is a section of $\tau^{*}{\mathcal O}_B(\xi)$ satisfying 
\be
s^{'}(\tau^{-1}(\xi))=0.
\ee
Hence, the curve, or homology cycle, representing $\tau^{*}{\mathcal O}_B(\xi) $ is $\tau^{-1}(\xi)$. That is,
\be
\tau^{*}{\mathcal O}_B (\xi) \cong  {\mathcal O}_B(\tau^{-1}(\xi))\cong  {\mathcal O}_B(\tau(\xi)),
\ee
where we have used the fact that, since $\tau$ is an involution, $\tau^{-1}=\tau$. We can now answer the above question definitely. The action $\xi \to \tau(\xi)$ of an involution on $H_2(B,{\mathbb Z})$ induces the action
\be\label{mappic}
{\mathcal O}_B(\xi)\to {\mathcal O}_B(\tau^{-1}(\xi))\cong {\mathcal O}_B(\tau(\xi))
\ee
on $Pic(B)$. We will use this relationship to determine the transformation laws of $e_1, e_2, e_4$ and $e_7$ under translations.

To do this, we first need to review some relevant facts about line bundles. Recall that $B$ is an elliptic fibration $\beta: B \to \cp{1}$. We want to discuss the group structure of both the smooth elliptic fibers, $T^2$, and the space of sections, $\Gamma(B)$, in terms of line bundles on $T^{2}$ and $B$ respectively. First, we consider a smooth elliptic curve $T^{2}$ with a fixed zero point, which we call $e$. Each point $p \in T^2$ determines a line bundle, namely
\be
p \to {\mathcal O}_{T^2}(p),
\ee
which is of degree one since it is defined by a single point. This specifies a map $T^2 \to Pic^1(T^2)$, where $Pic^1(T^2)\subset Pic(T^2)$  is the space of degree one line bundles on $T^2$ up to isomorphism. Note that since $Pic^1(T^2)$ is not closed under tensor product multiplication, it does not form a group. However, exploiting the fact that we have a marked point $e \in T^2$, we can, alternatively, define a map for each point $p$ to a degree zero line bundle, namely
\be
p \to {\mathcal O}_{T^2}(p-e).
\ee
${\mathcal O}_{T^2}(p-e)$ is of degree zero since it is defined as the difference of two points. This specifies a map $T^2 \to Pic^0(T^2)$, which can be shown to be an isomorphism. $Pic^0(T^2)$ is the space of all degree zero line bundles on $T^2$ up to isomorphism. It is clear that the bundle associated to the zero point $e$ under this map is the trivial bundle. Furthermore, $Pic^0(T^2)$ is closed under tensor product multiplication and forms a group. Note that the map $T^2 \to Pic^0(T^2)$ is a group homomorphism. This means that for any given three points $p, p^{'},p^{''} \in T^2$ which obey
\be
p{\stackrel{.}{+}}p^{'}=p^{''},
\ee
we have
\be
{\mathcal O}_{T^2}(p-e)\otimes {\mathcal O}_{T^2}(p^{'}-e)\cong{\mathcal O}_{T^2}(p^{''}-e).
\ee

Thus far, we have considered points in a smooth fiber $T^2$.
Using the zero section $e=e_9$ of $B$, we can try to extend our discussion to a family of elliptic curves, that is, to sections of our elliptic fibration $\beta: B \to \cp{1}$. Consider any section $\xi$ in $\Gamma(B)$. We can, as mentioned above, define a map
\be\label{0a}
\xi \to {\mathcal O}_B(\xi).
\ee
This specifies a map from $ \Gamma(B)\to Pic(B)$. But similarly to the discussion above, we can, alternatively, use the zero section $e$ to  define, for any section $\xi$, the mapping
\be
\xi \to {\mathcal O}_B(\xi-e).
\ee
This map has a similar property to the case of one elliptic curve. That is, for any three sections $\xi,\xi^{'},\xi^{''} \in \Gamma(B)$ which obey
\be\label{a}
\xi{\stackrel{.}{+}}\xi^{'}=\xi^{''}
\ee
we have 
\be\label{b}
{\mathcal O}_B(\xi-e)|_f\otimes {\mathcal O}_B(\xi^{'}-e)|_f= {\mathcal O}_B(\xi^{''}-e)|_f,
\ee
where the equation only holds for the restriction to any smooth fiber $f$.  However, care has to be taken at the singular fibers.

Let us now calculate the class $t_{e_6}(\xi)$ for any section $\xi$. To do this,  we have to  calculate the line bundle ${\mathcal O}_B(t_{e_6}^{-1}(\xi))$, as discussed in (\ref{mappic}). Using the map (\ref{0a}), this line bundle corresponds to the section
\be
\xi{\stackrel{.}{+}}({\stackrel{.}{-}}e_6)=\xi{\stackrel{.}{-}}e_6=t^{-1}_{e_6}(\xi).
\ee
Using equations (\ref{a}) and (\ref{b}), we find that
\be\label{great}
 {\mathcal O}_B(\xi-e)|_f\otimes {\mathcal O}^{-1}_B(e_6-e)|_f={\mathcal O}_B(t_{e_6}^{-1}(\xi)-e)|_f
\ee
where we have used ${\mathcal O}^{-1}_B(e_6-e)$ as the second factor since, in this case, $\xi^{'}={\stackrel{.}{-}}e_6$.
Tensoring (\ref{great}) with ${\mathcal O}_B(e)$ gives an equation in terms of divisors, namely
\be\label{4.5}
t_{e_6}(\xi)= \xi - e_{6} + e_9 +
a_{1}n_{1} + a_{2}n_{2} + a_{3}n_{3}+ a_{4}n_{4}+ a_{5}n_{5}+ a_{6}n_{6}+ a_0f,
\ee
where $a_0,\dots,a_6$ are integers. Note that this equation holds for both smooth and singular fibers. It even holds  without restriction to any fiber at all.  We have assured this by adding the fiber components. The specific coefficients $a_0,\dots,a_6$ will now be determined.

To begin with, intersect both sides of (\ref{4.5}) with the exceptional divisors $n_{i}$ for $i=1,...,6$. Taking into account the results of Table~\ref{tI2}, that is,  $(t_{e_6})(n_{k}) = o_{k},\;k=1,...,4$, $(t_{e_6})(n_{k}) = n_{k},\;k=5,6$, the intersection numbers and the fact that $n_i\cdot t_{e_6}(\xi)=t_{e_6}(n_i)\cdot\xi $, we obtain 
\be
\begin{split}
o_1\cdot \xi &= n_1\cdot \xi +1 -2a_1 \\
o_2\cdot \xi &= n_2\cdot \xi +1 -2a_2  \\
o_3\cdot \xi &= n_3\cdot \xi -1 -2a_3  \\
o_4\cdot \xi &= n_4\cdot \xi -1 -2a_4  \\ 
n_5\cdot \xi &= n_5\cdot \xi -2a_5  \\
n_6\cdot \xi &= n_6\cdot \xi -2a_6.  \\
\end{split}
\ee
Hence, 
\be
a_5=a_6=0
\ee
independently of the choice of section $\xi$. Recall that $e_1,e_2,e_4$ and $e_7$ are sections of $B$, that is, they intersect each $I_2$ fiber exactly once. Furthermore,  $e_1$ and $e_2$ intersect none of the exceptional divisors, the only exceptional divisors which are intersected by $e_4$ are $n_5$ and $n_6$, and  $e_7$ intersects only $n_2,n_4$ and $n_6$. Using these facts, we can determine the coefficients $a_1,...,a_4$ for $e_1,e_2,e_4$ and $e_7$. The results are presented in the first four columns of Table~\ref{te6exept}. Note that these coefficients are identical for $e_1, e_2$ and $e_4$, but differ for $e_7$.

\begin{table}[!ht]
\begin{center}
\begin{tabular}{|l||c|c|c|c|c|} \hline
 & $a_1 $ & $a_2 $ & $a_3 $ & $a_4 $ & $a_0$\\ \hline \hline
$ e_1 $ & $0 $ & $0$ & $-1$ & $-1 $ & $1$\\ \hline
$ e_2 $ & $0 $ & $0$ & $-1$ & $-1 $ & $1$\\ \hline
$ e_4 $ & $0 $ & $0$ & $-1$ & $-1 $ & $1$\\ \hline
$ e_7 $ & $0 $ & $1$ & $-1$ & $0  $ & $0$\\ \hline
\end{tabular}
\end{center}
\caption{The coefficients $a_0,...,a_4$ for the expansion of the pull-back of the sections  $e_1,e_2,e_4$ and $e_7$ under the action of $t_{e_6}$.}\label{te6exept}
\end{table}

To determine the remaining coefficients $a_0$, we use the fact that $t_{e_6}({\xi})$ is again a section and, hence, $t_{e_6}({\xi})\cdot t_{e_6}({\xi})=-1 $. Let us begin with $e_1, e_2$ and $e_4$. Using (\ref{4.5}) and the results in  Table~\ref{te6exept}, we have 
\be\label{t6i}
t_{e_6}{(e_i)}=e_i-e_6+e_9-n_3-n_4+a_0f\;\;\;\;i=1,2,4.
\ee
Squaring this expression then tells us that
 $a_0=1$ for $e_1,e_2$ and $e_4$. Hence, we have at long last determined the transformation laws 
\be
t_{e_6}(e_i)=f-l+e_i+e_6+e_7+e_9\;\;\;\;i=1,2,4,
\ee
where we used the definition of $n_3$ and $n_4$ in term of the sections $e_1,\dots,e_9$ and $l$ given in (\ref{def}). Similarly, using the expansion of $t_{e_6}(e_7)$ determined by (\ref{4.5}) and  Table~\ref{te6exept}, we find that $a_0=0$ for $e_7$. Hence,
\be
t_{e_6}(e_7)=l-e_5-e_8.
\ee
Our results for the coefficients $a_0$ are listed in the last column of Table~\ref{te6exept}.
We can now use $t_{e_6}(n_5)$ determined in the previous sub-section  and (\ref{t6i}) to show that
\be
\begin{split}
t_{e_6}(e_{3}) &= t_{e_6}(e_{4} + n_{5}) =f-l+e_4+e_6+e_7+e_9+n_5=f-l+e_3+e_6+e_7+e_9.  \\
\end{split}
\ee
\noindent

Finally, using the transformation laws of $f$ and $e_1,\dots,e_9$ it is easy to read off the transformation of $l$. We find that
\be
t_{e_6}(l)=2f-e_5+2e_6+e_7-e_8+2e_9.
\ee
A general summary of our findings for $t_{e_6}$ is presented in Table~\ref{table-auts}.

\subsection{Transformations of the Canonical Generators under $t_{e_4}$}

The calculation of the transformation laws of the canonical set of generators of $H_2(B,{\mathbb Z})$ under $t_{e_4}$ closely follows the calculation of the transformations under  $t_{e_6}$. First, we determine the transformations of $e_9, e_4, e_8$ and $e_3$. Using the definition of a translation, and the facts that $e=e_9$ is the zero section and that $e_4$ is a section of order two, we find
\be\label{g}
t_{e_4}(e_{9})=e_9{\stackrel{.}{+}}e_4 =e_{4}
\ee
and
\be\label{f}
t_{e_4}(e_4)  = e_{4}{\stackrel{.}{+}} e_{4} = e_{9}.
\ee
Using this, (\ref{def}) and Table~\ref{tI2}, it follows that
\be
t_{e_4}(e_{8}) = t_{e_4}(e_{9} + n_{1}) = e_4 + o_1 = e_4+f-n_1 = f+e_4-e_8+e_9
\ee
and
\be
t_{e_4}(e_{3}) = t_{e_4}(e_{4} + n_{5}) = e_9 + o_5 = e_9+f-n_3 = f-e_3+e_4+e_9.
\ee
To finish our calculation of the transformation laws of the canonical set of generators of $H_2(B,{\mathbb Z})$ under $t_{e_4}$, we have to use a formula similar to (\ref{4.5}) which, for $t_{e_4}$, is given by
\be\label{te4}
t_{e_4}(\xi) = \xi - e_{4} + e_{9} + a_{1}n_{1} + a_{2}n_{2} + a_{3}n_{3}+ a_{4}n_{4}+ a_{5}n_{5}+ a_{6}n_{6}+ a_0f.
\ee
To determine the coefficients for specific sections $\xi $, we  intersect both sides of (\ref{te4}) with $n_{i}$ for $i=1,...,6$.  Taking into account the results of Table~\ref{tI2}, that is, $(t_{e_4})(n_{k}) = o_{k}$ for $k=1,2,5,6$ and $(t_{e_4})(n_{k}) = n_{k}$ for $\;k=3,4$, we obtain
\be
\begin{split}
o_1\cdot \xi &= n_1\cdot \xi +1 -2a_1 \\
o_2\cdot \xi &= n_2\cdot \xi +1 -2a_2  \\
n_3\cdot \xi &= n_3\cdot \xi  -2a_3  \\
n_4\cdot \xi &= n_4\cdot \xi  -2a_4  \\ 
o_5\cdot \xi &= n_5\cdot \xi -1 -2a_5  \\
o_6\cdot \xi &= n_6\cdot \xi -1 -2a_6.  \\
\end{split}
\ee
Hence, 
\be
a_3=a_4=0
\ee
independently of the section $\xi$. Recall that $e_1,e_2,e_6$ and $e_7$ are sections, that is, they intersect each $I_2$ fiber exactly once. Furthermore,  $e_1$ and $e_2$ intersect none of the exceptional divisors and  the only exceptional divisors which are intersected by $e_6$ are $n_3$ and $n_4$. Using these facts,  we can determine the coefficients $a_1,...,a_4$ for $e_1,e_2,e_6$ and $e_7$. The first four columns of Table~\ref{te4exept} summarize our findings for the coefficients $a_1,\dots,a_4$.

\begin{table}[!ht]
\begin{center}
\begin{tabular}{|l||c|c|c|c|c|} \hline
 & $a_1 $ & $a_2 $ & $a_3 $ & $a_4 $ & $a_0$ \\ \hline \hline
$ e_1 $ & $0 $ & $0$ & $-1$ & $-1$  & $1$ \\ \hline
$ e_2 $ & $0 $ & $0$ & $-1$ & $-1$  & $1$ \\ \hline
$ e_6 $ & $0 $ & $0$ & $-1$ & $-1$  & $1$ \\ \hline
$ e_7 $ & $0 $ & $1$ & $-1$ & $0 $  & $0$ \\ \hline
\end{tabular}
\end{center}
\caption{The coefficients $a_0,...,a_4$ for the expansion of the pull-back of the sections  $e_1,e_2,e_6$ and $e_7$ under the action of $t_{e_4}$.}\label{te4exept}
\end{table}
To determine the remaining coefficients $a_0$, we use the fact that $t_{e_4}({\xi})$ is again a section.  Hence $t_{e_4}({\xi})\cdot t_{e_4}({\xi})=-1 $. Using this, and squaring the expansion 
\be
t_{e_4}{(e_i)}=e_i-e_4+e_9-n_5-n_6+af\;\;\;\;i=1,2,6,
\ee
we find $a_0=1$ for $e_1,e_2$ and $e_6$. Hence,
\be
t_{e_4}(e_i)=f-l+e_i+e_4+e_7+e_9\;\;\;\;i=1,2,6,
\ee
where we used the definitions of $n_5$ and $n_6$ given in (\ref{def}). Using the expansion of $t_{e_4}(e_7)$,  we find $a_0=0$ for $e_7$ and, hence, that
\be
t_{e_4}(e_7)=l-e_3-e_8.
\ee
Our results for the coefficients $a_0$ are listed in the last column of Table~\ref{te4exept}.
We can now show
\be
\begin{split}
t_{e_4}(e_{5}) &= t_{e_4}(e_{6} + n_{3}) = f-l+e_6+e_4+e_7+e_9 +n_3= f-l+e_4+e_5+e_7+e_9. \\
\end{split}
\ee

Using the transformation laws of $f$ and $e_1,\dots,e_9$ it is  easy to read of the transformation of $l$. We find that
\be
t_{e_4}(l)=2f-e_3+2e_4+e_7-e_8+2e_9.
\ee
A general summary of our findings for $t_{e_4}$ are presented in Table~\ref{table-auts}.

\section{The $\tau_{B1}$, $\tau_{B2}$ Action on $H_2(B,{\mathbb Z})$ and Invariant Classes}\label{comp}

Having determined the transformation laws of the canonical set of generators $e_1,\dots,e_9$ and $l$ of $H_2(B,{\mathbb Z})$ under $\alpha_B$ in Section~\ref{alpha} and $t_{e_4}$, $t_{e_6}$, in Section~\ref{trans}, we can now compose them using $\tau_{B1}=t_{e_6}\circ \alpha_B$ and $\tau_{B2}=t_{e_4}\circ \alpha_B$ to determine the action of $\tau_{B1}$ and $\tau_{B2}$. The results are presented in the fifth and sixth columns of Table~\ref{table-auts} respectively. The table also summarizes our results for $(-1)_B$, $\alpha_B$, $t_{e_6}$ and $t_{e_4}$.
\begin{table}[!ht]
\begin{center}
\begin{tabular}{|l||p{2cm}|p{2cm}|p{2cm}|p{2cm}|p{2cm}|p{2cm}|} \hline
 & $\hspace{0.5cm}(-1)_{B}$ & $\hspace{0.7cm}\alpha_{B}$ & $\hspace{0.7cm}t_{e_6}$ & $\hspace{0.7cm}t_{e_4}$ & $\hspace{0.7cm}\tau_{B1}$  & $\hspace{0.7cm}\tau_{B2} $ \\ \hline \hline
$f$   & $f$            & $f$                                  &  $f $               & $f$           & $f$        & $f$     \\  \hline 
$e_1$ & $l-e_7-e_1$  & $l -e_7-e_1 $  & $f-l+e_1+e_6+e_7+e_9$ & $f-l+e_1+e_4+e_7+e_9$ & $f-e_1+e_6+e_9$ & $f-e_1+e_4+e_9$  \\ \hline 
$e_2$ & $l-e_7-e_2$  & $e_2$ & $f-l+e_2+e_6+e_7+e_9 $ & $f-l+e_2+e_4+e_7+e_9 $ & $f-l+e_2+e_6+e_7+e_9$ & $f-l+e_2+e_4+e_7+e_9$ \\  \hline 
$e_3$ & $e_3$ & $ l-e_3-e_7$ & $f-l+e_3+e_6+e_7+e_9 $ & $f-e_3+e_4+e_9$ & $f-e_3+e_6+e_9$ & $f-l+e_3+e_4+e_7+e_9$  \\  \hline 
$e_4$ & $e_4$ & $e_4$  & $f-l+e_4+e_6+e_7+e_9$ & $e_9 $ & $f-l+e_4+e_6+e_7+e_9$& $e_9$  \\  \hline 
$e_5$ & $e_5$ & $l-e_ 5-e_7$ & $f-e_5+e_6+e_9$ & $f-l+e_5+e_4+e_7+e_9$ & $f-l+e_5+e_6+e_7+e_9$ & $f-e_5+e_4+e_9$  \\  \hline 
$e_6$ & $e_6$ & $e_6 $ & $e_9$ & $f-l+e_6+e_4+e_7+e_9$ & $e_9$ & $f-l+e_6+e_4+e_7+e_9$ \\  \hline 
$e_7$ & $l-e_1-e_2$ & $ 2l-(e_1+e_3+e_5+e_7+e_8)$ & $l-e_5-e_8$ & $l-e_3-e_8$ & $l-e_1-e_3$ & $l-e_1-e_5$ \\  \hline 
$e_8$ & $e_8$ & $l-e_7-e_8  $ & $f+e_6-e_8+e_9$ & $f+e_4-e_8+e_9$ & $f-l+e_6+e_7+e_8+e_9$ &  $f-l+e_4+e_7+e_8+e_9$ \\  \hline 
$e_9$ & $e_9$ & $e_9$ & $e_6$ & $e_4$ & $e_6$ & $e_4$  \\  \hline  \hline
$l $ & $2l-(e_1+e_2+e_7)$ & $3l-(e_1+e_3+e_5+2e_7+e_8)$ & $2f-e_5+2e_6+e_7-e_8+2e_9$ & $2f-e_3+2e_4+e_7-e_8+2e_9$ & $2f-e_1-e_3+2e_6+e_7+2e_9$ & $2f-e_1+2e_4-e_5+e_7+2e_9$  \\  \hline
\end{tabular}
\end{center}
\caption{The action of $(-1)_{B}$, $\alpha_{B}$, $t_{e_6}$, $t_{e_4}$, $\tau_{B1} $ and  $\tau_{B2} $  on $H_{2}(B,{\mathbb Z})$. Note that $f$ can be expressed as $f=3l-\sum_{i=1}^{9}e_i$.}\label{table-auts}
\end{table}

We now want to find all classes in $H_2(B,{\mathbb Z})$ that are invariant under the ${\mathbb Z}_2\times{\mathbb Z}_2 $ automorphism group. Let us denote the subspace of all such invariant classes by $H_2(B,{\mathbb Z})^{inv}$. We will show in \cite{dopr-ii} that 
\be\label{rankinv}
\rank H_2(B,{\mathbb Z})^{inv}=4.
\ee
Hence, there exist four generators of $H_2(B,{\mathbb Z})^{inv}$. One of these generators is trivial to find, namely
\be
f=3l-\sum_{i=1}^9 e_i.
\ee
This follows from the first line of Table~\ref{table-auts}, which reflects the statement, shown in \cite{opr}, that $f$ is preserved by any automorphism of $B$.

A second invariant generator, consisting only of fiber components of $\beta: B \to \cp{1}$, is also easy to find. We see using Table~\ref{tI2} that
\be
n_1+o_2
\ee
is preserved under the ${\mathbb Z}_2\times{\mathbb Z}_2 $ automorphism group. Using (\ref{def}), this can be expressed in  terms of the canonical set of generators as
\be
n_1+o_2=2l-e_1-e_2-e_3-e_4-e_5-e_6+e_8-e_9.
\ee
To find the third invariant generator, we exploit the fact that, for any $\xi \in H_2(B,{\mathbb Z})$, the class
\be\label{txi}
\xi+\tau_{B1}(\xi)+\tau_{B2}(\xi)+\tau_{B1}\circ \tau_{B2}(\xi)
\ee
is invariant under ${\mathbb Z}_2\times{\mathbb Z}_2 $. Choosing $\xi=e_9$, we take as the third generator of $H_2(B,{\mathbb Z})^{inv}$
\be
i\equiv e_9+\tau_{B1}(e_9)+\tau_{B2}(e_9)+\tau_{B1}\circ \tau_{B2}(e_9).
\ee
Using the results in Table~\ref{table-auts}, we find that
\be\label{invi}
i=2e_6+2e_4-n_1+o_2.
\ee
This can be expressed in terms of the canonical basis as
\be
i=2l-e_1-e_2-e_3+e_4-e_5+e_6-e_8+e_9.
\ee
Finally, we find the fourth invariant generator as a result of enlightened trial and error. The result is
\be\label{invM}
M\equiv 2e_2-2e_9-n_1-n_2,
\ee
which, using (\ref{def}), can be written as
\be
M=-l+2e_2+e_7.
\ee
The invariance of $M$ under ${\mathbb Z}_2\times{\mathbb Z}_2 $ is easily checked using the results in Table~\ref{table-auts}. It is straightforward to show that $f, n_1+o_2, i$ and $M$ are linearly independent classes.  We conclude that
\be\label{iM}
f,\;\;\;n_1+o_2,\;\;\;i=2e_6+2e_4-n_1+o_2,\;\;\;M=2e_2-2e_9-n_1-n_2
\ee
form a set of generators \footnote{To be precise, the set $\{f,n_1+o_2,i,M\}$ generates a sub-lattice of maximal rank in $H_2(B,{\mathbb Z})^{inv}$. They are, in fact, generators of $H_2(B,{\mathbb Q})^{inv}$ and not $H_2(B,{\mathbb Z})^{inv}$. Be that as is may, for our purposes in this paper and in \cite{dopr-iii,dopr-ii}, these classes are the most convenient. First of all, since they generate a lattice of the same rank as $H_2(B,{\mathbb Z})^{inv}$, they demonstrate that $H_2(B,{\mathbb Z})^{inv}$ is non-empty and of rank four. Secondly, they have convenient intersection numbers which simplify the construction of holomorphic vector bundles in \cite{dopr-iii,dopr-ii}. Since this subtlety does not play a role in either this paper or \cite{dopr-iii,dopr-ii}, here and henceforth we  refer to them simply as the generators of $H_2(B,{\mathbb{Z}})^{inv}$.} of $H_2(B,{\mathbb Z})^{inv}$. Their  intersection numbers are displayed in Table~\ref{inter}.
\begin{table}[!ht]
\begin{center}
\begin{tabular}{|c||c|c|c|c|} \hline
                  & $i$  & $f$ & $n_1+o_2$& $M$  \\ \hline\hline
$i$               & $-4$ & $4$     & $4$              & $0$  \\ \hline
$f$           & $4$  & $0$     & $0$              & $0$  \\ \hline
$n_1+o_2$ & $4$  & $0$     & $-4$             & $0$  \\ \hline
$M$               & $0$  & $0$     & $0$              & $-4$ \\ \hline
\end{tabular}
\end{center} 
\caption{The intersection numbers of the generators of $H_2(B,{\mathbb Z})^{inv}.$}\label{inter}
\end{table}
This choice of generators will turn out to be very convenient when we solve the numerical conditions on the rank four vector bundles $V$ on the Calabi-Yau space $X$ imposed by particle physics phenomenology \cite{dopr-iii,dopr-ii}. Note that it is very easy to find other invariant classes in $H_2(B,{\mathbb Z})$. For example, it is immediately obvious from Table~\ref{tI2} that $n_2+o_1$ is an invariant class. Furthermore, as we stated, (\ref{txi}) is an invariant class for any $\xi \in H_2(B,{\mathbb Z})$. However, it follows from (\ref{rankinv}) and the linear independence of $f, n_1+o_2, i$ and $M$ that any other invariant class must be a linear combination of these four generator classes. For example, one  easily  finds using (\ref{def}) that
\be
n_2+o_1=2f-(n_1+o_2).
\ee
Finally, we denote the set of generators for invariant curve classes on $B^{'} $ by 
\be\label{iM'}
f^{'},\;\;\;n^{'}_1+o^{'}_2,\;\;\;i^{'},\;\;\;M^{'},
\ee 
where $i^{'}$ and $M^{'}$ are defined similarly to (\ref{invi}) and (\ref{invM}).

\section{The Homology of $X$ and $Z$}\label{XZ}

Recall from \cite{opr} and Section~\ref{Z} that our analysis involves  two different types of Calabi-Yau threefolds. The first are the Calabi-Yau spaces $X$ which are simply connected and admit a freely acting automorphism group ${\mathbb Z}_2\times {\mathbb Z}_2$. The second are the smooth quotient spaces
\be\label{11.1}
Z=X/({\mathbb Z}_2\times {\mathbb Z}_2)
\ee
with fundamental group $\pi_1(Z)={\mathbb Z}_2\times {\mathbb Z}_2$. In this section, we will calculate the relevant parts of the homology rings $H_{*}(X,{\mathbb Z})$ and $H_{*}(Z,{\mathbb Z})$ of these two types of threefolds. More precisely, we will  first determine
\be\label{poinc}
\dim_{\mathbb C} H_{i}(X,{\mathbb C})=\dim_{\mathbb C} (H_{i}(X,{\mathbb Z})\otimes {\mathbb C}),\;\;\;i=0,\dots,6
\ee
and then show how to find a set of generators of $H_{4}(X,{\mathbb Z})$. A subset of these generators consists of classes that are invariant under the action of ${\mathbb Z}_2\times {\mathbb Z}_2$. These span the subgroup of invariant homology classes, which we denote by $H_{4}(X,{\mathbb Z})^{inv} \subset H_{4}(X,{\mathbb Z})$. We will explicitly compute a complete set of invariant generators. Using these results, we  then determine
\be
\dim_{\mathbb C} H_{i}(Z,{\mathbb C})= \dim_{\mathbb C}( H_{i}(Z,{\mathbb Z})\otimes {\mathbb C}),\;\;\;i=0,\dots,6
\ee
and explicitly find a set of generators of $H_{4}(Z,{\mathbb Z})$. Both quantities, the dimensions of $H_{i}(Z,{\mathbb C}),\;i=0,\dots,6$ and the generators of $H_{4}(Z,{\mathbb Z})$, are of physical importance. The dimension of the homology groups of $Z$ are a vital ingredient in determining the particle spectrum of string theories compactified on the Calabi-Yau threefold $Z$. The elements of $H_{4}(Z,{\mathbb Z})$ are divisors in $Z$ which clearly descend, via the modding out (\ref{11.1}), from the invariant divisors on $X$, that is, the elements of $H_{4}(X,{\mathbb Z})^{inv}$. The classes in $H_{4}(X,{\mathbb Z})^{inv}$  will be used in \cite{dopr-i} and \cite{dopr-ii} to construct invariant stable vector bundles on $X$ which, in turn, descend to stable vector bundles on $Z$. These bundles admit connections which correspond to solutions of the hermitian Yang-Mills equations on $Z$.

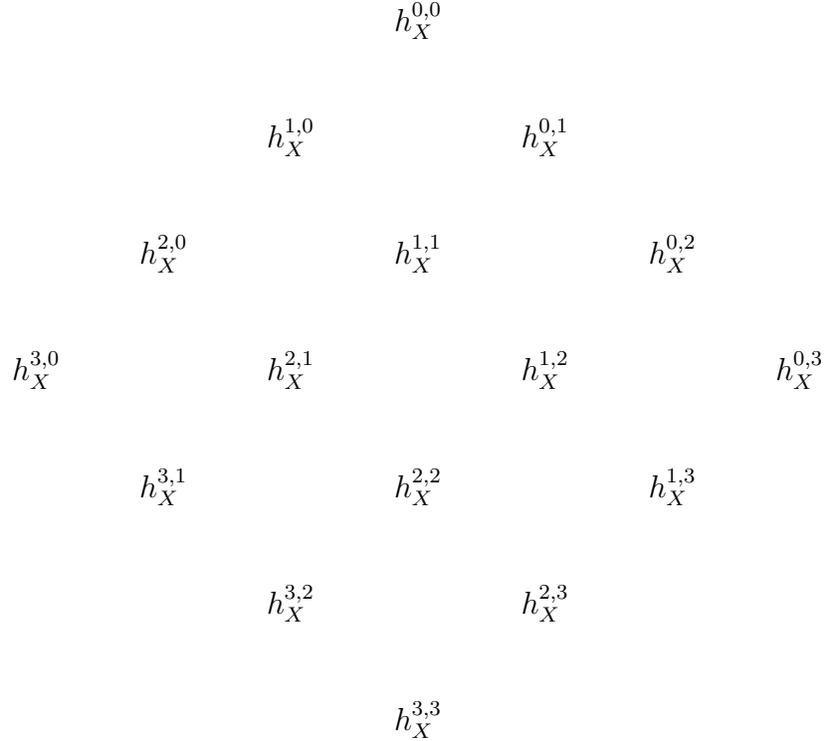
\begin{figure}[!ht]
\[
\xymatrix{
 && & h^{0,0}_X &  & &  \\
 && h^{1,0}_X &  &h^{0,1}_X  & &  \\
 &h^{2,0}_X & & h^{1,1}_X &  &h^{0,2}_X &  \\
 h^{3,0}_X & &h^{2,1}_X &  &h^{1,2}_X  & &h^{0,3}_X  \\
 &h^{3,1}_X & & h^{2,2}_X &  &h^{1,3}_X &  \\
 & &h^{3,2}_X & &h^{2,3}_X  & &  \\
 & & & h^{3,3}_X &  & &  \\
}
\]
\caption{The Hodge diamond of the Kahler manifold $X$.}
\label{HX}
\end{figure}

\begin{figure}[!ht]
\[
\xymatrix{
 & & & 1 &  & &  \\
 & & 0 &  &0  & &  \\
 &0 & & h^{1,1}_X &  &0 &  \\
 1 & &h^{2,1}_X &  &h^{2,1}_X  & &1  \\
 &0 & & h^{1,1}_X &  &0 &  \\
 & &0 & &0  & &  \\
 & & & 1 &  & &  \\
}
\]
\caption{The Hodge diamond of the Calabi-Yau threefold $X$.}
\label{hCY}
\end{figure}
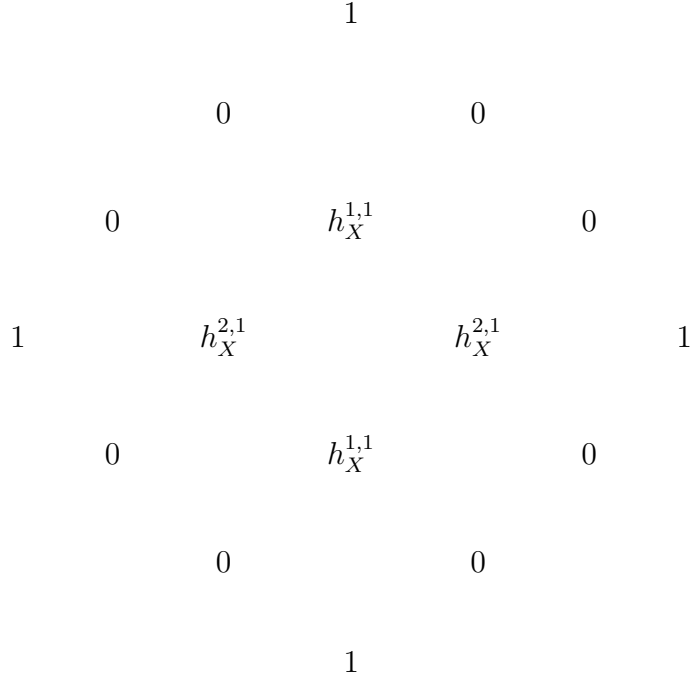

Let us begin with the Calabi-Yau threefold $X$. It is convenient to use Poincar\'e duality, that is
\be
H_i(X,{\mathbb Z})\cong H^{6-i}(X,{\mathbb Z}),\;\;\;0=1,\dots,6.
\ee
Furthermore, the fact that $X$ is a  Kahler manifold allows us to write the  Hodge diamond of $X$ as in Figure~\ref{HX}. Here
\be\label{hij}
h^{a,b}_X=\dim_{\mathbb C}H_{\bar{\partial}}^{a,b}(X),\;\;\;a,b=0,\dots,3
\ee
denotes the dimension of the $a$-th Dolbeaux cohomology of holomorphic $b$-forms on $X$.
It is well-known that for Kahler manifolds
\be
H^{k}(X,{\mathbb C})=\sum_{a+b=k}H_{\bar{\partial}}^{a,b}(X)
\ee
and, therefore, that
\be\label{kahler}
\dim_{\mathbb C} H^{k}(X,{\mathbb C})=\sum_{a+b=k}h^{a,b}.
\ee
For  simply connected Calabi-Yau threefolds, the Hodge diamond simplifies to the highly symmetric form given in  Figure~\ref{hCY}. Hence, to determine the dimension of the homology groups of $X$, we need to determine $h^{1,1}_X$ and $h^{2,1}_X$ only. To do this, we begin by calculating the elements of $H_{\bar{\partial}}^{1,1}(X)$ explicitly.
Since $X$ is simply connected, we have
\be\label{12.10}
H_{\bar{\partial}}^{1,1}(X)\cong Pic(X)\otimes {\mathbb C}\cong H^2(X,{\mathbb Z})\otimes {\mathbb C}.
\ee
Therefore, if we can determine  the generators of all line bundles on $X$, we will have determined all generators of $H^2(X,{\mathbb Z})$ and, hence, of $H_{\bar{\partial}}^{1,1}(X)$. Recall that $X$ is the fiber product of two rational elliptic surfaces $B$ and $ B^{'}$. Hence, we have the inclusion map
\be
i: B \times_{\cp{1}} B^{'} \to B \times B^{'}.
\ee
In terms of cohomology, this defines the pull-back map
\be
i^{*}: H^2(B\times B^{'},{\mathbb Z} ) \to H^2(B \times_{\cp{1}} B^{'},{\mathbb Z} ).
\ee
Using the Kunneth formula, we find
\be
H^2(B\times B^{'},{\mathbb Z} )\cong H^2(B,{\mathbb Z})\oplus H^2(B^{'},{\mathbb Z}).
\ee
With the help of the the isomorphisms
\be
H^2(B\times_{\cp{1}} B^{'},{\mathbb Z} )\cong Pic(B\times_{\cp{1}} B^{'}),\;\;\;H^2(B,{\mathbb Z} )\cong Pic(B),\;\;\;H^2( B^{'},{\mathbb Z} )\cong Pic(B^{'}),
\ee
we have
\be\label{Pic}
i^{*}:Pic(B) \oplus Pic(B^{'}) \to Pic(B \times_{\cp{1}} B^{'}).
\ee
To see how the map $i^{*}$ works, recall that for $X$ we had the natural projections
\begin{equation}
\xymatrix{
& X \ar[dl]_-{\pi'} \ar[dr]^-{\pi} && \ar@{}[ddr]|-{.} & \\
B \ar[dr]_-{\beta} & & B' \ar[dl]^-{\beta'}&&& \\
& \cp{1} &&&&
}
\end{equation}
Consider any line bundle $L$ on $B$ determined by the divisor $D$. That is,
\be
L=\oy(D).
\ee
Then, clearly
\be
\pi^{'*}L=\pi^{'*}\oy(D)=\ox(\pi^{'-1}(D))=\ox(D\times_{\cp{1}}B^{'}).
\ee
Similarly, for a line bundle $L^{'}$ on $B^{'}$ determined by the divisor $D^{'}$, that is
\be
L^{'}=\oz(D^{'}),
\ee
we obtain
\be
\pi^{*}L^{'}=\pi^{*}\oz(D^{'})=\ox(\pi^{-1}(D^{'}))=\ox(B\times_{\cp{1}}D^{'}).
\ee
Hence, the map (\ref{Pic}) is given by
\be\label{LL}
i^{*}:(L,L^{'})\to \pi^{'*}L \otimes \pi^{*}L^{'}.
\ee
It is not to difficult to show that the map $i^{*}$ is surjective, that is, all line bundles on $B\times_{\cp{1}}B^{'}$ arise from line bundles on $B$ and $B^{'}$ via (\ref{LL}). But what is the kernel of $i^{*}$? It follows from  (\ref{LL}) that $(L,L^{'})$ will be in the kernel of $i^{*}$ if and only if 
\be
\pi^{'*}L=(\pi^{*}L^{'})^{-1}.
\ee
Hence, the divisors $\pi^{'-1}(D)$ and $\pi^{-1}(D^{'})$ associated with $\pi^{'*}L$ and $\pi^{*}L^{'}$ respectively must satisfy
\be\label{11.21}
\pi^{'-1}(D)=-\pi^{-1}(D^{'}).
\ee
This can only happen if $L$ and $L^{'}$ are pull-back bundles from $\cp{1}$ of the form
\be
L=\beta^{*}\ocp(a),\;\;\;L^{'}=\beta^{'*}\ocp(-a),\;\;\;a \in {\mathbb Z}.
\ee
Then
\be
\pi^{'*}L=\ox(a(f\times f^{'})),\;\;\;\pi^{*}L^{'}=\ox(-a(f\times f^{'}))
\ee
and, hence, (\ref{11.21}) is satisfied.
Noting that $\ocp(a),\;a \in {\mathbb Z}$ spans $Pic(\cp{1})\cong H^2(\cp{1},{\mathbb Z})$, it follows that
\be\label{gh2}
H^{2}(X,{\mathbb Z})\cong Pic(B \times_{\cp{1}}B^{'})\cong \frac{Pic(B)\oplus Pic(B^{'})}{Pic(\cp{1})}\cong \frac{H^{2}(B,{\mathbb Z})\oplus H^{2}(B^{'},{\mathbb Z})} { H^{2}(\cp{1},{\mathbb Z})}.
\ee
This result allows us to determine all of the relevant quantities on $X$, namely, the Hodge numbers $h^{1,1}_X, h^{2,1}_X $ and the generators of  $H_{4}(X,{\mathbb Z})$.

First consider the Hodge numbers. Using (\ref{hij}), (\ref{12.10}), (\ref{gh2}) and the fact that $\rank H^2(B,{\mathbb Z})=10$, we find
\be\label{h11}
h^{1,1}_X=\dim_{\mathbb C} H^{1,1}_{\bar{\partial}}(X)=\dim_{\mathbb C} H^{2}(X,{\mathbb C}) =10+10-1=19.
\ee
To determine $h^{2,1}_X$,  note that for complex threefolds the topological Euler characteristic can be calculated as
\be
\chi(X)=\sum_{i=0}^{6}(-1)^i \dim_{\mathbb C}{H^i(X,{\mathbb C})}.
\ee
Furthermore, it can be shown that
\be
\chi(X)=c_3(X).
\ee
But, as discussed in \cite{opr}, $c_3(X)=0$. Using this and the Hodge diamond of Figure~\ref{hCY},  we find
\be
0=\chi(X)=1+h^{1,1}_X-1-2h^{2,1}_X-1+h^{1,1}_X+1.
\ee
It then follows, using (\ref{h11}), that
\be
h^{2,1}=19.
\ee
Note that using (\ref{kahler}), the Hodge diamond of Figure~\ref{hCY} and $h_X^{1,1}=h_X^{2,1}=19$ we can determine $\dim H^i(X,{\mathbb C})$ and, hence, by Poincar\'e duality $\dim H_i(X,{\mathbb C})$ for all $i=0,\dots,6$.
 
Now, recall from Poincar\'e duality (\ref{poinc}) that
\be
H_4(X,{\mathbb Z})\cong H^2(X,{\mathbb Z}).
\ee
Using this and (\ref{gh2}), we can determine the nineteen generators of $H_4(X,{\mathbb Z})$. First, note that if $D$ is a divisor in $B$, that is, $D \in H_2(B,{\mathbb Z})$, then
\be\label{12.33}
\pi^{'-1}(D)=D\times_{\cp{1}} B^{'} \in H_4(X,{\mathbb Z}).
\ee
Similarly, if $D^{'} \in H_2(B^{'},{\mathbb Z})$ then
\be\label{12.34}
\pi^{-1}(D^{'})=B\times_{\cp{1}} D^{'} \in H_4(X,{\mathbb Z}).
\ee
Taking $\{D\}$ to be the canonical set of generators $\{e_1,\dots,e_9,l\}$ of $H_2(B,{\mathbb Z})$ and $\{D^{'}\}$ to be $\{e_1^{'},\dots,e_9^{'},l^{'}\}$ in $H_2(B^{'},{\mathbb Z})$, we see from (\ref{gh2}) that the set
\be\label{12.35}
\{D\}\times_{\cp{1}} B^{'},\;\;\;B\times_{\cp{1}} \{D^{'}\}
\ee
generates $H_4(X,{\mathbb Z})$. Note that there are only nineteen, and not twenty, classes in (\ref{12.35}) since
\be\label{12.36}
f\times_{\cp{1}}B^{'}=B\times_{\cp{1}} f^{'}=f\times f^{'}.
\ee
This is the reason for the $H^2(\cp{1},{\mathbb Z} )$ denominator in (\ref{gh2}). We conclude that (\ref{12.35}), subject to (\ref{12.36}), gives an explicit set of nineteen generators of $H_4(X,{\mathbb Z})$.

So far, we have analyzed the Hodge diamond and fourth homology group of the Calabi-Yau threefold $X$. We now turn our attention to the Calabi-Yau threefold $Z$ defined in (\ref{11.1}). Here, given the previous results, it is easiest to proceed in reverse, first determining the generators of  $H_4(Z,{\mathbb Z})$ and then the Hodge numbers. To to this, note that any class in $H_4(Z,{\mathbb Z})$ must descend under modding out from a ${\mathbb Z}_2 \times {\mathbb Z}_2 $ invariant class in $H_4(X,{\mathbb Z})$. If we denote the subspace of invariant classes by $H_4(X,{\mathbb Z})^{inv} \subset H_4(X,{\mathbb Z})$, then there exists an isomorphism
\be
H_4(X,{\mathbb Z})^{inv}\cong H_4(Z,{\mathbb Z}).
\ee
What are the generators of $H_4(X,{\mathbb Z})^{inv}$? Recall from (\ref{invX}) that the generators of the  ${\mathbb Z}_2\times{\mathbb Z}_2 $ automorphism group on $X$ are
\be
\tau_{Xi}=\tau_{Bi} \times_{\cp{1}}\tau_{B^{'}i},\;\;\;i=1,2.
\ee
Acting with these involutions on (\ref{12.33}) and (\ref{12.34}) we have 
\be\label{12.38}
\tau_{Xi}(D \times_{\cp{1}} B^{'})=\tau_{Bi}(D)\times_{\cp{1}} B^{'}
\ee
and  
\be\label{12.39}
\tau_{Xi}(B \times_{\cp{1}} D^{'})=B \times_{\cp{1}} \tau_{B^{'}i}(D^{'})
\ee
respectively for $i=1,2$. It follows from (\ref{12.38}) that $D \times_{\cp{1}} B^{'}$ will be in $H_4(X,{\mathbb Z})^{inv}$ if and only if
\be
\tau_{Bi}(D)=D,\;\;\;i=1,2.
\ee
That is, if and only if
\be
D \in H_2(B,{\mathbb Z})^{inv}.
\ee
From the results of the previous section, we know that the classes $f, n_1+o_2, i$ and $M$ given in (\ref{iM}) generate $H_2(B,{\mathbb Z})^{inv}$. Similarly, it follows from (\ref{12.39}) that $B \times_{\cp{1}} D^{'}$ will be in $H_4(X,{\mathbb Z})^{inv}$ if and only if
\be
\tau_{B^{'}i}(D^{'})=D^{'},\;\;\;i=1,2.
\ee
That is
\be
D^{'} \in H_2(B^{'},{\mathbb Z})^{inv},
\ee
which is generated by the classes $f^{'}, n_1^{'}+o_2^{'}, i^{'}$ and $M^{'}$ given in (\ref{iM'}). It is clear, then, that a set of generators of $H_4(X,{\mathbb Z})^{inv}$ is given by
\be\label{12.44}
\{D_{inv}\}\times_{\cp{1}} B^{'},\;\;\;B \times_{\cp{1}} \{D_{inv}^{'}\},
\ee
where
\be
\{D_{inv}\}=\{f, n_1+o_2, i, M\}
\ee
and
\be
\{D_{inv}^{'}\}=\{f^{'}, n^{'}_1+o^{'}_2, i^{'}, M^{'}\}.
\ee
Note that it follows from (\ref{12.36}) that there are only seven, not eight, classes in (\ref{12.44}). Another convenient way of writing the generators (\ref{12.44}) is as
\be
\pi^{'-1}(f),\;\pi^{'-1}(n_1+o_2),\;\pi^{'-1}(i),\;\pi^{'-1}(M),\;\pi^{-1}(n_1^{'}+o_2^{'}),\;\pi^{-1}(i^{'}),\;\pi^{-1}(M^{'}).
\ee
These classes all descend under the modding out of $X$ by ${\mathbb Z}_2 \times {\mathbb Z}_2 $ to form a set of seven generators of $H_4(Z,{\mathbb Z})$. We conclude that 
\be
\rank H_4(Z, {\mathbb Z})=7
\ee
and, therefore, that $Z$ admits the homology classes required to produce anomaly free, three family particle physics theories.
 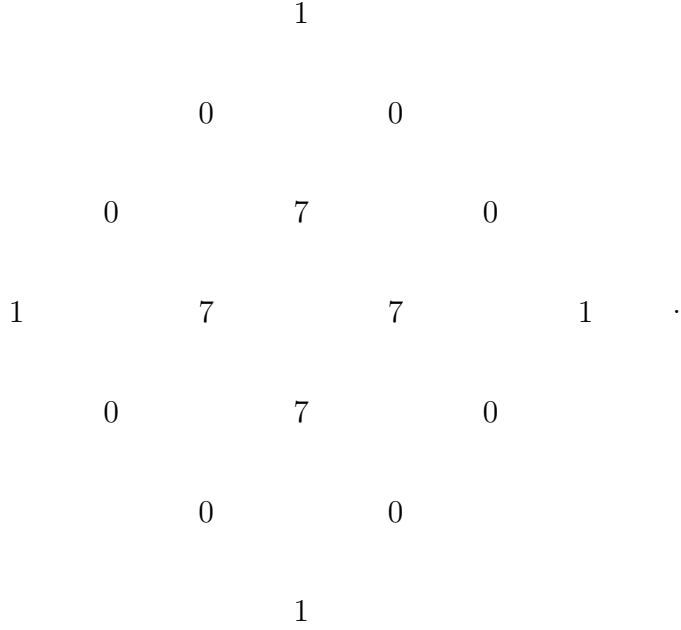
\begin{figure}[!ht]
\[
\xymatrix{
 & & & 1 &  & &  \\
 & & 0 &  &0  & &  \\
 &0 & & 7 &  &0 &  \\
 1 & &7 &  &7  & &1 &. \\
 &0 & & 7 &  &0 &  \\
 & &0 & &0  & &  \\
 & & & 1 &  & &  \\
}
\]
\caption{The Hodge diamond of the Calabi-Yau threefold $Z$.}
\label{hZ}
\end{figure}
From this result, it is straightforward to compute the Hodge numbers of $Z$. First, note that
\be\label{12.47}
H_{\bar{\partial}}^{1,1}(Z)\cong H^2(Z,{\mathbb Z})\otimes {\mathbb C}\cong H_4(Z,{\mathbb Z})\otimes {\mathbb C}.
\ee
Then,
\be\label{12.48}
h^{1,1}_Z=\dim_{\mathbb C} H_{\bar{\partial}}^{1,1}(Z)= \dim_{\mathbb C} H_4(Z,{\mathbb C})=\rank H_4(Z,{\mathbb Z})^{inv}=7.
\ee
The remaining Hodge number, $h_Z^{2,1}$, can be computed using
\be
\chi(Z)=\sum_{i=0}^{6} (-1)^i \dim_{\mathbb C}H^i(Z,{\mathbb C}),
\ee
the fact that
\be
\chi(Z)=c_3(Z)=\frac{c_3(X)}{4}=0
\ee
and the structure of the Hodge diamond. We find that
\be
0=1+h^{1,1}_Z-1-2h^{2,1}_Z-1+h^{1,1}_Z+1
\ee
which, using (\ref{12.48}), implies 
\be
h^{2,1}_Z=7.
\ee
Therefore, the Hodge diamond of $Z$ has the form given in  Figure~\ref{hZ}. Using this, Poincar\'e duality and (\ref{kahler}) it is now easy to read off the dimensions of $H_i(Z,{\mathbb Z})$ for $i=0,\dots,6$.

\section*{Acknowledgments}

Ren\'e Reinbacher and Burt Ovrut are supported in part by DOE under contract No. DE-AC02-76-ER-03071 and the NSF Focused Research Grant DMS 0139799. In addition, Ren\'e Reinbacher wishes to acknowledge partial support from an SAS Dissertation Fellowship and a Daimler-Benz Fellowship. Tony Pantev is supported in part by NSF grants DMS 0099715 and DMS 0139799 and an A.P. Sloan Research Fellowship.

%% file: figure3.pstex_t
\begin{picture}(0,0)%
\includegraphics{figure3.pstex}%
\end{picture}%
\setlength{\unitlength}{2368sp}%
\begingroup\makeatletter\ifx\SetFigFont\undefined%
\gdef\SetFigFont#1#2#3#4#5{%
  \reset@font\fontsize{#1}{#2pt}%
  \fontfamily{#3}\fontseries{#4}\fontshape{#5}%
  \selectfont}%
\fi\endgroup%
\begin{picture}(11583,8613)(1051,-8086)
\put(1051,-61){\makebox(0,0)[lb]{\smash{\SetFigFont{29}{34.8}{\rmdefault}{\mddefault}{\updefault}$\cp{1}$}}}
\put(8326,-8086){\makebox(0,0)[lb]{\smash{\SetFigFont{14}{16.8}{\rmdefault}{\mddefault}{\updefault}$a$}}}
\put(3001,-2611){\makebox(0,0)[lb]{\smash{\SetFigFont{14}{16.8}{\rmdefault}{\mddefault}{\updefault}$T=(1,4)$}}}
\put(8776,-286){\makebox(0,0)[lb]{\smash{\SetFigFont{14}{16.8}{\rmdefault}{\mddefault}{\updefault}$r=(1,0)$}}}
\put(2026,-3886){\makebox(0,0)[lb]{\smash{\SetFigFont{14}{16.8}{\rmdefault}{\mddefault}{\updefault}$p_2$}}}
\put(6976,-7261){\makebox(0,0)[lb]{\smash{\SetFigFont{14}{16.8}{\rmdefault}{\mddefault}{\updefault}$p_1$}}}
\put(11926,-7261){\makebox(0,0)[lb]{\smash{\SetFigFont{29}{34.8}{\rmdefault}{\mddefault}{\updefault}$\cp{1}$}}}
\end{picture}

%% file: figure11.pstex_t
\begin{picture}(0,0)%
\includegraphics{figure11.pstex}%
\end{picture}%
\setlength{\unitlength}{2368sp}%
\begingroup\makeatletter\ifx\SetFigFont\undefined%
\gdef\SetFigFont#1#2#3#4#5{%
  \reset@font\fontsize{#1}{#2pt}%
  \fontfamily{#3}\fontseries{#4}\fontshape{#5}%
  \selectfont}%
\fi\endgroup%
\begin{picture}(11583,8538)(1051,-8011)
\put(1051,-61){\makebox(0,0)[lb]{\smash{\SetFigFont{29}{34.8}{\rmdefault}{\mddefault}{\updefault}$\cp{1}$}}}
\put(8776,-286){\makebox(0,0)[lb]{\smash{\SetFigFont{14}{16.8}{\rmdefault}{\mddefault}{\updefault}$r=(1,0)$}}}
\put(2026,-3886){\makebox(0,0)[lb]{\smash{\SetFigFont{14}{16.8}{\rmdefault}{\mddefault}{\updefault}$p_2$}}}
\put(11926,-7261){\makebox(0,0)[lb]{\smash{\SetFigFont{29}{34.8}{\rmdefault}{\mddefault}{\updefault}$\cp{1}$}}}
\put(8326,-8011){\makebox(0,0)[lb]{\smash{\SetFigFont{14}{16.8}{\rmdefault}{\mddefault}{\updefault}$a$}}}
\put(6076,-7261){\makebox(0,0)[lb]{\smash{\SetFigFont{14}{16.8}{\rmdefault}{\mddefault}{\updefault}$p_1$}}}
\put(3451,-5086){\makebox(0,0)[lb]{\smash{\SetFigFont{14}{16.8}{\rmdefault}{\mddefault}{\updefault}$s=(0,1)$}}}
\put(6076,-3736){\makebox(0,0)[lb]{\smash{\SetFigFont{14}{16.8}{\rmdefault}{\mddefault}{\updefault}$p^{'}$}}}
\put(3376,-3061){\makebox(0,0)[lb]{\smash{\SetFigFont{14}{16.8}{\rmdefault}{\mddefault}{\updefault}$i=(0,1)$}}}
\put(9151,-5761){\makebox(0,0)[lb]{\smash{\SetFigFont{14}{16.8}{\rmdefault}{\mddefault}{\updefault}$p$}}}
\put(4426,-961){\makebox(0,0)[lb]{\smash{\SetFigFont{14}{16.8}{\rmdefault}{\mddefault}{\updefault}$j=(0,1)$}}}
\put(3451,-1636){\makebox(0,0)[lb]{\smash{\SetFigFont{14}{16.8}{\rmdefault}{\mddefault}{\updefault}$p^{''}$}}}
\put(10876,-6136){\makebox(0,0)[lb]{\smash{\SetFigFont{14}{16.8}{\rmdefault}{\mddefault}{\updefault}$t^{''}=(1,1)$}}}
\put(7801,-5086){\makebox(0,0)[lb]{\smash{\SetFigFont{14}{16.8}{\rmdefault}{\mddefault}{\updefault}$p_2$}}}
\put(8776,-4411){\makebox(0,0)[lb]{\smash{\SetFigFont{14}{16.8}{\rmdefault}{\mddefault}{\updefault}$p_1$}}}
\put(8776,-2986){\makebox(0,0)[lb]{\smash{\SetFigFont{14}{16.8}{\rmdefault}{\mddefault}{\updefault}$p_3$}}}
\put(8701,-961){\makebox(0,0)[lb]{\smash{\SetFigFont{14}{16.8}{\rmdefault}{\mddefault}{\updefault}$p_4$}}}
\end{picture}

%% file: figure8.pstex_t
\begin{picture}(0,0)%
\includegraphics{figure8.pstex}%
\end{picture}%
\setlength{\unitlength}{1973sp}%
\begingroup\makeatletter\ifx\SetFigFont\undefined%
\gdef\SetFigFont#1#2#3#4#5{%
  \reset@font\fontsize{#1}{#2pt}%
  \fontfamily{#3}\fontseries{#4}\fontshape{#5}%
  \selectfont}%
\fi\endgroup%
\begin{picture}(6975,7043)(901,-6984)
\put(5476,-3511){\makebox(0,0)[lb]{\smash{\SetFigFont{25}{30.0}{\rmdefault}{\mddefault}{\updefault}$\cong$}}}
\put(7876,-3511){\makebox(0,0)[lb]{\smash{\SetFigFont{25}{30.0}{\rmdefault}{\mddefault}{\updefault}$I_2$}}}
\put(901,-586){\makebox(0,0)[lb]{\smash{\SetFigFont{25}{30.0}{\rmdefault}{\mddefault}{\updefault}$o$}}}
\put(3451,-661){\makebox(0,0)[lb]{\smash{\SetFigFont{25}{30.0}{\rmdefault}{\mddefault}{\updefault}$n$}}}
\end{picture}

%% file: sectWM.pstex_t
\begin{picture}(0,0)%
\includegraphics{sectWM.pstex}%
\end{picture}%
\setlength{\unitlength}{1973sp}%
\begingroup\makeatletter\ifx\SetFigFont\undefined%
\gdef\SetFigFont#1#2#3#4#5{%
  \reset@font\fontsize{#1}{#2pt}%
  \fontfamily{#3}\fontseries{#4}\fontshape{#5}%
  \selectfont}%
\fi\endgroup%
\begin{picture}(10974,7674)(139,-7123)
\put(3301,-1486){\makebox(0,0)[lb]{\smash{\SetFigFont{12}{14.4}{\rmdefault}{\mddefault}{\updefault}$\pi^{-1}(p^{''})$}}}
\put(3601,-3286){\makebox(0,0)[lb]{\smash{\SetFigFont{12}{14.4}{\rmdefault}{\mddefault}{\updefault}$\pi^{-1}(p^{'})$}}}
\put(976,-1486){\makebox(0,0)[lb]{\smash{\SetFigFont{12}{14.4}{\rmdefault}{\mddefault}{\updefault}$\pi^{-1}(j)$}}}
\put(976,-3511){\makebox(0,0)[lb]{\smash{\SetFigFont{12}{14.4}{\rmdefault}{\mddefault}{\updefault}$\pi^{-1}(i)$}}}
\put(1051,-5236){\makebox(0,0)[lb]{\smash{\SetFigFont{12}{14.4}{\rmdefault}{\mddefault}{\updefault}$\pi^{-1}(s)$}}}
\put(8026,-5236){\makebox(0,0)[lb]{\smash{\SetFigFont{12}{14.4}{\rmdefault}{\mddefault}{\updefault}$\pi^{-1}(p)$}}}
\put(8026,-6886){\makebox(0,0)[lb]{\smash{\SetFigFont{12}{14.4}{\rmdefault}{\mddefault}{\updefault}$\pi^{-1}(t^{''})$}}}
\put(7501,-1486){\makebox(0,0)[lb]{\smash{\SetFigFont{12}{14.4}{\rmdefault}{\mddefault}{\updefault}$\pi^{-1}(p_4)$}}}
\put(7501,-3436){\makebox(0,0)[lb]{\smash{\SetFigFont{12}{14.4}{\rmdefault}{\mddefault}{\updefault}$\pi^{-1}(p_3)$}}}
\put(7426,-4711){\makebox(0,0)[lb]{\smash{\SetFigFont{12}{14.4}{\rmdefault}{\mddefault}{\updefault}$\pi^{-1}(p_1)$}}}
\put(5851,-5986){\makebox(0,0)[lb]{\smash{\SetFigFont{12}{14.4}{\rmdefault}{\mddefault}{\updefault}$\pi^{-1}(p_2)$}}}
\put(2101,-61){\makebox(0,0)[lb]{\smash{\SetFigFont{12}{14.4}{\rmdefault}{\mddefault}{\updefault}$f_0$}}}
\put(6751,-286){\makebox(0,0)[lb]{\smash{\SetFigFont{12}{14.4}{\rmdefault}{\mddefault}{\updefault}$\tilde{p}_2^{-1}(a)$}}}
\end{picture}

%% file: psibranch.pstex_t
\begin{picture}(0,0)%
\includegraphics{psibranch.pstex}%
\end{picture}%
\setlength{\unitlength}{2368sp}%
\begingroup\makeatletter\ifx\SetFigFont\undefined%
\gdef\SetFigFont#1#2#3#4#5{%
  \reset@font\fontsize{#1}{#2pt}%
  \fontfamily{#3}\fontseries{#4}\fontshape{#5}%
  \selectfont}%
\fi\endgroup%
\begin{picture}(11583,8121)(1051,-7594)
\put(1051,-61){\makebox(0,0)[lb]{\smash{\SetFigFont{29}{34.8}{\rmdefault}{\mddefault}{\updefault}$\cp{1}$}}}
\put(11926,-7261){\makebox(0,0)[lb]{\smash{\SetFigFont{29}{34.8}{\rmdefault}{\mddefault}{\updefault}$\cp{1}$}}}
\put(2026,-3886){\makebox(0,0)[lb]{\smash{\SetFigFont{14}{16.8}{\rmdefault}{\mddefault}{\updefault}$\bar{p}_2$}}}
\put(6976,-7261){\makebox(0,0)[lb]{\smash{\SetFigFont{14}{16.8}{\rmdefault}{\mddefault}{\updefault}$\bar{p}_1$}}}
\put(9376,-211){\makebox(0,0)[lb]{\smash{\SetFigFont{14}{16.8}{\rmdefault}{\mddefault}{\updefault}$\Delta^{-1}(t^{''})$}}}
\put(10051,-1111){\makebox(0,0)[lb]{\smash{\SetFigFont{14}{16.8}{\rmdefault}{\mddefault}{\updefault}$\Delta^{-1}(s)$}}}
\put(9751,-2611){\makebox(0,0)[lb]{\smash{\SetFigFont{14}{16.8}{\rmdefault}{\mddefault}{\updefault}$\Delta^{-1}(i)$}}}
\put(1126,-1336){\makebox(0,0)[lb]{\smash{\SetFigFont{14}{16.8}{\rmdefault}{\mddefault}{\updefault}$a$}}}
\put(1126,-2011){\makebox(0,0)[lb]{\smash{\SetFigFont{14}{16.8}{\rmdefault}{\mddefault}{\updefault}$x$}}}
\put(1126,-2911){\makebox(0,0)[lb]{\smash{\SetFigFont{14}{16.8}{\rmdefault}{\mddefault}{\updefault}$b$}}}
\put(1126,-4411){\makebox(0,0)[lb]{\smash{\SetFigFont{14}{16.8}{\rmdefault}{\mddefault}{\updefault}$c$}}}
\put(1126,-6136){\makebox(0,0)[lb]{\smash{\SetFigFont{14}{16.8}{\rmdefault}{\mddefault}{\updefault}$b_i$}}}
\put(9226,-4111){\makebox(0,0)[lb]{\smash{\SetFigFont{14}{16.8}{\rmdefault}{\mddefault}{\updefault}$\Delta^{-1}(j)$}}}
\end{picture}

%% file: deltafibers.pstex_t
\begin{picture}(0,0)%
\includegraphics{deltafibers.pstex}%
\end{picture}%
\setlength{\unitlength}{2368sp}%
\begingroup\makeatletter\ifx\SetFigFont\undefined%
\gdef\SetFigFont#1#2#3#4#5{%
  \reset@font\fontsize{#1}{#2pt}%
  \fontfamily{#3}\fontseries{#4}\fontshape{#5}%
  \selectfont}%
\fi\endgroup%
\begin{picture}(12847,7619)(-3524,-7283)
\put(1201,-3211){\makebox(0,0)[lb]{\smash{\SetFigFont{14}{16.8}{\rmdefault}{\mddefault}{\updefault}$2e_9$}}}
\put(1201,-2311){\makebox(0,0)[lb]{\smash{\SetFigFont{14}{16.8}{\rmdefault}{\mddefault}{\updefault}$e_2$}}}
\put(1201,-1561){\makebox(0,0)[lb]{\smash{\SetFigFont{14}{16.8}{\rmdefault}{\mddefault}{\updefault}$e_1$}}}
\put(4651,-2461){\makebox(0,0)[lb]{\smash{\SetFigFont{14}{16.8}{\rmdefault}{\mddefault}{\updefault}$n_1$}}}
\put(6376,-2461){\makebox(0,0)[lb]{\smash{\SetFigFont{14}{16.8}{\rmdefault}{\mddefault}{\updefault}$n_2$}}}
\put(4501,-4261){\makebox(0,0)[lb]{\smash{\SetFigFont{14}{16.8}{\rmdefault}{\mddefault}{\updefault}$n_3$}}}
\put(6376,-4261){\makebox(0,0)[lb]{\smash{\SetFigFont{14}{16.8}{\rmdefault}{\mddefault}{\updefault}$n_4$}}}
\put(4501,-5761){\makebox(0,0)[lb]{\smash{\SetFigFont{14}{16.8}{\rmdefault}{\mddefault}{\updefault}$n_5$}}}
\put(6376,-5761){\makebox(0,0)[lb]{\smash{\SetFigFont{14}{16.8}{\rmdefault}{\mddefault}{\updefault}$n_6$}}}
\put(1201,-4861){\makebox(0,0)[lb]{\smash{\SetFigFont{14}{16.8}{\rmdefault}{\mddefault}{\updefault}$2e_6$}}}
\put(1201,-6361){\makebox(0,0)[lb]{\smash{\SetFigFont{14}{16.8}{\rmdefault}{\mddefault}{\updefault}$2e_4$}}}
\put(8476,-286){\makebox(0,0)[lb]{\smash{\SetFigFont{14}{16.8}{\rmdefault}{\mddefault}{\updefault}$B$}}}
\put(-899,-3361){\makebox(0,0)[lb]{\smash{\SetFigFont{14}{16.8}{\rmdefault}{\mddefault}{\updefault}$\delta$}}}
\put(-3224,-3286){\makebox(0,0)[lb]{\smash{\SetFigFont{14}{16.8}{\rmdefault}{\mddefault}{\updefault}$a$}}}
\put(-3224,-5086){\makebox(0,0)[lb]{\smash{\SetFigFont{14}{16.8}{\rmdefault}{\mddefault}{\updefault}$b$}}}
\put(-3224,-6661){\makebox(0,0)[lb]{\smash{\SetFigFont{14}{16.8}{\rmdefault}{\mddefault}{\updefault}$c$}}}
\put(-3224,-2236){\makebox(0,0)[lb]{\smash{\SetFigFont{14}{16.8}{\rmdefault}{\mddefault}{\updefault}$b_2$}}}
\put(-3224,-1111){\makebox(0,0)[lb]{\smash{\SetFigFont{14}{16.8}{\rmdefault}{\mddefault}{\updefault}$b_1$}}}
\put(-3524,-136){\makebox(0,0)[lb]{\smash{\SetFigFont{14}{16.8}{\rmdefault}{\mddefault}{\updefault}$\cp{1}$}}}
\end{picture}

%% file: intersect1.pstex_t
\begin{picture}(0,0)%
\includegraphics{intersect1.pstex}%
\end{picture}%
\setlength{\unitlength}{2368sp}%
\begingroup\makeatletter\ifx\SetFigFont\undefined%
\gdef\SetFigFont#1#2#3#4#5{%
  \reset@font\fontsize{#1}{#2pt}%
  \fontfamily{#3}\fontseries{#4}\fontshape{#5}%
  \selectfont}%
\fi\endgroup%
\begin{picture}(3162,3915)(226,-5911)
\put(826,-5911){\makebox(0,0)[lb]{\smash{\SetFigFont{14}{16.8}{\rmdefault}{\mddefault}{\updefault}$n_1$}}}
\put(2101,-5911){\makebox(0,0)[lb]{\smash{\SetFigFont{14}{16.8}{\rmdefault}{\mddefault}{\updefault}$o_1$}}}
\put(226,-4336){\makebox(0,0)[lb]{\smash{\SetFigFont{14}{16.8}{\rmdefault}{\mddefault}{\updefault}$e_6$}}}
\put(2701,-3961){\makebox(0,0)[lb]{\smash{\SetFigFont{14}{16.8}{\rmdefault}{\mddefault}{\updefault}$e$}}}
\end{picture}

%% file: intersect2.pstex_t
\begin{picture}(0,0)%
\includegraphics{intersect2.pstex}%
\end{picture}%
\setlength{\unitlength}{2368sp}%
\begingroup\makeatletter\ifx\SetFigFont\undefined%
\gdef\SetFigFont#1#2#3#4#5{%
  \reset@font\fontsize{#1}{#2pt}%
  \fontfamily{#3}\fontseries{#4}\fontshape{#5}%
  \selectfont}%
\fi\endgroup%
\begin{picture}(2192,3915)(226,-5911)
\put(226,-4336){\makebox(0,0)[lb]{\smash{\SetFigFont{14}{16.8}{\rmdefault}{\mddefault}{\updefault}$e_6$}}}
\put(826,-5911){\makebox(0,0)[lb]{\smash{\SetFigFont{14}{16.8}{\rmdefault}{\mddefault}{\updefault}$n_5$}}}
\put(2101,-5911){\makebox(0,0)[lb]{\smash{\SetFigFont{14}{16.8}{\rmdefault}{\mddefault}{\updefault}$o_5$}}}
\put(301,-3061){\makebox(0,0)[lb]{\smash{\SetFigFont{14}{16.8}{\rmdefault}{\mddefault}{\updefault}$e$}}}
\end{picture}

%% file: latex.bbl
\begin{thebibliography}{OPR}

\bibitem{hw1} 
P.~Ho\v rava and E.~Witten,
\newblock  Heterotic and Type I String Dynamics from Eleven Dimensions,
\newblock  {\em Nucl. Phys.} {B460} (1996) 506.

\bibitem{hw2}
P.~Ho\v rava and E.~Witten,
\newblock  Eleven-Dimensional Supergravity on a Manifold with Boundary,
\newblock {\em Nucl. Phys.} {B475} (1996) 94.

\bibitem{w1}
E.~Witten,
\newblock Strong Coupling Expansion Of Calabi-Yau Compactification,
\newblock {\em Nucl. Phys.} {B471} (1996) 135.



\bibitem{losw1}
A.~Lukas, B.~A. Ovrut, K.S.~Stelle and D.~Waldram,
\newblock   The Universe as a Domain Wall,
\newblock {\em Phys.Rev.} D59 (1999) 086001.

\bibitem{losw2}
A.~Lukas, B.~A. Ovrut, K.S. Stelle and D. Waldram,
\newblock  Heterotic M--theory in Five Dimensions,
\newblock  {\em Nucl.Phys.} B552 (1999) 246-290.

\bibitem{low1}
A.~ Lukas, B.~ Ovrut, and D.~Waldram,
\newblock Non-standard embedding and five-branes in heterotic M-Theory,
\newblock {\em Nucl.Phys.} B552 (1999) 246-290.

\bibitem{adhm}
M. F.~Atiyah, N.J.~Hitchin, V. G.~Drinfeld, Y. I.~Manin,
\newblock Construction of instantons
\newblock {\em Phys. Lett.} A65 (1996) 185.

\bibitem{fmw1}
R.~Friedman, J.~Morgan, E.~Witten,
\newblock Vector Bundles And F Theory,
\newblock {\em Commun.Math.Phys.} 187 (1997) 679-743

\bibitem{fmw2}
R.~Friedman, J.~Morgan, E.~Witten
\newblock Vector Bundles over Elliptic Fibrations,
\newblock alg-geom/9709029.

\bibitem{low2}
R.~Donagi, A.~ Lukas, B.~ Ovrut, and D.~Waldram,
\newblock Non-Perturbative Vacua and Particle Physics in M-Theory,
\newblock {\em JHEP} 9905 (1999) 018.


\bibitem{dlow}
R.~Donagi, A.~ Lukas, B.~ Ovrut, and D.~ Waldram,
\newblock Holomorphic Vector Bundles and Non-Perturbative Vacua in M-Theory,
\newblock {\em  JHEP}, 9906 (1999) 034.




\bibitem{dopw-i}
R.~Donagi, B.~Ovrut, T.~Pantev, and D.~Waldram,
\newblock Spectral involutions on rational elliptic surfaces,
\newblock {\em Adv.Theor.Math.Phys.} 5 (2002) 499-561,  math.AG/0008011.

\bibitem{dopw-ii}
R.~Donagi, B.~Ovrut, T.~Pantev, and D.~Waldram,
\newblock Standard-{M}odel bundles,
\newblock {\em  Adv.Theor.Math.Phys.} 5 (2002) 563-615, math.AG/0008010.

\bibitem{dopw-iii}
R.~Donagi, B.~Ovrut, T.~Pantev, and D.~Waldram,
\newblock Standard-{M}odel bundles on non-simply connected {C}alabi-{Y}au threefolds,
\newblock {\em JHEP}, 0108 (2001) 053,   hep-th/0008008.

\bibitem{dopw-iv}
R.~Donagi, B.~Ovrut, T.~Pantev, and D.~Waldram,
\newblock Standard Models from Heterotic M-theory,
\newblock {\em Adv.Theor.Math.Phys.}, 5 (2002) 93-137.

\bibitem{schoenCY}
C.~Schoen,
\newblock On fiber products of rational elliptic surfaces with
section,
\newblock {\em Math. Z.}, 197(2):177--199, 1988.


\bibitem{w2}
E.~Witten,
\newblock Symmetry Breaking Patterns in Superstring Theory,
\newblock {\em Nucl. Physics} B258 (1985) 75-100.

\bibitem{w3}
E.~Witten,
\newblock private communication.

\bibitem{rt}
R.~Thomas,
\newblock Examples of bundles on Calabi-Yau 3-folds for string theory compactifications,
\newblock {\em Adv.Theor.Math.Phys.} 4 (2000) 231-247.

\bibitem{opr}
B.~Ovrut, T.~Pantev, R.~Reinbacher,
\newblock Torus-Fibered Calabi-Yau Threefolds with Non-Trivial Fundamental Group,
\newblock UPR 1015-T, hep-th/0212221.



\bibitem{dopr-i}
R.~Donagi, B.~Ovrut, T.~Pantev, R.~Reinbacher,
\newblock Holomorphic Vector Bundles, Calabi-Yau Spaces with Non-Trivial Homotopy and the Standard Model,
\newblock {in preparation}.

\bibitem{dopr-iii}
R.~Donagi, B.~Ovrut, T.~Pantev, R.~Reinbacher.
\newblock Standard-like Models, Nucleon Decay and SU(4) Instantons,
\newblock {in preparation}.

\bibitem{dlor}
R.~Donagi,  A.~Lukas, B.~Ovrut,, R.~Reinbacher.
Symmetry breaking pattern by Wilson loops in heterotic M-theory,
\newblock {in preparation}.

\bibitem{dopr-ii}
R.~Donagi, B.~Ovrut, T.~Pantev, R.~Reinbacher,
\newblock Stable Vector Bundles on Non-Simply Connected Calabi-Yau Spaces,
\newblock{UPR 1018-T}.

\bibitem{gh}
Griffiths \& Harris,
\newblock Principles of Algebraic Geometry,
\newblock {\em Wiley Classics Library Edition}, Published 1994.


\bibitem{kodaira-casIII}
K.~Kodaira,
\newblock On compact analytic surfaces,
\newblock {\em Ann. of Math.}, 78:1--40, 1963.


\end{thebibliography}
